\documentclass[12pt]{article}
\usepackage{epsfig,amsmath,amsfonts,amssymb,makeidx,ifthen}
\newtheorem{theorem}{Theorem}[section]
\newtheorem{lemma}[theorem]{Lemma}

\makeatletter
\@addtoreset{equation}{section}
\makeatother

\makeatletter
\long\def\@makecaption#1#2{{\small
\advance\leftskip1cm
\advance\rightskip1cm
\vskip\abovecaptionskip
\sbox\@tempboxa{#1: #2}%
\ifdim \wd\@tempboxa >\hsize
 #1: #2\par
\else
\global \@minipagefalse
\hb@xt@\hsize{\hfil\box\@tempboxa\hfil}%
\fi
\vskip\belowcaptionskip}}
\makeatother
\def\eq#1\en{\begin{equation}#1\end{equation}}  
\def\eqa#1\ena{\begin{align}#1\end{align}}
\def\eqg#1\eng{\begin{gather}#1\end{gather}}
\newcommand{\lb}[1]{\label{e:#1}}
\newcommand{\rlb}[1]{\eqref{e:#1}} 
\newcommand{\nl}{\notag\\}

\newcommand{\normi}[1]{\bigl\Vert#1\bigr\Vert_\infty}

\newcommand{\sqbk}[1]{\left[#1\right]}

\newcommand{\sbkt}[1]{\langle#1\rangle}
\newcommand{\bbkt}[1]{\bigl\langle#1\bigr\rangle}

\newcommand{\sumtwo}[2]%
{\mathop{\sum_{#1}}_{#2}}
\newcommand{\sumthree}[3]%
{\mathop{\mathop{\sum_{#1}}_{#2}}_{#3}}
\newcommand{\sumfour}[4]%
{\mathop{\mathop{\mathop{\sum_{#1}}_{#2}}_{#3}}_{#4}} 
\newcommand{\mintwo}[2]%
{\mathop{\min_{#1}}_{#2}}
\newcommand{\maxtwo}[2]%
{\mathop{\max_{#1}}_{#2}}
\newcommand{\maxthree}[3]%
{\mathop{\mathop{\max_{#1}}_{#2}}_{#3}}
\newcommand{\limtwo}[2]%
{\mathop{\lim_{#1}}_{#2}}
\newcommand{\suptwo}[2]%
{\mathop{\sup_{#1}}_{#2}}
\newcommand{\supthree}[3]%
{\mathop{\mathop{\sup_{#1}}_{#2}}_{#3}}
\newcommand{\supfour}[4]%
{\mathop{\mathop{\mathop{\sup_{#1}}_{#2}}_{#3}}_{#4}} 
\newcommand{\inftwo}[2]%
{\mathop{\inf_{#1}}_{#2}}
\newcommand{\infthree}[3]%
{\mathop{\mathop{\inf_{#1}}_{#2}}_{#3}}
\newcommand{\inffour}[4]%
{\mathop{\mathop{\mathop{\inf_{#1}}_{#2}}_{#3}}_{#4}} 

\newcommand\calB{{\cal B}}

\newcommand\calD{{\cal D}}

\newcommand\calS{{\cal S}}
\newcommand\calT{{\cal T}}

\newcommand{\tiS}{\tilde{S}}

\newcommand{\bsp}{\boldsymbol{p}}
\newcommand{\bsq}{\boldsymbol{q}}

\newcommand{\bsv}{\boldsymbol{v}}

\newcommand{\bsdelta}{\boldsymbol{\delta}}

\newcommand{\bsrho}{\boldsymbol{\rho}}

\newcommand{\sfA}{\mathsf{A}}

\newcommand{\sfF}{\mathsf{F}}

\newcommand{\sfR}{\mathsf{R}}

\newcommand{\sfW}{\mathsf{W}}

\newcommand{\bbR}{\mathbb{R}}

\newcommand{\up}{\uparrow}
\newcommand{\dn}{\downarrow}

\newcommand{\Di}{\mathit{\Delta}}
\newcommand{\qedm}{\rule{1.5mm}{3mm}}
\newcommand{\fp}[2]{\dfrac{\partial #1}{\partial #2}}

\newcommand{\fd}[2]{\dfrac{d#1}{d#2}}

\newcommand{\fdt}[1]{\fd{#1}{t}}

\newcommand{\Ra}{R^\alpha}
\newcommand{\bn}{{(\beta,\nu)}}
\newcommand{\Rbn}{R^\bn}
\newcommand{\Hn}{H^\nu}
\newcommand{\la}{\lambda^\alpha}
\newcommand{\thetaa}{\theta^\alpha}
\newcommand{\psia}{\psi^\alpha}
\newcommand{\Rat}{R^{\alpha(t)}}
\newcommand{\lat}{\lambda^{\alpha(t)}}
\newcommand{\tao}{\tau_\mathrm{o}}
\newcommand{\hatn}{{\hat{\nu}}}
\newcommand{\hatnd}{{\hat{\nu}^\dagger}}
\newcommand{\hata}{{\hat{\alpha}}}
\newcommand{\hatad}{{\hat{\alpha}^\dagger}}
\newcommand{\ca}{{(\alpha)}}
\newcommand{\xty}{{x\to y}}
\newcommand{\ytx}{{y\to x}}
\newcommand{\hatx}{\hat{x}}
\newcommand{\hatxd}{\hat{x}^\dagger}
\newcommand{\Dx}{\calD\hatx}
\newcommand{\sts}{\mathrm{st}}
\newcommand{\too}{\to}
\newcommand{\ttt}{{t\in[-\tau,\tau]}}
\newcommand{\lti}{\lim_{\tau\up\infty}}
\newcommand{\tR}{\tilde{\sfR}}
\newcommand{\tF}{\tilde{\sfF}}
\newcommand{\vone}{\vec{1}}
\newcommand{\vxi}{\vec{\xi}}
\newcommand{\bsphi}{\boldsymbol{\varphi}}
\newcommand{\vdelta}{\vec{\delta}}

\newcommand{\oesd}{O(\epsilon^2\delta)}

\newcommand{\beot}{{(\beta_1,\beta_2)}}
\newcommand{\Db}{\Di\beta}
\newcommand{\vecr}{\vec{r}}
\newcommand{\vecp}{\vec{p}}

\begin{document}
\noindent
{\bf\Large Exact equalities and thermodynamic relations \\for nonequilibrium steady states}
\par\bigskip

\noindent
Teruhisa S. Komatsu\footnote{
Laboratory for Computational Molecular Design, RIKEN QBiC, Kobe 650-0047, Japan
}, 
Naoko Nakagawa\footnote{
College of Science, 
Ibaraki University, Mito, Ibaraki 310-8512, Japan
}, 
Shin-ichi Sasa\footnote{
Department of Physics, Kyoto University, Kyoto, 606-8502, Japan
},
and Hal Tasaki\footnote{
Department of Physics, Gakushuin University, Mejiro, Toshima-ku, 
Tokyo 171-8588, Japan
}

\begin{abstract}
We study thermodynamic operations which  bring a nonequilibrium steady state (NESS) to another NESS in physical systems under nonequilibrium conditions.
We model the system by a suitable Markov jump process, and treat thermodynamic operations as protocols according to which the external agent varies parameters of the Markov process.
Then we prove, among other relations, a NESS version of the Jarzynski equality and the extended Clausius relation.
The latter can be a starting point of thermodynamics for NESS.
We also find that the corresponding nonequilibrium entropy has a microscopic representation in terms of symmetrized Shannon entropy in systems where the microscopic description of states involves ``momenta''. 
All the results in the present paper are mathematically rigorous.
\end{abstract}


\tableofcontents

\section{Introduction}
\subsection{Background and motivation}

General Properties of physical systems in thermal equilibrium are relatively well understood both from physical and mathematical points of view.
Thermodynamics characterizes macroscopic properties of equilibrium states, and poses strong constraints on possible transitions between equilibrium sates, especially when an outside agent makes operation to the system.
Statistical mechanics provides a probabilistic description of equilibrium states based on the microscopic mechanical description of the system.

To develop similar universal theories for systems out of equilibrium has been a major remaining challenge in theoretical physics.
Among various classes of nonequilibrium states, nonequilibrium steady states (which we shall abbreviated as NESS throughout the paper), which have no macroscopic changes but have nonvanishing flows, may be a promising ground for developing such theories.

Looking back into the history of equilibrium physics, we find that the thermodynamic definition  of entropy (and free energy) in terms of thermodynamic operation was a fundamental starting point.
The notion and the properties of the entropy and the free energy were  essential guides for the later development of equilibrium statistical mechanics.
If we trust the analogy to the history, one possible route toward universal theories for NESS may be to start from a thermodynamics for NESS and pin down relevant thermodynamic functions.
In \cite{OP}, Oono and Paniconi made a proposal of an operational thermodynamics for NESS and coined the term ``steady state thermodynamics" (SST).

We should point out however that the notion of entropy for NESS is much more subtle than one might imagine in the beginning.
One reason is that the entropy in equilibrium physics plays several essentially different roles; it is a thermodynamic function whose derivatives correspond to physically observable quantities, it is a quantitative measure of which adiabatic process is possible and which is not, it is also a large deviation functional governing the fluctuation of physical quantities.
There is a possibility that this ``degeneracy'' is an accident observed only in equilibrium states, and the degeneracy is immediately lifted when one goes out of equilibrium.
If this is the case we shall encounter more than one ``nonequilibrium entropies'' each of which characterizing different physical aspect of a NESS.
It will then be important to clarify which physics of NESS is represented by which extension of entropy.

In our own attempt to develop SST and to extend the notion of entropy to NESS \cite{KNST1,KNST2}, we have concentrated on the aspect which gave  birth to the concept of entropy, namely, the Clausius relation in thermodynamics.
We started from a microscopic description of a heat conducting NESS, and showed that a very natural generalization of the Clausius relation, in which the heat is replaced with its ``renormalized'' counterpart, is valid  when the ``order of nonequilibrium'' $\epsilon$ is sufficiently small.
This was a realization of the early phenomenological discussions by Landauer \cite{Landauer} and by Oono and Paniconi \cite{OP}, and an extension of the similar result by Ruelle \cite{Ruelle} for models with Gaussian thermostat.
We also found that, in systems where microscopic states have no time-reversal symmetry (i.e., the microscopic description of states involves ``momenta''), the microscopic representation of the entropy differs from the traditional  Gibbs-Shannon form, and requires further symmetrization with respect to the time-reversal transformation.

This approach to SST was later extended to quantum systems \cite{SaitoTasaki}.
A geometric interpretation of the thermodynamic relations and corresponding exact relations were discussed in \cite{SagawaHayakwa,YugeSagawaSugitaHayakawa} both for classical and quantum systems.

Other schemes for SST, which are distinct from those in \cite{KNST1,KNST2}, have been proposed  \cite{HatanoSasa,BGJLL2012,BGJLL2013,MaesNetocny2012,Sasa2013}.
See the end of section~\ref{s:SSTexample} for details.
See also \cite{SasaTasaki} for an earlier attempt at approaching SST from a phenomenological point of view, and \cite{PradhanAmannSeifert2010,PradhanAmannSeifert2011,DickmanMotai} for discussions about the ``zeroth law'' in SST.
Closely related problem of heat capacity in NESS is discussed in \cite{Boksenbojm,Mandal}.
See \cite{Nakagawa2014} for a unified treatment of thermodynamics in NESS and adiabatic pumping in equilibrium.

Among other various promising attempt at discussing entropy (or a related quantity) in NESS, let us refer to the macroscopic fluctuation theory developed by Bertini, De Sole, Gabrielli, Jona-Lasinio, and Landim 
\cite{BSGJLL01,BSGJLL06,BSGJLL09,JonaLasinio10},
exact solution of the large deviation functional by Derrida, Lebowitz, and Speer \cite{DLS01,DLS03},
the proposal of the additivity principle by Bodineau and Derrida \cite{BD},
and the recent interesting proposal based on adiabatic accessibility by Lieb and Yngvason \cite{LiebYngvason}.
As for closely related approaches based on ``fluctuation theorems'' see, e.g.,  the recent review \cite{Seifert12}.
\bigskip

In the present paper, which is the first mathematical paper in our series of works on SST, we  treat a general class of Markov jump processes that model  nonequilibrium systems, and prove physically important relations including the NESS version of the Jarzynski equality and the extended Clausius relation.
Although most of the results have been announced before, they were all derived heuristically.
We here present mathematically rigorous results for the first time.

Let us describe the organization of the present paper\footnote{
This part can be read as a summary of the whole paper.
}.

The following section~\ref{s:SSTexample} has been prepared for the readers who are not familiar with our approach (and other related approaches) to nonequilibrium thermodynamics.
We shall motivate our study by briefly describing the standard Clausius relation in equilibrium thermodynamics in a simple setting, and explaining the difficulties one encounters when trying to extending it naively to NESS.
We then describe our scheme of ``renormalization'' and introduce the  extended Clausius relation for NESS.
Finally we compare our approach with other proposals of thermodynamics for NESS.

In section~\ref{s:Setupanddef}, we shall give almost complete definitions necessary in the present paper.
In sections~\ref{s:Markov} and \ref{s:path}, we define the Markov jump process and the corresponding description in terms of paths.
These definitions are standard.
Of particular importance are the notion of protocol $\hata$ and the expectation value $\sbkt{f}^{\hata}_{\sts\too}$ defined in \rlb{fss}.
In section~\ref{s:entropyprod}, we define essential ``thermodynamic'' quantities, namely, the entropy production $\Theta$, its nonequilibrium part $\Psi$, and the work $W$.  We also discuss the experimental measurability of these quantities.
We shall discuss some examples in section~\ref{s:examples}.

In section~\ref{s:JarNESS} we discuss the Jarzynski type equality \rlb{SSTJ1} that holds for thermodynamic operations between two NESS.
The equality is exact and rigorous, and will be the basis of our main result, namely the extended Clausius relation.
Let us stress that our Jarzynski type equality \rlb{SSTJ1} for NESS is distinct from existing exact equalities for general stochastic processes in that it contains only (almost) measurable thermodynamic type quantities.
It is challenging to design experimental verification of the equality \rlb{SSTJ1} with modern techniques in calorimetry.

In section~\ref{s:SST}, we shall be heuristic, and describe thermodynamic relations that can be derived from the rigorous equality \rlb{SSTJ1}.
In section~\ref{s:exClEnt}, we start from the most important extended Clausius relation, and also discuss a different representation of the relation in terms of the excess entropy production.
We further discuss higher order ``thermodynamic'' relations in section~\ref{s:SSThigher}, and present a heuristic estimate of error terms in section~\ref{s:heuristicerror}.

In section~\ref{s:SSTrigorous}, which is a core of the present paper, we discuss rigorous versions of the thermodynamic relations without going into the proofs.
After fixing the class of models for simplicity, we state Theorem~\ref{t:SandS} which allows us to identify (with a certain precision) our nonequilibrium entropy with the Shannon entropy.
Then in Theorems~\ref{t:SSTstep} and \ref{t:SSTsmooth}, we state the extended Clausius relation and its higher order generalizations for the step protocol and the quasi-static protocol, respectively.
The extended Clausius relation written in terms of the excess entropy production is stated in Theorem~\ref{t:SSTexcess}.
Finally in Theorem~\ref{t:ClausiusIneq}, we state an inequality corresponding to the extended Clausius relation.

Sections~\ref{s:proof1}, \ref{s:matrix}, \ref{s:KNS}, and \ref{s:ClIneq} are devoted to the proofs of the theorems.

In section~\ref{s:proof1}, we present arguments based on time-reversal symmetry to prove exact equalities discussed in section~\ref{s:JarNESS}.
The arguments are basically standard, but our Jarzynski type equality for NESS is proved by using a new statement which we call ``splitting Lemma'' (Lemma~\ref{l:split}).

In section~\ref{s:matrix} we present totally different approach based on the method of modified rate matrix.
This method is used in section~\ref{s:splitproof} to prove the splitting lemma, and in section~\ref{s:SSTproof} to justify the heuristic estimates in section~\ref{s:heuristicerror} of the error in the thermodynamic relations.
This completes the proof of the extended Clausius relation and the related relations stated in Theorems~\ref{t:SSTstep}, \ref{t:SSTsmooth}, and \ref{t:SSTexcess}.

In section~\ref{s:KNS}, we shall use the results from the previous sections to prove Theorem~\ref{t:KNrep} about a useful and suggestive representation (first written down by two of us, T.S.K. and N.N.) of the probability distribution of NESS.
Then this representation is used to prove Theorem~\ref{t:SandS} about the nonequilibrium entropy.

In section~\ref{s:ClIneq}, we prove Theorem~\ref{t:ClausiusIneq} about the extended Clausius {\em in}\/equality for NESS.
The proof makes use of the standard argument based on the relative entropy and a rigorous version of the linear response formula stated as Lemma~\ref{l:LRrep}.

In the final section~\ref{s:Ssym}, we discuss a slightly different class of models which include ``momenta''.
We show that essentially all the results (except for the extended Clausius {\em in}\/equality) automatically extend to this situation if one replaces the Shannon entropy with a new quantity  \rlb{Ssym} called the symmetrized Shannon entropy.
The symmetrized Shannon entropy has a very suggestive form and might be a key for further understanding of the essential properties of NESS in systems with momenta.
In section~\ref{s:toy}, we discuss a simple toy model which  illustrates the need of the symmetrized Shannon entropy.

\subsection{SST in a typical example}
\label{s:SSTexample}
Here we shall briefly discuss the essence of SST, i.e., the extended Clausius relation in the simplest example of heat conducting system.
We also try to place our work in the broader context of thermodynamics and statistical mechanics by discussing relevant background.

\begin{figure}[btp]
\begin{center}
\includegraphics[width=10cm]{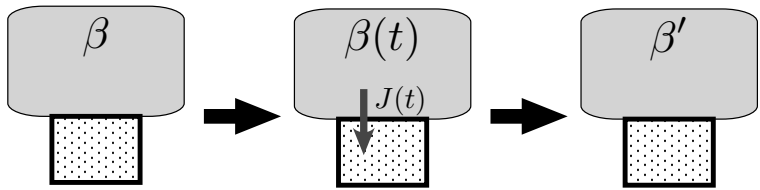}
\end{center}
\caption[dummy]{
Thermodynamic operation in an equilibrium state.
The bigger gray box denotes the heat bath whose inverse temperature can be controlled, and the smaller box filled with dots denote the system that we are interested in.
We start from the equilibrium state with the inverse temperature $\beta$, and slowly change the inverse temperature to $\beta'$.
The inverse temperature of the bath and the heat flux from the bath to the system at time $t$ are denoted as $\beta(t)$ and $J(t)$, respectively.
The total entropy production in the bath $-\int_{-\tau}^\tau dt\,\beta(t)J(t)$ plays an essential role in the Clausius relation \rlb{EqCl}.
}
\label{fig:OPEQ}
\end{figure}

\paragraph*{Clausius relation for operation between equilibrium states:}
To motivate our extended Clausius relation, we first review the standard Clausius relation, which is the starting point of equilibrium thermodynamics.
Consider a physical system (which can be basically anything) attached to a single heat bath whose temperature can be controlled.
See Fig.~\ref{fig:OPEQ}.
When the inverse temperature of the bath is fixed at $\beta$, the system settles to the equilibrium state corresponding to $\beta$ after a sufficiently long time.
Recall that a physical system in equilibrium exhibits no macroscopic changes, and has no macroscopic flows (of, e.g., matter or energy).

We next consider a thermodynamic operation.
We start from a situation where the inverse temperature of the bath is $\beta$ and the system is in the corresponding equilibrium.
Then we change the inverse temperature of the bath according to a protocol  fixed in advance, i.e., a smooth function $\beta(t)$ of time $\ttt$ where $\beta(-\tau)=\beta$ and $\beta(\tau)=\beta'$.
We shall assume that $\tau$ is large and $\beta(t)$ varies slowly.
Let $J(t)$ be the heat flux (the energy that flows within a unit time) from the bath to the system.
Then the well known Clausius relation is
\eq
S(\beta')-S(\beta)\simeq\int_{-\tau}^\tau dt\,\beta(t)J(t),
\lb{EqCl}
\en
where  $S(\beta)$ is the entropy of the system in the equilibrium state corresponding to $\beta$.
(More precisely, the entropy is a function of the equilibrium state.)
The relation \rlb{EqCl} becomes an exact equality in the limit  $\tau\up\infty$ where $\beta(t)$ varies infinitesimally slowly.
The equality can be proved mathematically in various setting.

We note that $(-1)$ times the right-hand side of \rlb{EqCl} is interpreted as the total entropy production in the heat bath.
To see this, take a short time interval from $t$ to $t+\Di t$.
The energy (heat) that flows into the bath during this interval is $\Di Q=-\Di t\,J(t)$.
Then the corresponding increase (or production) of entropy in the bath is given by the standard relation $\Di S_{\rm bath}=\beta(t)\,\Di Q=-\Di t\,\beta(t)\,J(t)$, which becomes the minus of \rlb{EqCl} after integration over the whole process.

Although the entropy $S(\beta)$ was introduced above as a purely thermodynamic (or macroscopic) quantity, there is a neat expression in terms of microscopic probability distribution.
If one represents the equilibrium state in terms of the canonical distribution $\rho^\beta_x=e^{-\beta H_x}/Z(\beta)$ (see section~\ref{s:Markov} for the notation), the same entropy is written as
\eq
S(\beta)=-\sum_x\rho^\beta_x\log\rho^\beta_x,
\lb{Sstandard}
\en
where the right-hand side is nothing but the Shannon entropy of the probability distribution $\rho^\beta_x$.

\paragraph*{Extended Clausius relation for operation between NESS:}
In our approach to thermodynamics for NESS, we wish to focus on possible extensions of the Clausius relation \rlb{EqCl} and the expression \rlb{Sstandard} of the entropy.
The hope is that proper extensions might be a starting point of a full-fledged thermodynamics.

\begin{figure}[btp]
\begin{center}
\includegraphics[width=10cm]{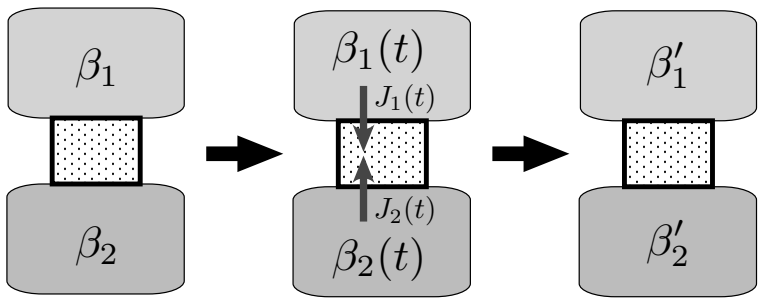}
\end{center}
\caption[dummy]{
Thermodynamic operation in a NESS (nonequilibrium steady state).
There are two heat baths and the system of interest in between them.
We fix the inverse temperatures of the baths to $\beta_1$ and $\beta_2$, and assume that the system is in the corresponding NESS.
Then we slowly change the inverse temperatures of the baths to $\beta'_1$ and $\beta'_2$.
The inverse temperatures of the baths at time $t$ are denoted as $\beta_1(t)$ and $\beta_2(t)$, and the heat flux from the baths to the system as $J_1(t)$ and $J_2(t)$.

Now the total entropy production (in the baths) $-\bigl\{\int_{-\tau}^\tau dt\,\beta_1(t)J_1(t)+\int_{-\tau}^\tau dt\,\beta_2(t)J_2(t)\bigr\}$ is proportional to the total time $2\tau$, and cannot appear in a thermodynamic relation as it is.
In fact our extended Clausius relation \rlb{exCl01} is written in terms of its ``renormalized'' counterpart.
}
\label{fig:OPNESS}
\end{figure}

To be specific, we focus on a  NESS in a  heat conducting system, which is a typical nonequilibrium setting.
Consider a system which is attached to two large heat baths whose temperatures can be controlled.
See Fig.~\ref{fig:OPNESS}.

Suppose first that the inverse temperatures of the baths are fixed at $\beta_1$ and $\beta_2$, respectively.
It is expected that, after a sufficiently long time, the system settles to a stationary state which has a steady temperature gradient and a constant heat current through it.
Such a state, which has no macroscopically observable changes, but has a nonvanishing flow of energy, is a typical example of NESS.

As in the case of equilibrium, we consider a thermodynamic operation to NESS.
We start from the situation where the two heat baths have fixed inverse temperatures $\beta_1$ and $\beta_2$, and the system is in the corresponding NESS.
Then we change the inverse temperatures of the baths according to a fixed protocol, i.e.,  functions $\beta_1(t)$ and $\beta_2(t)$ of time $\ttt$.
We write $\beta_1=\beta_1(-\tau)$, $\beta_2=\beta_2(-\tau)$, $\beta_1'=\beta_1(\tau)$, and $\beta_2'=\beta_2(\tau)$.

We wish to ask whether a relation analogous to the Clausius relation \rlb{EqCl} holds in this setting.
Since there are two heat baths, we need to consider two heat currents separately.
By $J_k(t)$, where $k=1,2$, we denote the heat flux from the $k$-th bath to the system at time $t$.
Then a naive analogue to \rlb{EqCl} is
\eq
S(\beta_1',\beta_2')-S(\beta_1,\beta_2)\stackrel{?}{\simeq}
\sum_{k=1,2}\int_{-\tau}^\tau dt\,\beta_k(t)\,J_k(t),
\lb{exCl00}
\en
where the minus of the right-hand side is the total entropy production in the two heat baths.
Here $S(\beta_1,\beta_2)$ is a certain function of two inverse temperatures $\beta_1$, $\beta_2$, which should be called the nonequilibrium entropy.

But it turns out that a relation like \rlb{exCl00} can never be valid.
This is most clearly seen by examining the right-hand side in the case where both $\beta_1(t)=\beta_1$ and $\beta_2(t)=\beta_2$ are independent of $t$.
Suppose that $\beta_1<\beta_2$.
In the NESS characterized by the two inverse temperatures $\beta_1$ and $\beta_2$, there is a  steady heat flux $J_\mathrm{st}>0$ from the bath 1 to the bath 2 through the system.
We thus have $J_1(t)=J_\mathrm{st}$ and $J_2(t)=-J_\mathrm{st}$ for any $\ttt$.
The right-hand side of \rlb{exCl00} is thus equal to $2\tau(\beta_1-\beta_2)J_\mathrm{st}$, which is negative and grows proportionally with $\tau$.
The left-hand side, on the other hand, is vanishing since $\beta'_k=\beta_k$.
The relation  \rlb{exCl00} is clearly invalid.

More generally, fix the initial inverse temperature $\beta_k$ and the final  inverse temperature $\beta_k'$ (for $k=1,2$), and take reference functions $\tilde{\beta}_k(s)$ of $s\in[-1,1]$ such that $\tilde{\beta}_k(-1)=\beta_k$ and  $\tilde{\beta}_k(1)=\beta'_k$.
For a given $\tau>0$, we choose our protocol as $\beta_k(t)=\tilde{\beta}_k(t/\tau)$.
Note that, when $\tau$ becomes large, the right-hand side of \rlb{exCl00} diverges (roughly) proportionally to $\tau$ because there always is a heat current going through the system.
On the other hand the left-hand side is independent of $\tau$, because $S(\beta_1,\beta_2)$ should be a function of the two inverse temperatures.
We again conclude that the relation  \rlb{exCl00} cannot be valid.

\begin{figure}[btp]
\begin{center}
\includegraphics[width=8cm]{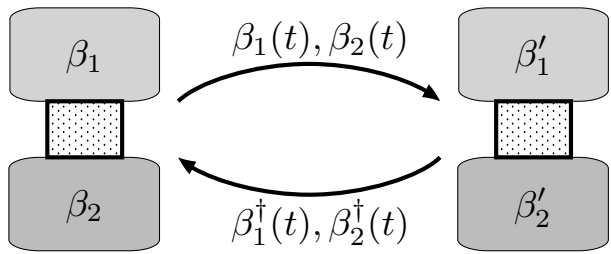}
\end{center}
\caption[dummy]{
We consider the reverse protocol in which the inverse temperatures are varied from $(\beta'_1,\beta'_2)$ back to $(\beta_1,\beta_2)$ in exactly the reverse manner.
Note that we are {\em not}\/ proposing reverse time-evolutions of any kind, but considering the ordinary time-evolution under the reverse protocol.
By taking the difference between the total entropy productions (in the baths) in the original and in the reverse protocols, we get a finite quantity which appears in the extended Clausius relation \rlb{exCl01}.
}
\label{fig:RevProt}
\end{figure}

\paragraph*{Extended Clausius relation:}
We need to find a way to ``renormalize'' the divergence in the right-hand side  of \rlb{exCl00} to get a finite quantity.
One strategy is to introduce the reverse operation as follows.
See Fig.~\ref{fig:RevProt}.
We start from the situation where the two heat baths have fixed inverse temperatures $\beta'_1$ and $\beta'_2$, and the system is in the corresponding NESS.
Then we change the inverse temperatures of the baths according to the reverse protocol defined by the functions $\beta^\dagger_k(t):=\beta_k(-t)$ for $k=1,2$.
Again we denote by $J^\dagger_k(t)$, where $k=1,2$, the heat flux from the $k$-th bath to the system
at time $t$ in this process.
We expect $J^\dagger_k(t)\simeq J_k(-t)$ when the operation is slow enough\footnote{%
To be precise, this is true when there always is a nonvanishing temperature difference.
When $\beta_1(t)=\beta_2(t)$, the currents are very small and we rather have $J^\dagger_k(t)\simeq -J_k(-t)$.
}. But we don't  have  the exact equality $J^\dagger_k(t)= J_k(-t)$ in general since the currents at a given moment may depend on the history of the system.
The subtle difference between $J^\dagger_k(t)$ and $J_k(-t)$ can be a key to understand the nature of NESS.

Since $J^\dagger_k(t)\simeq J_k(-t)$, the total entropy production in the baths  
\newline$-\sum_{k=1,2}\int_{-\tau}^\tau dt\,\beta^\dagger_k(t)\,J^\dagger_k(t)$ for the reverse protocol should diverge as $\tau\up\infty$ in the same manner as that in the original protocol, i.e., the minus of the right-hand side of \rlb{exCl00}.
This observation suggests that their difference $\sum_{k=1,2}\int_{-\tau}^\tau dt\,\beta_k(t)\,J_k(t)-\sum_{k=1,2}\int_{-\tau}^\tau dt\,\beta^\dagger_k(t)\,J^\dagger_k(t)$ may be finite in the limit $\tau\up\infty$, and may play a meaningful role.
This is indeed the case, and we shall prove, for a class of models close to equilibrium, that the extended Clausius relation
\eq
S(\beta_1',\beta_2')-S(\beta_1,\beta_2)\simeq
\frac{1}{2}\sum_{k=1,2}\Bigl\{\int_{-\tau}^\tau dt\,\beta_k(t)\,J_k(t)-
\int_{-\tau}^\tau dt\,\beta^\dagger_k(t)\,J^\dagger_k(t)\Bigr\},
\lb{exCl01}
\en
holds in the limit $\tau\up\infty$.
See Theorem~\ref{t:SSTsmooth} for the precise statement.
Note that the unwanted divergence is properly ``renormalized'' by considering the difference between the total entropy productions (in the baths) in the original and the time-reversed operations.
The same relation can be written by using the notion of excess entropy production as in \cite{KNST1,KNST2}.
See Theorem~\ref{t:SSTexcess}.

The nonequilibrium entropy in \rlb{exCl01} satisfies $S(\beta,\beta)=S(\beta)$, where $S(\beta)$ is the equilibrium entropy.
Likewise the extended Clausius relation \rlb{exCl01} reduces to the original Clausius relation \rlb{EqCl} if the temperatures of the baths are always identical with each other, i.e., $\beta_1(t)=\beta_2(t)$ for any $\ttt$.
We can say that our relation \rlb{exCl01} is a natural extension of the original Clausius relation \rlb{EqCl} to operations between NESS.
We also stress that the right-hand side of \rlb{exCl01} can be, in principle,  measured experimentally;
one needs to perform a pair of experiments for the original and the reverse protocols, and measure the heat currents from the two baths.

There are however (at least) two serious drawbacks in our theory.
First the extended Clausius relation \rlb{exCl01} is an approximate relation which is meaningful only when the system is close to the equilibrium.
Our theory says nothing about systems which are very far from equilibrium.
Secondly the extended Clausius relation \rlb{exCl01} holds only for protocols where the parameters are varied very slowly.
In the equilibrium thermodynamics, the Clausius {\em in}\/equality is known to hold for processes which are not necessarily slow.
We can also prove an inequality corresponding to \rlb{exCl01} (only for models without ``momenta''), but it contains an error term which is not perfectly under control.
See Theorem~\ref{t:ClausiusIneq} and section~\ref{s:Ssym}.

\paragraph*{Other schemes of ``renormalization'':}
Let us note that the above procedure is certainly not the unique way of ``renormalizing'' the divergent entropy production.
For the moment, at least three other schemes of renormalization are known.

The scheme by Hatano and Sasa \cite{HatanoSasa} developed for the overdamped Langevin system was the first realization of SST based on microscopic (or mesoscopic) dynamics.
A very close, but slightly different, scheme based on macroscopic fluctuation theory was recently proposed by Bertini, Gabrielli, Jona-Lasinio, and Landim \cite{BGJLL2012,BGJLL2013}.
The scheme due to Maes and Netocny \cite{MaesNetocny2012} makes a full use of the large deviation analysis.
See \cite{MaesNetocny2012,Sasa2013} for discussions about the relations between different schemes.

Unlike our scheme, all these three schemes lead to extended Clausius relations (or analogous equivalent relations) which are exact for systems arbitrarily far away from equilibrium.
Moreover, these equalities are accompanied by corresponding {\em in}\/equalities which are valid for general processes.
These are clear advantages of the three schemes.

On the other hand, the renormalization in these three schemes requires subtraction of rather involved quantities which are not directly observable in experiments.
In this sense our scheme, which uses only directly measurable quantities, has an advantage.
We also note that our scheme applies to a larger class of models than the others.
Although the Maes-Netocny scheme is based on microscopic (or mesoscopic) dynamics, it does not apply to models with inertia (momenta) as it is.
As for the Hatano-Sasa scheme it has been pointed out \cite{SpinnyFord21012} that a consistent thermodynamic interpretation is impossible once the momentum degrees of freedom is introduced. See also \cite{Sasa2013}.
Among the four, ours seems to be the only scheme which provides a consistent thermodynamic relation in models including momenta (although we lack inequalities).
See section~\ref{s:Ssym}.

For the moment we cannot say anything definite about which (or, even any) of the four schemes is most promising.
We believe that further investigations from mathematical, theoretical, and experimental points of view are necessary.


\section{Setup and definitions}
\label{s:Setupanddef}

Here we introduce general Markov jump processes that we study, and fix the notation.
Quantities specific to our approach to nonequilibrium physics are introduced in section~\ref{s:entropyprod}.
We also describe typical examples in section~\ref{s:examples}.

\subsection{Markov jump process}
\label{s:Markov}
Let the state space $\calS$ be a finite set.
The elements $x,y,\ldots\in\calS$ are states (in a suitable mesoscopic  description) of the system.
The probability distribution is denoted in vector notation as $\bsp=(p_x)_{x\in\calS}$ where $p_x$ is the probability to find the system in a state $x$.

\begin{figure}[btp]
\begin{center}
\includegraphics[width=9cm]{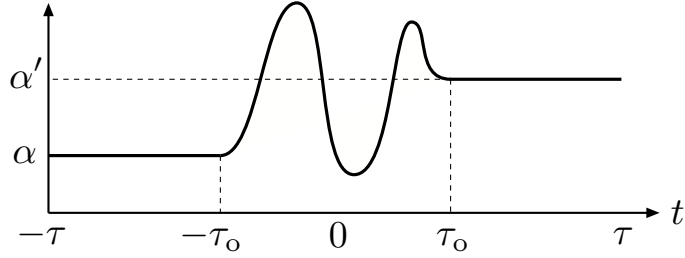}
\end{center}
\caption[dummy]{
The protocol $\hata=(\alpha(t))_{\ttt}$, which brings the model parameters from $\alpha$ to $\alpha'$.
The parameters $\alpha(t)$ stay constant in the initial and the final stages.
}
\label{fig:protocol}
\end{figure}

We assume that there is a set of parameters $\alpha$ which characterizes the system.
For concreteness we assume that $\alpha$ takes its values in a compact subset of $\bbR^n$ for some $n>1$.
Fix an arbitrary time scale $\tao>0$.
During the time interval $[-\tao,\tao]$, an external agent performs an operation to the system by controlling $\alpha$ according to a protocol (i.e., a function of time $t$) $\alpha(t)$ (with $t\in[-\tao,\tao]$) which is fixed in advance.
The function $\alpha(t)$ need not be continuous.
We write the initial and the final values of the parameters as $\alpha=\alpha(-\tao)$ and $\alpha'=\alpha(\tao)$, respectively.
We also take $\tau$ which is much larger than $\tao$, and consider the time evolution of the system in the longer time interval $[-\tau,\tau]$.
We extend the protocol to the whole time interval by simply setting $\alpha(t)=\alpha$ for $t\in[-\tau,-\tao]$ and $\alpha(t)=\alpha'$ for $t\in[\tao,\tau]$.
See Figure~\ref{fig:protocol}.
We denote the whole protocol as $\hata=(\alpha(t))_{\ttt}$.
A special protocol in which $\alpha(t)$ takes a constant value $\alpha$ throughout $[-\tau,\tau]$ is denoted as $(\alpha)$.

\bigskip

We consider a Markov jump process characterized by a protocol $\hata$.

For given parameters $\alpha$ and $x,y\in\calS$ such that $x\ne y$, let $\Ra_\xty\ge0$ be the transition rate from the state $x$ to $y$.
Physically speaking transitions in our system is caused by interactions between the system and heat baths attached to it.
We assume that $\Ra_\xty\ne0$ implies $\Ra_\ytx\ne0$ for any $x\ne y$.
We also assume that the whole state space $\calS$ is ``connected'' by nonvanishing $\Ra_\xty$.
More precisely, for any $x,y\in\calS$ with $x\ne y$, one can take a sequence $x_0,x_1,\ldots,x_n$ such that $x_0=x$, $x_n=y$, and $\Ra_{x_{j-1}\to x_j}\ne0$ for any $j=1,2,\ldots,n$.
We also define the escape rate at $x\in\calS$ by
\eq
\la_x:=\sumtwo{y\in\calS}{(y\ne x)}\Ra_\xty.
\lb{ladef}
\en

The Markov jump process corresponding to the protocol $\hata=(\alpha(t))_\ttt$ is defined by the master equation
\eq
\fdt{p_x(t)}=-\lat_x\,p_x(t)+\sumtwo{y\in\calS}{(y\ne x)}p_y(t)\,\Rat_\ytx,
\lb{master}
\en
for any $x\in\calS$ and $\ttt$, where $p_x(t)$ is the probability to find the system in  $x$ at time $t$.
The equation \rlb{master} is neatly rewritten in the vector notation as
\eq
\fdt{\bsp(t)}=\sfR^{\alpha(t)}\bsp(t),
\lb{master2}
\en
where $\bsp(t)=(p_x(t))_{x\in\calS}$ is regarded as a column vector.
The transition rate matrix\footnote{%
The standard generator of a stochastic process is given by the transpose of $\sfR$.
} $\sfR^\alpha$ is defined by specifying its entries as $(\sfR^\alpha)_{yx}=\Ra_\xty$ for $x\ne y$ and $(\sfR^\alpha)_{xx}=-\la_x$.
The formal solution of \rlb{master2} is written as
\eq
\bsp(t)=\exp_{\leftarrow}\Bigl[\int_{-\tau}^t ds\,\sfR^{\alpha(s)}\Bigr]\,\bsp^\mathrm{init},
\lb{formalsol}
\en
where $\bsp^\mathrm{init}$ is the initial distribution given at $-\tau$.
The time-ordered exponential is defined by
\eqa
&\exp_{\leftarrow}\Bigl[\int_{-\tau}^t ds\,\sfR^{\alpha(s)}\Bigr]
\nl&\ \ :=\lim_{N\up\infty}
\exp\Bigl[\frac{(t+\tau)\sfR^{\alpha(s_{N-1})}}{N}\Bigr]
\exp\Bigl[\frac{(t+\tau)\sfR^{\alpha(s_{N-2})}}{N}\Bigr]
\,\cdots\,\,
\exp\Bigl[\frac{(t+\tau)\sfR^{\alpha(s_0)}}{N}\Bigr],
\lb{Torder}
\ena
with $s_j=\{(t+\tau)/N\}j-\tau$.
When $\alpha(t)$ is time-independent, \rlb{Torder} coincides with the usual exponential $\exp[(t+\tau)\sfR^\alpha]$.

It is a well known consequence of the Perron-Frobenius theorem that, for any parameter $\alpha$, one has
\eq
\lim_{s\up\infty}\exp[s\,\sfR^\alpha]\,\bsp^\mathrm{init}=\bsrho^\alpha,
\en
where $\bsp^\mathrm{init}$ is an arbitrary initial probability distribution.
Here  $\bsrho^\alpha=(\rho^\alpha_x)_{x\in\calS}$ is the unique stationary probability distribution characterized by the condition $\sfR^\alpha\bsrho^\alpha=0$.  It is also known that $\rho^\alpha_x>0$ for any $x$.
Physically speaking $\bsrho^\alpha$ is the probability distribution for the nonequilibrium steady state (NESS) of the system with constant parameters $\alpha$.

\subsection{Description in terms of paths}
\label{s:path}
\begin{figure}[btp]
\begin{center}
\includegraphics[width=7cm]{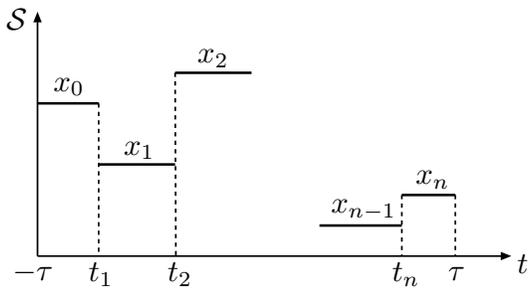}
\end{center}
\caption[dummy]{
A schematic picture of a path $\hatx$.
}
\label{fig:path}
\end{figure}

It is sometimes more convenient to describe the Markov jump process in terms of a path (or a history) $\hatx$ of the state.
A path is naturally identified with a piecewise constant function $\hatx=(x(t))_{\ttt}$, but we shall often specify it in terms of the history of jumps as
\eq
\hatx=(n,(x_0,x_1,\ldots,x_n),(t_1,t_2,\ldots,t_n)),
\lb{xpath}
\en
where $n=0,1,2,\ldots$ is the total number of jumps, $x_0,\ldots,x_n\in\calS$ (such that $x_{j-1}\ne x_{j}$ for $j=1,\ldots,n$) are the states that the system has visited, and $t_j$ is the time at which the jump $x_{j-1}\to x_j$ took place.
They are ordered as $-\tau<t_1<t_2<\ldots<t_n<\tau$, and we often write $t_0=-\tau$ and $t_{n+1}=\tau$.
We also write $x_0$ and $x_n$ as $x(-\tau)$ and $x(\tau)$, respectively.
See Figure~\ref{fig:path}.

Then the weight (more precisely, the transition probability density) associated with a path $\hatx$ is
\eq
\calT^{\hata}[\hatx]:=
\prod_{j=1}^nR^{\alpha(t_j)}_{x_{j-1}\to x_j}\,
\prod_{j=0}^n\exp\Bigl[-\int_{t_j}^{t_{j+1}}dt\,\lambda_{x_j}^{\alpha(t)}\Bigr].
\lb{Wdef}
\en
The weight is normalized so that
\eq
\int\Dx\,\delta_{x(-\tau),x}\,\calT^{\hata}[\hatx]=1
\en
for any initial state $x\in\calS$, where the ``integral'' over all the paths is defined by
\eq
\int\Dx(\cdots):=
\sum_{n=0}^\infty
\ \sumtwo{x_0,\ldots,x_n\in\calS}{(x_{j-1}\ne x_j)}\ 
\int_{-\tau}^\tau dt_1\int_{{t_1}}^\tau dt_2\int_{{t_2}}^\tau dt_3\cdots
\int_{{t_{n-1}}}^\tau dt_n(\cdots).
\en
In this language, the general solution \rlb{formalsol} is written as
\eq
p_x(t)=\int\Dx\,p_{x(-\tau)}^\mathrm{init}\,\delta_{x(t),x}\,\calT^{\hata}[\hatx].
\lb{formalsolpath}
\en
One way to see the equivalence of \rlb{formalsol} and \rlb{formalsolpath} is to write the matrix product explicitly (in terms of the sums over $\calS$) in \rlb{Torder}.

Let $f[\hatx]$ be an arbitrary function of $\hatx$.
We define the expectation value of $f[\hatx]$ by
\eq
\sbkt{f}^{\hata}_{\sts\too}:=
\int\Dx\,f[\hatx]\,\rho^{\alpha(-\tau)}_{x(-\tau)}\,\calT^{\hata}[\hatx],
\lb{fss}
\en
where the subscript ``$\sts\too$'' indicates that the system starts from the steady state for parameter $\alpha(-\tau)$, and nothing is specified for the final condition.
We stress that this is a physically natural expectation, which can be realized  experimentally.

\subsection{Entropy production, work, and time-reversal}
\label{s:entropyprod}
Let us further specify our problem, and also introduce some important quantities.

We assume that each state $x\in\calS$ is associated with its energy $\Hn_x\in\bbR$.
Here $\nu$ is a parameter (or a set of parameters) that characterizes the Hamiltonian $\Hn_x$, and is a component of $\alpha$.
See \rlb{alphacomp1} and \rlb{alphacomp2} for examples.
We assume that $\nu$ takes its value in a compact subset of $\bbR^{n'}$ for some $n'\ge1$.

\paragraph*{Entropy production:}
For any $x,y\in\calS$ such that $\Ra_\xty\ne0$, we define the entropy production in the heat baths\footnote{%
Throughout the present paper, the entropy production always means the entropy production in the heat baths.
}  associated with the transition $\xty$ by
\eq
\thetaa_\xty:=\log\frac{\Ra_\xty}{\Ra_\ytx},
\lb{thetadef0}
\en
or, equivalently, by the ``local detailed balance condition''
\eq
\Ra_\ytx=e^{-\thetaa_\xty}\,\Ra_\xty.
\lb{thetadef}
\en
Clearly one has $\thetaa_\xty=-\thetaa_\ytx$.
Mathematically speaking \rlb{thetadef0} is a mere definition.
With this definition of $\thetaa_\xty$, we can justifies the ``detailed fluctuation theorem'' \rlb{DFT}, which will be a basis of the present work.
To give the quantity a physical interpretation as entropy production, we need some preparations.

A class of  processes called equilibrium (stochastic) dynamics describe  a system attached to heat baths with a single temperature and free from any non-conservative forces (see footnote \ref{fn:noncons}).
Such a system approaches the corresponding equilibrium state after a sufficiently long time\footnote{
The approach to equilibrium is indeed a {\em non}\/equilibrium phenomenon that can be studied in the framework of equilibrium dynamics.
}.
The transition rates $\Rbn_\xty$ (where we have written $\alpha=\bn$) in an equilibrium dynamics satisfy the detailed balance condition
\eq
e^{-\beta \Hn_x}\Rbn_\xty=e^{-\beta \Hn_y}\Rbn_\ytx,
\lb{dbeq}
\en
for any $x,y\in\calS$.
Here $\beta$ is the single inverse temperature of the heat baths.
It is well known (and easy to prove) that the condition \rlb{dbeq} ensures that the corresponding stationary distribution is the canonical distribution $\rho^\bn_x=e^{-\beta\Hn_x}/Z_\nu(\beta)$, where $Z_\nu(\beta)=\sum_{x\in\calS}e^{-\beta\Hn_x}$ is the normalization constant.

Under the detailed balance condition \rlb{dbeq}, the entropy production \rlb{thetadef0} becomes
\eq
\theta^\bn_\xty=\beta\,(\Hn_x-\Hn_y)=-\beta\,q_\xty,
\lb{thetabH}
\en
where $q_\xty=\Hn_y-\Hn_x$ is the change in the energy of the system, which is equal to the heat transferred from the baths to the system.
The final expression in \rlb{thetabH} is nothing but the well-known formula for the change (or the production) of entropy in equilibrium thermodynamics.

The main subject of the present work is {\em non}\/-equilibrium stochastic dynamics, for which the detailed balance condition \rlb{dbeq} can never be satisfied for any choice of $\beta$ and $\Hn_x$.
We nevertheless assume here that the entropy production $\thetaa_\xty$ is written as 
\eq
\thetaa_\xty=-\beta_\mathrm{B}\,q_\xty,
\lb{thetabq}
\en
where $\beta_\mathrm{B}$ is the inverse temperature of the single heat bath\footnote{
We assume here (and in what follows) that each transition is associated with only a single bath.
See also section~\ref{s:examples}.
} that is relevant to the transition $\xty$, and $q_\xty$ is the energy (heat) transferred from the bath to the system during the transition.
The idea behind the identification \rlb{thetabq} is that heat baths are always in equilibrium states so that we can use the relation from equilibrium thermodynamics for each transition, even when the system never settles to equilibrium\footnote{
In a formulations based on a stochastic process (as in the present work), the relation \rlb{thetabq} is nothing more than an interpretation.
In more microscopic formulations based on mechanics, one may justify such relations.  See, e.g., \cite{KNST3}.
}.

Throughout the present work we assume that the nonequilibrium system can be interpreted as a perturbation to an equilibrium system.
As for the entropy production we write
\eq
\thetaa_\xty=\psia_\xty+\beta\,(\Hn_x-\Hn_y),
\lb{psidef}
\en
where $\beta>0$ is a certain reference inverse temperature, which may not be unique.
The quantity $\psia_\xty$ should be called the nonequilibrium part of entropy production.
It also satisfies  $\psia_\xty=-\psia_\ytx$.

For a given path $\hatx$ as in \rlb{xpath}, we can define the total entropy production in $\hatx$ as
\eq
\Theta^{\hata}[\hatx]=\sum_{j=1}^n{\theta}^{\alpha(t_j)}_{x_{j-1}\to x_{j}},
\lb{Thetadef}
\en
and its nonequilibrium part as
\eq
\Psi^{\hata}[\hatx]=\sum_{j=1}^n{\psi}^{\alpha(t_j)}_{x_{j-1}\to x_{j}}.
\lb{Psidef}
\en
For any subinterval $[\tau_1,\tau_2]\subset[-\tau,\tau]$, we define partial entropy productions by
\eqg
\Theta^{[\tau_1,\tau_2],\hata}[\hatx]=
\sum_{j=1}^n\chi\bigl[t_j\in[\tau_1,\tau_2]\bigr]
\,{\theta}^{\alpha(t_j)}_{x_{j-1}\to x_{j}},
\lb{Theta12}\\
\Psi^{[\tau_1,\tau_2],\hata}[\hatx]=
\sum_{j=1}^n\chi\bigl[t_j\in[\tau_1,\tau_2]\bigr]
\,{\psi}^{\alpha(t_j)}_{x_{j-1}\to x_{j}},
\lb{Psi12}
\eng
where $\chi[\text{true}]=1$ and $\chi[\text{false}]=0$.

\paragraph*{Work:}
Let us write the protocol for the parameter of the Hamiltonian as $\hatn:=(\nu(t))_{\ttt}$, which is a component of the full protocol $\hata$.
For a path $\hatx$, we define
\eq
W^\hatn[\hatx]:=
\sum_{j=0}^n(H^{\nu(t_{j+1})}_{x_j}-H^{\nu(t_{j})}_{x_j})
=\sum_{j=0}^n\int_{t_j}^{t_{j+1}}dt\,
\fdt{\nu(t)}\sqbk{\fp{H_{x(t)}^\nu}{\nu}}_{\nu=\nu(t)},
\lb{Workdef}
\en
where the final expression is valid only when $\nu(t)$ is differentiable.
Note that $H^{\nu(t_{j+1})}_{x_j}-H^{\nu(t_{j})}_{x_j}$ is the change in the energy of the system during the interval $(t_j,t_{j+1})$, in which the state of the system is always $x_j$.
Since this change in the energy is caused solely by the change in $\nu$, we can identify it with the work done by the external agent who operates on the system.
Therefore \rlb{Workdef} is the total work done by the external agent to the system in the path $\hatx$.
Note that, since $\nu(t)$ varies only for $t\in[-\tao,\tao]$, the summand in \rlb{Workdef} vanishes if $(t_j,t_{j+1})\cap[-\tao,\tao]=\emptyset$.

It is obvious from \rlb{psidef} that $\Theta$, $\Psi$, and $W$ are related with each other.
In fact by summing up  \rlb{psidef} for all the transitions in $\hatx$, we see
\eqa
\Theta^{\hata}[\hatx]&=\Psi^{\hata}[\hatx]
+\beta\sum_{j=1}^n\bigl(H^{\nu(t_j)}_{x_{j-1}}-H^{\nu(t_j)}_{x_{j}}\bigr)
\nl&=\Psi^{\hata}[\hatx]
+\beta\Bigl\{
\sum_{j=0}^n\bigl(H^{\nu(t_{j+1})}_{x_j}
-H^{\nu(t_{j})}_{x_j}\bigr)-H_{x_n}^{\nu(t_{n+1})}
+H_{x_0}^{\nu(t_0)}\Bigr\}
\nl&=\Psi^{\hata}[\hatx]+\beta\,W^\hatn[\hatx]
+\beta\bigl(H^{\nu(-\tau)}_{x(-\tau)}-H^{\nu(\tau)}_{x(\tau)}\Bigr).
\lb{TWPH}
\ena

\paragraph*{Time-reversal:}
For a path $\hatx$ as in \rlb{xpath}, we define its time-reversal $\hatxd$ as
\eq
\hatxd=(n,(x_n,x_{n-1},\ldots,x_0),(-t_n,-t_{n-1},\ldots,-t_2,-t_1)).
\lb{xdagger}
\en
If we use the language of function and denote original path as $\hatx=(x(t))_{\ttt}$, the time-reversed path is $\hatxd=(x(-t))_{\ttt}$.
Similarly for a protocol $\hata=(\alpha(t))_{\ttt}$ and its component $\hatn=(\nu(t))_\ttt$, we define their time-reversal as $\hatad=(\alpha(-t))_{\ttt}$ and  $\hatn=(\nu(-t))_{\ttt}$, respectively.

One easily finds that the total entropy production, its nonequilibrium part, and the work are antisymmetric with respect to the time-reversal, i.e.,
\eq
\Theta^{\hata}[\hatx]=-\Theta^{\hatad}[\hatxd],\quad
\Psi^{\hata}[\hatx]=-\Psi^{\hatad}[\hatxd],\quad
W^\hatn[\hatx]=-W^{\hatnd}[\hatxd].
\lb{Thetasym}
\en

\paragraph*{Measurability of the quantities:}
In operational approaches to thermodynamics we believe it essential to distinguish between physical quantities which are experimentally measurable (at least in principle) and which are not.
In what follows we assume that a path $\hatx$ has been realized, and ask whether the quantities $W^\hatn[\hatx]$, $\Theta^{\hata}[\hatx]$, and $\Psi^{\hata}[\hatx]$ are measurable.
This corresponds to the measurability of these quantities in a single experiment.

As in most treatments of equilibrium thermodynamics, we assume that the total work $W^\hatn[\hatx]$ is measurable.
The work is a purely mechanical quantity, and the external agent can, in principle, always determine it by precisely measuring the (generalized) force and the displacement.

We next argue that the total entropy production (in the baths) $\Theta^{\hata}[\hatx]$ is also measurable.
Suppose that the system is in touch with $n$ heat baths, where the inverse temperature of the $j$-th bath is $\beta_j$.
Let $Q_j^{\hata}[\hatx]$ be the total amount of heat that flowed into the system from the $j$-th bath during the experiment, i.e., the sum of $q_\xty$ for every transition (which involve the $j$-th bath) in the path $\hatx$ . 
We assume that the total heat $Q_j^{\hata}[\hatx]$ can be measured for  each $j$.
This may not be a trivial assumption, but in principle we can think of carefully designed heat baths where heat flux can be monitored accurately\footnote{
Such measurements are indeed possible in modern calorimetry.
}.
Since the relation \rlb{thetabq} means that the total entropy production is written as $\Theta^{\hata}[\hatx]=-\sum_{j=1}^n\beta_jQ_j^{\hata}[\hatx]$, we conclude that $\Theta^{\hata}[\hatx]$ is measurable.

The measurability of the nonequilibrium part $\Psi^{\hata}[\hatx]$  of the total entropy production is more subtle.
We argue that $\Psi^{\hata}[\hatx]$ is measurable or semi-measurable depending on the model.
(See the next section for details of the models.)
In the models for heat conduction, where the system exchanges energy only with heat baths, we find $\Psi^{\hata}[\hatx]=-\sum_{j=1}^n(\beta_j-\beta)Q_j^{\hata}[\hatx]$ from \rlb{psiaHC}.
This means that $\Psi^{\hata}[\hatx]$ is determined from the measurable total heat $Q_j^{\hata}[\hatx]$. 
In the models of systems driven by an external non-conservative force, on the other hand, the system exchanges energy with the external field as well as the heat baths.
It then turns out (see \rlb{psibfd}) that $\Psi^{\hata}[\hatx]$ is identical to $\beta$ times the total work done to the system by the external field.
The work done by the external field may be measured in principle\footnote{
One strategy is to measure the back action from the system to the generator (such as a coil) of the field.
In a colloidal system it may be possible to determine the work done by the field by precisely measuring the positions of charged particles.
}, but the measurement seems to be extremely difficult in general.
We thus regard $\Psi^{\hata}[\hatx]$ as a semi-measurable quantity in this case.

\subsection{Examples}
\label{s:examples}

Although our theory applies to a large variety of physical models, it might be useful to have some concrete examples in mind.
Here we  define a standard class of equilibrium dynamics, and then describe two typical problems of nonequilibrium physics.

\paragraph*{Equilibrium dynamics:}
Before discussing nonequilibrium problems, let us discuss equilibrium dynamics, which will be the starting point.

To define transition rates, it is convenient to first specify the Hamiltonian $\Hn_x$ and the connectivity function $c(x,y)$ such that $c(x,y)=c(y,x)\ge0$ for any $x,y\in\calS$ with $x\ne y$.
We assume that the state space $\calS$ is connected via nonvanishing $c(x,y)$, or more precisely, for any $x,y\in\calS$ one can take a sequence $x_0,x_1,\ldots,x_n$ such that $x_0=x$, $x_n=y$, and $c(x_{j-1},x_j)\ne0$ for any $j=1,2,\ldots,n$.
We make no assumptions on the Hamiltonian $\Hn_x$ except that it is real.

Then the transition rates for the equilibrium dynamics at the inverse temperature $\beta$ may be defined, for example, as
\eq
\Rbn_\xty=c(x,y)\,e^{\beta\Hn_x},
\lb{Req1}
\en
or
\eq
\Rbn_\xty=c(x,y)\,e^{(\beta/2)(\Hn_x-\Hn_y)}.
\lb{Req2}
\en
It is clear that both the definitions satisfy the necessary conditions for transition rates including the detailed balance condition \rlb{dbeq}.

Suppose that one has $\Hn_x=\Hn_y$ and $c(x,y)\ne0$ for some $x\ne y$.
Then the rate \rlb{Req2} becomes $\Rbn_\xty=\Rbn_\ytx=c(x,y)$, and is independent of the inverse temperature $\beta$.

\bigskip

The abstract scheme discussed above applies to various concrete physical settings.
Let us describe a system of particles on a lattice.
Let the lattice $\Lambda$ be a finite set whose elements are denoted as $u,v\ldots\in\Lambda$.
We denote by $\calB$ the set of bonds on $\Lambda$. 
More precisely the element of $\calB$ is a pair $\{u,v\}=\{v,u\}$ with some  $u,v\in\Lambda$ such that $u\ne v$.
We assume that  $\Lambda$ is connected via the bonds in $\calB$.
The simplest example is the one-dimensional periodic lattice $\Lambda=\{1,2,\ldots,L\}$ with $\calB=\bigl\{\{x,x+1\}\,\bigr|\,x\in\Lambda\bigr\}$, where we identify $L+1$ with $1$.

\begin{figure}[btp]
\begin{center}
\includegraphics[width=4cm]{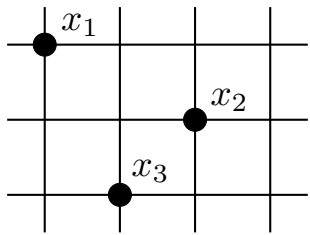}
\end{center}
\caption[dummy]{
A configuration $x=(x_1,x_2,x_3)$ of three particles on the lattice.
}
\label{fig:Config}
\end{figure}

We assume that there are $N$ particles on the lattice, and let $x$ denote a configuration of the particles on $\Lambda$.
More precisely, we set $x=(x_1,\ldots,x_N)$, where $x_j\in\Lambda$ is the position of the $j$-th particle ($j=1,\ldots,N$).
One may or may not impose the hard-core condition, i.e., $x_j\ne x_k$ whenever $j\ne k$.
See Figure~\ref{fig:Config}.

For any two configurations $x=(x_1,\ldots,x_N)$ and $y=(y_1,\ldots,y_N)$, we set $c(x,y)=1$ if $\{x_k,y_k\}\in\calB$ for some $k$ and $x_j=y_j$ for any $j$ such that $j\ne k$, and $c(x,y)=0$ otherwise.
In other words, $c(x,y)=1$ if and only if one can modify the configuration $x$ into $y$ by moving one particle along a bond in $\calB$.

As for the Hamiltonian, the standard choice is
\eq
H_x:=\sum_{j=1}^NV_1(x_j)+\sumtwo{j,k=1}{(j>k)}^NV_2(x_j,x_k),
\lb{Hameq}
\en
where the single particle potential $V_1(\cdot)$ and the two-particle interaction potential $V_2(\cdot,\cdot)$ are arbitrary real valued functions on $\Lambda$ and $\Lambda\times\Lambda$, respectively.

\paragraph*{Heat conduction:}
Let us discuss an idealized model of heat conduction.
We assume that the system  interacts with $n$ heat baths with different temperatures.
We label the baths by the index $j=1,2,\ldots,n$, and denote by $\beta_j$ the inverse temperature of the $j$-th bath.
The set of parameters that characterizes the model is 
\eq
\alpha=(\beta_1,\ldots,\beta_n,\nu).
\lb{alphacomp1}
\en

With any $x,y\in\calS$ such that $c(x,y)\ne0$ we associate a unique index $j(x,y)=j(y,x)\in\{1,\ldots,n\}$, which indicates that the $j(x,y)$-th bath is relevant for the transition between $x$ and $y$.
Then we define the transition rate as
\eq
R^\alpha_{\xty}=R^{(\beta_{j(x,y)},\nu)}_{\xty},
\lb{RforHeatConduction}
\en
for any $x, y\in\calS$ such that $x\ne y$, where the right-hand side is defined by \rlb{Req1} or \rlb{Req2}.

From the definition \rlb{thetadef0}, one finds
\eq
\theta_{\xty}^\alpha=\beta_{j(x,y)}\,(\Hn_x-\Hn_y)
=\Di\beta_{j(x,y)}\,(\Hn_x-\Hn_y)+\beta\,(\Hn_x-\Hn_y),
\en
where we have chosen the reference inverse temperature $\beta$ (somewhat arbitrarily), and wrote $\Di\beta_j:=\beta_j-\beta$.
Comparing with \rlb{psidef}, one finds 
\eq
\psi^\alpha_{\xty}=\Di\beta_{j(x,y)}\,(\Hn_x-\Hn_y),
\lb{psiaHC}
\en
which is indeed small when all the inverse temperatures $\beta_1,\ldots,\beta_n$ are close to each other, and $\beta$ is chosen properly\footnote{%
In most of realistic situations for heat conduction, only some small portions of the system is in touch with the heat baths.
To model such a situation by using a system of particles on a lattice, we assume that the energy of the system changes only when a particle hops within one of the portions which are in touch with the baths.
In other words, if an allowed transition $\xty$ is such that a particle hops outside the portions, then one must have $\Hn_x=\Hn_y$.
We further use the transition rule \rlb{Req2} so as to make the corresponding transition rate (which is indeed 1) independent of any inverse temperatures.
}.

\paragraph*{Driven system:}
We shall illustrate a system which is in contact with a single heat bath with the inverse temperature $\beta$, but is driven by a non-conservative external force.

For each pair $x,y\in\calS$ such that $c(x,y)\ne0$, we define a quantity $d_{\xty}\in\bbR$ which satisfies the antisymmetry $d_{\xty}=-d_{y\to x}$.
Physically, $d_{\xty}$ is interpreted as the displacement (in the direction of the non-conservative external force) of the particle associated with the transition $\xty$.
In the simplest example of particles on the one-dimensional periodic lattice, we set $d_{\xty}=1$ if a particle jumps to the right in the transition $\xty$, and $d_{\xty}=-1$ if a particle jumps to the left. 

We assume that the non-conservative\footnote{
\label{fn:noncons}
We define the force in this setting as $f_{\xty}=f d_\xty$.
The force $f_\xty$ is said to be conservative if one can write $f_{\xty}=U_x-U_y$ for any $x,y\in\calS$ with a suitable function (i.e., potential) $U_x$.
} external force $f$ is applied to the system.
The model is parameterized by 
\eq
\alpha=(\beta,\nu,f).
\lb{alphacomp2}
\en
We then define the transition rate by
\eq
R^\alpha_\xty=e^{\beta f d_\xty/2}\,R^\bn_\xty,
\lb{Rwithf}
\en
for any $x,y\in\calS$ such that $x\ne y$,
where the right-hand side is defined by \rlb{Req1} or \rlb{Req2}.

From the definitions \rlb{thetadef0} and \rlb{psidef}, one readily finds
\eq
\psi^\alpha_{\xty}=\beta f d_{\xty}.
\lb{psibfd}
\en
This means that the nonequilibrium part of the total entropy production $\Psi^\hata[\hatx]$ (see \rlb{Psidef}) can be interpreted as the total work done by the non-conservative external force  to the system (multiplied by $\beta$).

\section{Jarzynski-type equalities for NESS}
\label{s:JarNESS}

We start by presenting some exact equalities which are valid for general operations (i.e., protocols) to NESS.
They are reminiscent of the Jarzynski equality \rlb{Jarzynski}, which holds for operations to equilibrium states \cite{Jarzynski,Seifert12}.

We note, however, that the derivation of these equalities for NESS is not as straightforward as that of the original Jarzynski equality \cite{Jarzynski}.
One of the main difficulties is that we do not know the explicit form of the probability distribution $\bsrho^\alpha$ of NESS while the corresponding stationary distribution in the equilibrium case is the canonical distribution.

Our main equality is the following.
We here consider a general protocol $\hata$ introduced in the beginning of section~\ref{s:Markov}.  See, in particular,  Figure~\ref{fig:protocol}.
As we have discussed at the end of section~\ref{s:entropyprod}, we regard that the work $W^\hatn$ is measurable, and the nonequilibrium part $\Psi^{\hata}$ of the entropy production is measurable or  semi-measurable depending on the model.

\begin{theorem}
\label{t:1}
There exists a function (that we call the free energy) $F(\alpha)$ of the parameters $\alpha$ which coincides with the equilibrium free energy $-\beta^{-1}\log\sum_{x\in\calS}e^{-\beta \Hn_x}$ for an equilibrium system with $\alpha=(\beta,\nu)$, and we have for any protocol $\hata$ that
\eq
F(\alpha')-F(\alpha)=-\frac{1}{\beta}\lti
\log\frac{\bbkt{\,\exp[-(\beta\,W^\hatn+\Psi^{\hata})/2]\,}^{\hata}_{\sts\too}}
{\bbkt{\,\exp[-(\beta\,W^{\hatnd}+\Psi^{\hatad})/2]\,}^{\hatad}_{\sts\too}}.
\lb{SSTJ1}
\en
\end{theorem}

This theorem will be proved in section~\ref{s:t1proof}, using the results from section~\ref{s:matrix}.

We recall that $\nu$ is a component of $\alpha$ (as in \rlb{alphacomp1} and  \rlb{alphacomp2}), and likewise the protocol $\hatn=(\nu(t))_\ttt$ is a component of the full protocol   $\hata=(\alpha(t))_\ttt$.
Note that we fix the operation time scale $\tao$ when taking the limit $\tau\up\infty$.
It means that we are  treating an arbitrary operation, including very ``wild'' ones.

The equality \rlb{SSTJ1} expresses the difference of the (nonequilibrium) free energy in terms of the expectation values defined for nonequilibrium processes.
In this sense it may be regarded as a nonequilibrium version of the  Jarzynski equality 
\eq
F(\beta,\nu')-F(\beta,\nu)=-\frac{1}{\beta}
\log\bbkt{e^{-\beta W^\hatn}}^{(\beta,\hatn)}_{\mathrm{eq}\to}
\lb{Jarzynski}
\en
for equilibrium processes.
A fundamental difference of our equality from the original equality is that we must consider the expectation values for both the original protocol $\hata$ and its time-reversal $\hatad$.
Let us stress, however, that the expectation $\sbkt{\cdots}^{\hatad}_{\sts\too}$ is not at all unphysical; one simply executes the operation according to the protocol $\hatad$, and considers a natural time-evolution starting from the NESS corresponding to $\alpha'$.
Another difference is that here the expectation values involve the nonequilibrium part  $\Psi$  of the entropy production as well as the work $W$.

We recall that one can prove the minimum work principle 
$\sbkt{W^\hatn}^{(\beta,\hatn)}_{\mathrm{eq}\to}\ge F(\beta,\nu')-F(\beta,\nu)$, which is a representation of the second law of thermodynamics (in the standard equilibrium thermodynamics), 
by simply applying the Jensen inequality to the Jarzynski equality \rlb{Jarzynski}.
Unfortunately our equality \rlb{SSTJ1} does not lead directly to any inequalities since the right-hand side is the ratio of the two expectation values.

Below in Theorem~\ref{t:Nakagawa},
we see another equality which better resembles the original Jarzynski equality \rlb{Jarzynski}.
But our \rlb{SSTJ1} is of considerable importance especially because it is intimately related to thermodynamic relations in NESS, as we shall discuss in  section~\ref{s:SST}.

We note that various exact equalities which are valid for general stochastic processes (including ours) have been derived from the ``detailed fluctuation theorem'' in, e.g., \cite{Crooks99,Crooks00,Seifert05}.
Many similar equalities can be derived in the same manner.
We believe that our equality \rlb{SSTJ1} is essentially different from these equalities.
While the equalities derivable with the methods in \cite{Crooks99,Crooks00,Seifert05} contain quantities like $\log\rho^\alpha_x$ which depend explicitly on the unknown stationary distribution $\bsrho^\alpha$, our equality only contains $W$ and $\Psi$ which are (semi-)measurable thermodynamic type quantities.
The derivation of our new equality is based not only on the ``detailed fluctuation theorem'' but also on the new ``splitting lemma'' (Lemma~\ref{l:split}) which allows us to treat a process whose initial distribution is $\bsrho^\alpha$ without using $\bsrho^\alpha$ explicitly.

\bigskip
It may be inspiring to rewrite the quantity inside the limit in \rlb{SSTJ1} as
\eq
\log\frac{\sbkt{e^{-\beta\,W^\hatn/2}}^\hata_\mathrm{mod}}{\sbkt{e^{-\beta\,W^\hatnd/2}}^\hatad_\mathrm{mod}}
+
\log\frac{\bbkt{\,\exp[-\Psi^{\hata}/2]\,}^{\hata}_{\sts\too}}
{\bbkt{\,\exp[-\Psi^{\hatad}/2]\,}^{\hatad}_{\sts\too}},
\en
where we have defined the modified expectation by
\eq
\sbkt{f}^\hata_\mathrm{mod}:=
\frac{\bbkt{\,f\,\exp[-\Psi^{\hata}/2]\,}^{\hata}_{\sts\too}}
{\bbkt{\,\exp[-\Psi^{\hata}/2]\,}^{\hata}_{\sts\too}}.
\lb{trafficexpectation}
\en

We note that the extra weight $\exp[-\Psi^{\hata}[\hatx]/2]$ has an effect of canceling the nonequilibrium contribution $\psia_\xty$ in the transition rates.
This is in particular true for the example of driven system discussed in section~\ref{s:examples}.
Compare the transition rate \rlb{Rwithf} with the formula \rlb{psibfd} for $\psia_\xty$.

The physics described by the expectation $\sbkt{\cdots}^\hata_\mathrm{mod}$ is then expected to be close to that of equilibrium.  
But there still is considerable ``nonequilibrium effect'' coming from the escape rates $\lambda^\alpha_x$, which are untouched in the modification \rlb{trafficexpectation}.
Possible essential roles played by the escape rates in nonequilibrium states have been emphasized by Maes and his collaborators \cite{MaesNetocny03,BaiesiMaesWynants,BaertsBasuMaesSafaverdi}.
In the expectation  $\sbkt{\cdots}^\hata_\mathrm{mod}$, nonequilibrium flows are cancelled and we can focus on the effects from the escape rates.
We still do not know if this interpretation leads us to any new insights.

\paragraph*{Remark 1:}
The equality \rlb{SSTJ1} is one of the series of equalities which can be proved in the similar manner.
A general form includes an arbitrary constant $\kappa\in\bbR$, and is
\eq
F(\alpha')-F(\alpha)=-\frac{1}{\beta}\lti
\log\frac{\bbkt{\,\exp[-\kappa\,\beta\,W^\hatn-\Psi^{\hata}/2]\,}^{\hata}_{\sts\too}}
{\bbkt{\,\exp[-(1-\kappa)\,\beta\,W^\hatnd-\Psi^{\hatad}/2]\,}^{\hatad}_{\sts\too}}.
\lb{SSTJ2}
\en

\paragraph*{Remark 2:}
We can also prove the following exact equalities which involve the total entropy production $\Theta$, rather than its nonequilibrium part $\Psi$.
See the end of section~\ref{s:t1proof} for the proof.
There exists a function  $\tiS(\alpha)$ of the parameters $\alpha$, and we have for any protocol $\hata$ that
\eq
\tiS(\alpha')-\tiS(\alpha)=\lti
\log\frac{\bbkt{\,\exp[-\Theta^{\hata}/2]\,}^{\hata}_{\sts\too}}
{\bbkt{\,\exp[-\Theta^{\hatad}/2]\,}^{\hatad}_{\sts\too}}.
\lb{SSTJ4}
\en
Although one may be tempted to identify $\tiS(\alpha)$ as nonequilibrium entropy, this interpretation may not be adequate.
For parameters $\alpha$ which correspond to equilibrium, one finds that  $\tiS(\alpha)$ is (similar to but) not the same as the equilibrium entropy.
See \rlb{tiSeq}.

\bigskip
Another exact Jarzynski-type equality for NESS was derived by one of us (N.N.) in \cite{Nakagawa}.
Since we can prove this equality with the same machinery as the previous one, we shall briefly discuss it here.

For an arbitrary function $f[\hatx]$ of $\hatx$, we define its $\Psi$-modified expectation as
\eq
\sbkt{f}^{\hata}_{\text{$\Psi$-mod}}:=
\frac{\bbkt{\,f\,\exp[-(\Psi^{[-\tau,-\tau/2],(\alpha)}+\Psi^{[\tau/2,\tau],(\alpha')})/2]\,}^{\hata}_{\sts\too}}
{\bbkt{\,\exp[-(\Psi^{[-\tau,-\tau/2],(\alpha)}+\Psi^{[\tau/2,\tau],(\alpha')})/2]\,}^{\hata}_{\sts\too}},
\lb{Psimod}
\en
where $\Psi$ for restricted time intervals are defined in \rlb{Psi12}.
We have replaced the protocols by $(\alpha)$ and $(\alpha')$ in order to emphasize that $\alpha(t)$ is constant in these intervals (we here assume $\tau\ge2\tao$).
This is similar to the  modified expectation \rlb{trafficexpectation} introduced above, but now the modification factor presents only in the time intervals $[-\tau,-\tau/2]$ and $[\tau/2,\tau]$.
One can say that the system is in the modified nonequilibrium at the beginning and the end of the history, while it is in the full-fledged nonequilibrium in the middle.
This hybrid allows one to prove the following strong result.
\begin{theorem}
\label{t:Nakagawa}
Let $F(\alpha)$ be the nonequilibrium free energy introduced in Theorem~\ref{t:1}.
For any protocol $\hata$, one has
\eq
\lti
\sbkt{e^{-\beta\,W^\hatn}}^{\hata}_{\sts\too}
=e^{-\beta\{F(\alpha')-F(\alpha)\}}\,\lti\bbkt{\exp[-\Psi^{[-\tau/2,\tau/2],\hatad}]}^{\hatad}_{\text{\rm$\Psi$-mod}}.
\lb{Nakagawa}
\en
\end{theorem}

This theorem will be proved in section~\ref{s:tNakproof}, using the results from section~\ref{s:matrix}.

It is remarkable that the left-hand side of \rlb{Nakagawa} only includes the standard mechanical work $W^\hatn$ and the physically natural average $\sbkt{\cdots}^{\hata}_{\sts\too}$.
Although the right-hand side is a little more complicated, it was shown in \cite{Nakagawa} that it can also be measured experimentally at least when the ``degree of nonequilibrium'' is small enough.

By using the Jensen inequality, one can show from \rlb{Nakagawa} a ``second law''
\eq
\lti
\sbkt{W^\hatn}^{\hata}_{\sts\too}
\ge F(\alpha')-F(\alpha)-\frac{1}{\beta}\lti\log
\bbkt{\exp[-\Psi^{[-\tau/2,\tau/2],\hatad}]}^{\hatad}_{\text{$\Psi$-mod}}.
\lb{Nakagawa2}
\en

\section{Thermodynamic relations for NESS}
\label{s:SST}

We shall observe here that our main equality \rlb{SSTJ1} can be used to generate thermodynamic relations associated with operations that bring a NESS to a different NESS.
The simplest and the most important is the extended Clausius relation \rlb{ExCl}, \rlb{ExCled}, and \rlb{ExClexcess}, which was derived in our earlier works \cite{KNST1,KNST2}.
We concentrate on heuristic arguments in the present section, and discuss corresponding rigorous results in section~\ref{s:SSTrigorous}.

\subsection{Extended Clausius relation and entropy}
\label{s:exClEnt}
\paragraph*{Extended Clausius relation:}
Here we shall concentrate on a situation where the system is close to equilibrium and the change in the parameters during the operation is small.
It is then expected that the arguments of the two exponential functions in the right-hand side of the equality \rlb{SSTJ1} are small, because $\psia_\xty$ vanishes in an equilibrium system, and $W^\hatn$ is small if the change of the Hamiltonian is small.
By expanding in these quantities to the lowest order, we see that \rlb{SSTJ1} yields
\eq
F(\alpha')-F(\alpha)\simeq\frac{1}{2\beta}\Bigl\{
\sbkt{\beta\,W^\hatn+\Psi^\hata}^\hata
-\sbkt{\beta\,W^\hatnd+\Psi^\hatad}^\hatad
\Bigr\},
\lb{FF}
\en
where we have abbreviated $\sbkt{\cdots}^{\hata}_{\sts\too}$ as $\sbkt{\cdots}^{\hata}$ for simplicity.
We also assumed  $\tau$ is sufficiently large, and have omitted $\lti$.

Although \rlb{FF} may be interpreted as a thermodynamic relation, it is much better to rewrite it in terms of the total entropy production $\Theta$.
By substituting \rlb{TWPH}, this becomes
\eq
\beta\,\{F(\alpha')-F(\alpha)\}\simeq
\frac{\sbkt{\Theta^\hata}^\hata-\sbkt{\Theta^\hatad}^\hatad}{2}
+\beta\,\sbkt{H^{\nu'}}^{\alpha'}_{\sts}-\beta\,\sbkt{H^{\nu}}^{\alpha}_{\sts}.
\lb{ECL01}
\en
For an arbitrary function $g_x$ on $\calS$, we have defined its expectation value in the steady state with $\alpha$ as 
\eq
\sbkt{g}^{\alpha}_{\sts}:=\sum_{x\in\calS}g_x\,\rho^\alpha_x.
\lb{gastDef}
\en

If one introduces the nonequilibrium entropy through the ``familiar'' relation
\eq
S(\alpha):=\beta\bigl\{\sbkt{H^{\nu}}^{\alpha}_{\sts}-F(\alpha)\bigr\},
\lb{S=U-F}
\en
the relation \rlb{ECL01} becomes
\eq
S(\alpha')-S(\alpha)\simeq-\frac{\sbkt{\Theta^\hata}^\hata-\sbkt{\Theta^\hatad}^\hatad}{2},
\lb{ExCl}
\en
which is the extended Clausius relation obtained in \cite{KNST1,KNST2}.
This is essentially the same as \rlb{exCl01} in the introduction.
To be slightly more precise about the near equality, we can write the same relation as
\eq
S(\alpha')-S(\alpha)=-\frac{\sbkt{\Theta^\hata}^\hata-\sbkt{\Theta^\hatad}^\hatad}{2}+O(\epsilon^2\delta)+O(\delta^2),
\lb{ExCled}
\en
where $\epsilon$ denotes the ``degree of nonequilibrium'', and $\delta$ denotes the amount of change in the parameters (see section~\ref{s:SSTrigorous} for precise definitions).
The $O(\delta^2)$ term can be omitted for a quasi-static (i.e., smooth and slow) protocol.
In Theorems~\ref{t:SSTstep} and \ref{t:SSTsmooth}, we present corresponding rigorous estimates.

Consider an equilibrium protocol $\hata_\mathrm{eq}=(\alpha_\mathrm{eq}(t))_\ttt$, where $\alpha_\mathrm{eq}(t)$ for any $\ttt$ corresponds to an equilibrium system.
In such a case the standard adiabatic theorem implies that 
$\sbkt{\Theta^{\hata_\mathrm{eq}}}^{\hata_\mathrm{eq}}\simeq-\sbkt{\Theta^{\hata_\mathrm{eq}^\dagger}}^{\hata_\mathrm{eq}^\dagger}$ for a sufficiently slow and smooth process.
Then the relation \rlb{ExCl} becomes
\eq
S(\alpha'_\mathrm{eq})-S(\alpha_\mathrm{eq})
\simeq-\sbkt{\Theta^{\hata_\mathrm{eq}}}^{\hata_\mathrm{eq}},
\lb{StCl}
\en
which is nothing but the standard Clausius relation \rlb{EqCl}.
Since $S(\alpha_\mathrm{eq})$ coincides with the standard entropy, we find the the extended Clausius relation \rlb{ExCl} is an extension of the standard Clausius relation.

Note that, in a NESS, both $\sbkt{\Theta^\hata}^\hata$ and $\sbkt{\Theta^\hatad}^\hatad$ grow proportionally with the total time $\tau$ since there always is a nonvanishing entropy production.
The extended Clausius relation \rlb{ExCl} shows that their difference is a finite quantity independent of $\tau$ (provided that $\tau$ is long enough) and characterizes the effect of the operation.
It is crucial that the quantities  $\sbkt{\Theta^\hata}^\hata$ and $\sbkt{\Theta^\hatad}^\hatad$ can be measured by executing (at least) two experiments with the protocol $\hata$ and the corresponding reverse  protocol $\hatad$.

\paragraph*{Excess entropy production:}
The right-hand side of the extended Clausius relation \rlb{ExCl} or \rlb{ExCled} can also be written in terms of an interesting quantity called the excess entropy production.
We first define the entropy production rate $\sigma_\mathrm{st}^\alpha$ in the NESS with the parameters $\alpha$ by 
\eq
\sigma_\mathrm{st}^\alpha:=\frac{1}{2\tau}
\sbkt{\Theta^\ca}^\ca_{\sts\too},
\lb{sigmasta}
\en
where $\ca$ denotes the protocol in which $\alpha(t)=\alpha$ for any $\ttt$.
We shall prove in section~\ref{s:excess} that the definition is independent of $\tau$.
This independence may be intuitively apparent since the system is always in the NESS with $\alpha$, and $\Theta^\ca$ is the total entropy production.

For an arbitrary protocol $\hata=(\alpha(t))_\ttt$, we next define the corresponding house-keeping entropy production by
\eq
\Sigma^\hata_\mathrm{hk}:=\int_{-\tau}^\tau dt\,
\sigma_\mathrm{st}^{\alpha(t)}.
\lb{Thk}
\en
The house-keeping entropy production is indeed the main contribution to the total entropy production $\sbkt{\Theta^\hata}^\hata_{\sts\too}$, especially when $\tau$ is large and $\alpha(t)$ varies slowly.
The difference $\sbkt{\Theta^\hata}^\hata_{\sts\too}-\Sigma^\hata_\mathrm{hk}$ is called the {\em excess entropy production}\/.
It represents the intrinsic response of the system to the change of the parameters.

By using the excess entropy production, the extended Clausius relation \rlb{ExCled} is written as \cite{KNST1,KNST2}
\eq
S(\alpha')-S(\alpha)=
-\Bigl\{\sbkt{\Theta^\hata}^\hata-\Sigma^\hata_\mathrm{hk}\Bigr\}
+O(\epsilon^2\delta),
\lb{ExClexcess}
\en
where we omitted the $O(\delta^2)$ term assuming a quasi-static protocol.
We present a rigorous version of the relation in Theorem~\ref{t:SSTexcess}.

\paragraph*{Nonequilibrium entropy:}
Finally let us discuss the basic property of the entropy $S(\alpha)$.
For any probability distribution $\bsp=(p_x)_{x\in\calS}$, the corresponding Shannon entropy is defined by
\eq
S_\mathrm{Sh}[\bsp]:=-\sum_{x\in\calS}p_x\log p_x.
\lb{SSh}
\en
It is well-known that for an equilibrium parameter $\alpha_\mathrm{eq}$, one has $S(\alpha_\mathrm{eq})=S_\mathrm{Sh}[\bsrho^{\alpha_\mathrm{eq}}]$, i.e., the Shannon entropy of the stationary distribution (which is the canonical distribution) is exactly equal to the thermodynamic entropy.
This equality is no longer valid in nonequilibrium systems, but we can still show the near equality
\eq
S(\alpha)=S_\mathrm{Sh}[\bsrho^\alpha]+O(\epsilon^3).
\lb{S=SSh}
\en
The proof is based on the representation
\eq
\log\rho^\alpha_x=\beta\,F(\alpha)-\beta\,H^\alpha_x-\frac{1}{2}\lim_{\tau\up\infty}\Bigl\{
\bbkt{\Psi^{(\alpha)}}^{\tau,(\alpha)}_{x\to}
-\bbkt{\Psi^{(\alpha)}}^{\tau,(\alpha)}_{\sts\to x}
\Bigr\}+O(\epsilon^3),
\lb{theKN}
\en
for the stationary distribution $\bsrho^\alpha=(\rho^\alpha_x)_{x\in\calS}$, which was derived by Komatsu and Nakagawa \cite{KN}.
See Theorems~\ref{t:SandS} and \ref{t:KNrep} for rigorous versions.

\paragraph*{Remark:} Recall that we also have the equality \rlb{SSTJ4}, which directly deals with the entropy production $\Theta$.
Although one might suspect that \rlb{SSTJ4} leads us immediately to the extended Clausius relation \rlb{ExCl}, it is indeed not the case.
Unlike the nonequilibrium part $\Psi$, the total entropy production $\Theta$ does not  vanish in the limit of equilibrium.
The expansion in $\Theta$ is not justified even in a heuristic discussion.

\subsection{Higher order relations}
\label{s:SSThigher}
We continue to be heuristic, and discuss higher order contributions from the expansion of \rlb{SSTJ1} that we considered above.

Recall that for a general expectation $\sbkt{\cdots}$ and random variables $X_1,X_2,\cdots,X_k$, the corresponding cumulant is defined by
\eq
{}^\mathrm{c}\!\sbkt{X_1X_2\cdots X_k}:=\left.\frac{\partial}{\partial u_1}\frac{\partial}{\partial u_2}\cdots\frac{\partial}{\partial u_k}
\log\sbkt{\,e^{\sum_{i=1}^ku_iX_i}\,}\right|_{u_1=u_2=\cdots=u_k=0},
\lb{cumulant}
\en
which leads to the formal expansion
\eq
\log\sbkt{\,e^{X}\,}=\sum_{k=1}^\infty\frac{{}^\mathrm{c}\!
\sbkt{\,X^k\,}}{k!},
\lb{cumulant2}
\en
where ${}^\mathrm{c}\!\sbkt{\,X^k\,}:={}^\mathrm{c}\!\sbkt{\,\underbrace{X\cdots X}_{k}\,}$.

By applying \rlb{cumulant2} (formally) to the right-hand side of the equality \rlb{SSTJ1}, one finds
\eq
F(\alpha')-F(\alpha)=-\frac{1}{\beta}
\sum_{k=1}^\infty\frac{(-1)^k}{2^kk!}
\Bigl\{
{}^\mathrm{c}\!
\bbkt{(\beta\,W^\hatn+\Psi^{\hata})^k}^{\hata}
-{}^\mathrm{c}\!
\bbkt{(\beta\,W^{\hatnd}+\Psi^{\hatad})^k}^{\hatad}
\Bigr\}.
\lb{SSTJC}
\en
This is an improvement, which contains  infinitely many ``nonlinear'' terms, of the extended Clausius relation \rlb{ExCl} or \rlb{ExCled}.
But a thermodynamic relation with infinitely many terms may not be useful or enlightening.
It may be useful to have truncated versions of the nonlinear relation which are valid in certain limited situations.

Here we still concentrate on a situation where the system is close to equilibrium and the change in the parameters during the operation is small.
As before we denote by $\epsilon$ the ``degree of nonequilibrium'', and by $\delta$ the amount of change in the parameters.
We shall define these quantities precisely later in section~\ref{s:SSTrigorous}.

The error estimate in \rlb{ExCled} was derived heuristically (but not proved) in \cite{KNST1,KNST2}.
With a similar argument, we can show for any $n=1,2,\ldots$ that
\eq
{}^\mathrm{c}\!
\bbkt{(\beta\,W^\hatn+\Psi^{\hata})^{2n}}^{\hata}
-{}^\mathrm{c}\!
\bbkt{(\beta\,W^{\hatnd}+\Psi^{\hatad})^{2n}}^{\hatad}
=O(\epsilon^{2n}\,\delta)+O(\delta^2).
\lb{formalerrorestimate}
\en
We shall describe the derivation of this estimates in the next section~\ref{s:heuristicerror}.

By using \rlb{formalerrorestimate}, we can write the truncated version of the higher order relations, which is
\eqa
F(\alpha')-F(\alpha)=&-\frac{1}{\beta}
\sum_{k=1}^{2n-1}\frac{(-1)^k}{2^kk!}
\Bigl\{
{}^\mathrm{c}\!
\bbkt{(\beta\,W^\hatn+\Psi^{\hata})^k}^{\hata}
-{}^\mathrm{c}\!
\bbkt{(\beta\,W^{\hatnd}+\Psi^{\hatad})^k}^{\hatad}
\Bigr\}
\nl&
+O(\epsilon^{2n}\,\delta)+O(\delta^2).
\lb{SSTJC2n}
\ena
See Theorems~\ref{t:SSTstep} and \ref{t:SSTsmooth} for the corresponding rigorous estimates.

\paragraph*{Remark:}
In \cite{KNST2}, we have derived (again non-rigorously) a ``non-linear nonequilibrium thermodynamic relation'', which does not exactly fit into the above form.
Although we can derive the equality in \cite{KNST2} within the present framework, we won't discuss the derivation here\footnote{
A convenient derivation is to start from \rlb{SSTJ2} with $\kappa=1$, and apply the cumulant expansion as above.
}.

\subsection{Perturbative estimate of the error}
\label{s:heuristicerror}
Let us present a heuristic derivation of \rlb{formalerrorestimate}.
We shall make use of formal perturbative estimates freely without being bothered by the validity of the perturbation calculation.
The following estimate (as it is) can be  mathematically justified only for systems with a fixed time $\tau$ and extremely small $\epsilon$ and $\delta$.
Since the convergence estimate is not uniform in $\tau$ (in the present section), we have no rigorous control of the the limit $\tau\up\infty$, in which we are interested.
Nevertheless this estimate will be used as a part of rigorous argument later in section~\ref{s:SSTproof}.

Define $\Di\Phi^\hata$ and  $\Di\Phi^\hatad$, which are of $O(\delta)$, by
\eq
\Psi^\hata+\beta W^\hatn=\Psi^{(\alpha)}+\Di\Phi^\hata,\quad
\Psi^\hatad+\beta W^\hatnd=\Psi^{(\alpha)}+\Di\Phi^\hatad.
\en 
Then the quantity to be estimated is written as
\eqa
{}^\mathrm{c}\!
\bbkt{(\beta\,W^\hatn+\Psi^{\hata})^{2n}}^{\hata}
&
-{}^\mathrm{c}\!
\bbkt{(\beta\,W^{\hatnd}+\Psi^{\hatad})^{2n}}^{\hatad}
=
{}^\mathrm{c}\!
\bbkt{(\Psi^{(\alpha)})^{2n}}^{\hata}
-{}^\mathrm{c}\!
\bbkt{(\Psi^{(\alpha)})^{2n}}^{\hatad}
\nl&
+2n\,{}^\mathrm{c}\!
\bbkt{\Di\Phi^\hata(\Psi^{(\alpha)})^{2n-1}}^{\hata}
-2n\,{}^\mathrm{c}\!
\bbkt{\Di\Phi^\hatad(\Psi^{(\alpha)})^{2n-1}}^{\hatad}
+O(\delta^2).
\ena

To evaluate ${}^\mathrm{c}\!
\bbkt{(\Psi^{(\alpha)})^{2n}}^{\hata}
-{}^\mathrm{c}\!
\bbkt{(\Psi^{(\alpha)})^{2n}}^{\hatad}$,
we expand around the same expectation taken in the constant protocol $\sbkt{\cdots}^{(\alpha)}$.
To be precise, we note that one can write  $\sbkt{f}^\hata=\sbkt{f\,(1+\Di\Gamma^\hata)}^{(\alpha)}$ for any function $f[\hatx]$, where $\Di\Gamma[\hatx]=O(\delta)$.
Then the leading contribution is  
${}^\mathrm{c}\!
\bbkt{(\Psi^{(\alpha)})^{2n}}^{(\alpha)}
-{}^\mathrm{c}\!
\bbkt{(\Psi^{(\alpha)})^{2n}}^{(\alpha)}$, which is obviously vanishing.
The remainder is ${}^\mathrm{c}\!\bbkt{(\Psi^{(\alpha)})^{2n}(\Di\Gamma^\hata-\Di\Gamma^\hatad)}^{(\alpha)}$, which is of order $\Psi^{2n}\,O(\delta)=O(\epsilon^{2n}\,\delta)$.
We have also noted $\Psi=O(\epsilon)$.

The evaluation of the terms ${}^\mathrm{c}\!
\bbkt{\Di\Phi^\hata(\Psi^{(\alpha)})^{2n-1}}^{\hata}
-{}^\mathrm{c}\!
\bbkt{\Di\Phi^\hatad(\Psi^{(\alpha)})^{2n-1}}^{\hatad}$ is more delicate and essential.
Note that a naive order counting would show that these terms are of $O(\epsilon^{2n-1}\delta)+O(\delta^2)$.
This is of $O(\epsilon)$ worse than what we want.

To make an optimal estimate we expand around the same expectation taken in the equilibrium process $\sbkt{\cdots}^{\hata_\mathrm{eq}}$, where $\hata_\mathrm{eq}$ is a protocol\footnote{
In a system with parameters \rlb{alphacomp2} with a nonequilibrium protocol $(\beta,\nu(t),f(t))$, for example, we choose the equilibrium protocol as $(\beta,\nu(t),0)$.
} which always stays in equilibrium and close to $\hata$.
To be precise we note that $\sbkt{f}^\hata=\sbkt{f\,(1+\Di\Omega^\hata)}^{\hata_\mathrm{eq}}$ for any $f[\hatx]$, where $\Di\Omega^\hata[\hatx]=O(\epsilon)$.
Let us also note that the expectation in equilibrium processes satisfies   the symmetry relation
\eq
e^{-\beta F(\alpha_\mathrm{eq})}
\sbkt{\,e^{-\beta W^{\hatn}/2}\,f\,}^{\hata_\mathrm{eq}}
=
e^{-\beta F(\alpha'_\mathrm{eq})}
\sbkt{\,e^{-\beta W^{\hatn^\dagger}/2}\,f^\dagger\,}^{{\hata_\mathrm{eq}}^\dagger},
\lb{eqsym}
\en
where $\hatn$ is a component of the equilibrium protocol $\hata_\mathrm{eq}$, and the time-reversed function $f^\dagger$ is defined by $f^\dagger[\hatxd]=f[\hatx]$.
This can be easily derived from \rlb{DFTsym}.
Since $F(\alpha'_\mathrm{eq})-F(\alpha_\mathrm{eq})=O(\delta)$ and $W=O(\delta)$, we find that $\sbkt{\,f\,}^{\hata_\mathrm{eq}}=\sbkt{\,f^\dagger\,}^{(\hata_\mathrm{eq})^\dagger}+O(f\,\delta)$.

Now it is crucial for us that $(\Di\Phi^\hata)^\dagger=-\Di\Phi^\hatad$ and $(\Psi^{(\alpha)})^\dagger=-\Psi^{(\alpha)}$.
Therefore the leading order in the expansion is
${}^\mathrm{c}\!
\bbkt{\Di\Phi^\hata(\Psi^{(\alpha)})^{2n-1}}^{\hata_\mathrm{eq}}
-{}^\mathrm{c}\!
\bbkt{\Di\Phi^\hatad(\Psi^{(\alpha)})^{2n-1}}^{\hata_\mathrm{eq}^\dagger}
=O(\Di\Phi\,\Psi^{2n-1}\,\delta)=O(\epsilon^{2n-1}\,\delta^2)
$, which can be absorbed into $O(\delta^2)$ in \rlb{formalerrorestimate}.
The next contribution in the expansion is of order $\Di\Phi\,\Psi^{2n-1}\,\Di\Omega=O(\epsilon^{2n}\,\delta)$, which is the lowest order contribution.

\section{Rigorous results about thermodynamic relations}
\label{s:SSTrigorous}


Although our equality \rlb{SSTJ1} is rigorous and works for essentially arbitrary protocol $\hata$, the corresponding thermodynamic relations \rlb{ExCl}, \rlb{ExCled}, \rlb{ExClexcess}, and \rlb{SSTJC2n} have been derived only heuristically.
Here we present rigorous results which (at least partially) justify our claim.
See \cite{MaesNetocny10} for a related work for the linear response theory.


\paragraph*{Models:}
In order to state rigorous results, we shall specify the class of models that we consider.
Although our proof covers quite a general class of models, we shall be moderately concrete here.

As in section~\ref{s:examples}, we fix a finite state space $\calS$ and a corresponding connectivity function $c(x,y)$.
We also fix the reference inverse temperature $\beta$.

We consider an arbitrary real Hamiltonian $\Hn_x$ which depends smoothly on $\nu$, and satisfies $|\Hn_x|\le\bar{E}$ for any $x\in\calS$ and $\nu$ (in the allowed range) with a constant $\bar{E}>0$.
We also introduce a nonequilibrium function $\xi^\kappa_\xty$ for any $x,y\in\calS$ such that $c(x,y)\ne0$, where $\kappa$ is a parameter (or a set of parameters) which takes its value in a compact subset of $\bbR^{n''}$.
We assume that  $\xi^\kappa_\xty$ satisfies  $|\xi^\kappa_\xty|\le1$ for any $\kappa$.

We introduce the degree of nonequilibrium $\epsilon\ge0$, and characterize the system by 
\eq
\alpha:=(\beta,\epsilon,\nu,\kappa).
\en
We then define the corresponding transition rate by
\eq
\Ra_\xty=e^{\epsilon\,\xi^\kappa_\xty}\,\Rbn_\xty,
\lb{Rrigorous}
\en
where the rate $\Rbn_\xty$ for the equilibrium dynamics is defined by \rlb{Req1} or \rlb{Req2}.
Note that both the examples \rlb{RforHeatConduction} and \rlb{Rwithf} are written in the form \rlb{Rrigorous}.

\paragraph*{Nonequilibrium entropy:}
We start from the characterization of our nonequilibrium entropy.
The entropy $S(\alpha)$ is defined by ``thermodynamic'' relation \rlb{S=U-F} in terms of the free energy $F(\alpha)$, which will be defined later in \rlb{Fdef}.

We have already noted that $S(\alpha)$ coincides with the standard entropy in equilibrium, and is close to the Shannon entropy \rlb{SSh} of the stationary distribution $\bsrho^\alpha$ as in \rlb{S=SSh}.
The following is the corresponding rigorous statement.

\begin{theorem}[Nonequilibrium entropy and the Shannon entropy]
\label{t:SandS}
One has
\eq
\Bigl|S(\alpha)-S_\mathrm{Sh}[\bsrho^\alpha]\Bigr|\le A\,\epsilon^3
\lb{S=SSh2}
\en
for any $\alpha$, where $A$ is a positive constant which depends only on the class of models.
\end{theorem}

This theorem is proved in section~\ref{s:KNS} by using Theorem~\ref{t:KNrep}, which is a rigorous version of the representation \rlb{theKN}.

Like the constant $A$ above, all the constants in the following theorems depend only on the choice of the class of models.

\paragraph*{Extended Clausius relation for a step protocol:}
Let us discuss rigorous statements about the extended Clausius relation and the related higher order relations.
As a first step, we treat the step protocol, which is defined by
\eq
\alpha(t)=
\begin{cases}
\alpha=(\beta,\epsilon,\nu,\kappa)&t\in[-\tau,0],\\
\alpha'=(\beta,\epsilon,\nu',\kappa')&t\in(0,\tau].\\
\end{cases}
\lb{step}
\en

\begin{theorem}[Extended Clausius relation for the step protocol]
\label{t:SSTstep}
There are positive constants $B$, $C_n$, and $C'_n$ (with $n=1,2,\ldots$), and we have the following.
Define the amount of change characterizing the protocol \rlb{step} by
\eq
\delta=|\nu-\nu'|+B\epsilon\,|\kappa-\kappa'|.
\lb{deltadef}
\en
Then for the step protocol \rlb{step} with any $\epsilon$, $\nu$, and $\kappa$, one has
\eqa
&\Biggl|
F(\alpha')-F(\alpha)+\frac{1}{\beta}\lti
\sum_{k=1}^{2n-1}\frac{(-1)^k}{2^kk!}
\Bigl\{
{}^\mathrm{c}\!
\bbkt{(\beta\,W^\hatn+\Psi^{\hata})^k}^{\hata}_{\sts\too}
-{}^\mathrm{c}\!
\bbkt{(\beta\,W^{\hatnd}+\Psi^{\hatad})^k}^{\hatad}_{\sts\too}
\Bigr\}
\Biggr|
\nl&
\le C_n\,\epsilon^{2n}\,\delta+C'_n\,\delta^2,
\lb{SSTJCR}
\ena
for any positive integer $n$.
In particular, by setting $n=1$ (and $C=\beta C_1$, $C'=\beta C'_1$), one gets
\eq
\Biggl|
S(\alpha')-S(\alpha)
+\lti\frac{\sbkt{\Theta^\hata}^\hata_{\sts\too}-\sbkt{\Theta^\hatad}^\hatad_{\sts\too}}{2}
\Biggr|
\le C\epsilon^{2}\,\delta+C'\delta^2,
\lb{ECL03}
\en
which is a rigorous version of the extended-Clausius relation \rlb{ExCled}.
\end{theorem}

This theorem is proved in section~\ref{s:SSTproof}, where we make use of the exact relation \rlb{SSTJ1} in Theorem~\ref{t:1} combined with a rigorous perturbative argument.

\begin{figure}[btp]
\begin{center}
\includegraphics[width=9cm]{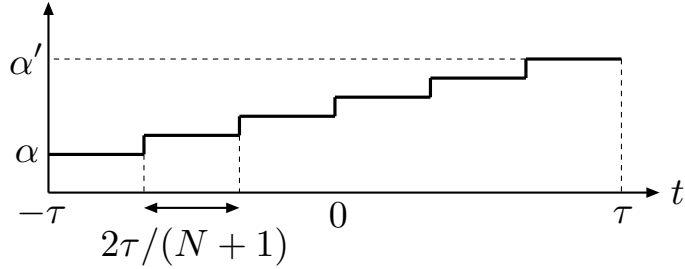}
\end{center}
\caption[dummy]{
The monotone stepwise protocol defined by \rlb{manysteps} and \rlb{manysteps2}.
There are $N$ steps separated by the temporal width $2\tau/(N+1)$.
We can prove the extended Clausius relation \rlb{ECL03B} in the limit where we take $\tau\up\infty$ and then $N\up\infty$.
}
\label{fig:protocolMono}
\end{figure}

\paragraph*{Extended Clausius relation for a quasi-static protocol:}
Although the above Theorem~\ref{t:SSTstep} applies only to protocols which consist of a single step, it is quite natural to expect that corresponding relations for slowly varying protocols are also valid.
Heuristically speaking, one may approximate a smooth slowly varying protocol by a sum of small step protocols, for which the relations \rlb{SSTJCR} or \rlb{ECL03} are valid.

Unfortunately a rigorous control of such quasi-static limit is not easy (probably) for technical reasons.
Instead, we shall here present a (much easier) result for monotone protocols which can be obtained as limits of protocols that consists of many small steps\footnote{
For simplicity we only consider protocols in which the reference inverse temperature $\beta$ is fixed.
It is not difficult to treat protocols where $\beta$ varies; one simply redefines $\beta$ properly when one decomposes the whole protocol into a sum of step protocols. 
}.
See Figure~\ref{fig:protocolMono}.

Define\footnote{%
To be consistent with the notation introduced in section~\ref{s:Markov}, we are here taking $\tau_0=\{(N-1)/(N+1)\}\tau$.
In this case $\tau_0$ also diverges when we let $\tau\up\infty$.
} $\hata=(\alpha(t))_\ttt$ by
\eq
\alpha(t)=\alpha_j\quad\text{for}\ t\in\Bigl[
-\tau+\frac{2\tau}{N+1}j,\,\,-\tau+\frac{2\tau}{N+1}(j+1)
\Bigr),
\lb{manysteps}
\en
where $j=0,\ldots,N$, and 
\eq
\alpha_j:=\Bigl(\beta,\epsilon,
\frac{(N-j)\nu+j\nu'}{N},\frac{(N-j)\kappa+j\kappa'}{N}\Bigr).
\lb{manysteps2}
\en

\begin{theorem}[Extended Clausius relation for a quasi-static protocol]
\label{t:SSTsmooth}
Under the same assumptions as in Theorem~\ref{t:SSTstep}, the following is valid for the protocol \rlb{manysteps}. 
One has, for $n=1,2,\ldots$
\eqa
&\Biggl|
F(\alpha')-F(\alpha)+\frac{1}{\beta}\lim_{N\up\infty}\lti
\sum_{k=1}^{2n-1}\frac{(-1)^k}{2^kk!}
\Bigl\{
{}^\mathrm{c}\!
\bbkt{(\beta\,W^\hatn+\Psi^{\hata})^k}^{\hata}_{\sts\too}
-{}^\mathrm{c}\!
\bbkt{(\beta\,W^{\hatnd}+\Psi^{\hatad})^k}^{\hatad}_{\sts\too}
\Bigr\}
\Biggr|
\nl&
\le C_n\,\epsilon^{2n}\,\delta,
\lb{SSTJCRB}
\ena
and, in particular,
\eq
\Biggl|
S(\alpha')-S(\alpha)
+\lim_{N\up\infty}\lti\frac{\sbkt{\Theta^\hata}^\hata_{\sts\too}-\sbkt{\Theta^\hatad}^\hatad_{\sts\too}}{2}
\Biggr|
\le C\epsilon^{2}\,\delta,
\lb{ECL03B}
\en
where $\delta$ is defined by \rlb{deltadef}.
\end{theorem}

\noindent
{\em Proof:}\/
We simply replace $\delta$ in Theorem~\ref{t:SSTstep} by $\delta/N$, and sum up the inequalities.~\qedm

\bigskip

Note that there are no terms proportional to $\delta^2$ in the right-hand sides of \rlb{SSTJCRB} and \rlb{ECL03B}, reflecting the quasi-static limit $N\up\infty$.

As an important variation, consider the protocol \rlb{manysteps} in which we keep the parameter $\nu$ for the Hamiltonian constant and only vary the parameter $\kappa$.
Since one has $\delta=B\epsilon\,|\kappa-\kappa'|$ by setting $\nu=\nu'$ in \rlb{deltadef}, the extended Clausius relation \rlb{ECL03B} becomes
\eq
\Biggl|
S(\alpha')-S(\alpha)
+\lim_{N\up\infty}\lti\frac{\sbkt{\Theta^\hata}^\hata_{\sts\too}-\sbkt{\Theta^\hatad}^\hatad_{\sts\too}}{2}
\Biggr|
\le C''\epsilon^{3}\,|\kappa-\kappa'|.
\lb{ECL03C}
\en
with $C''=BC$.

To see the implication of \rlb{ECL03C}, take the simple but important example where $\kappa$ takes its value in $[0,1]$, and $\xi^\kappa_\xty=\kappa\,\xi_\xty$ for some nonequilibrium function $\xi_\xty$.
Then by choosing $\kappa=1$ and $\kappa'=0$ in \rlb{ECL03C}, we get
\eq
\Biggl|
S(\beta,\nu)-S(\alpha)
+\lim_{N\up\infty}\lti\frac{\sbkt{\Theta^\hata}^\hata_{\sts\too}-\sbkt{\Theta^\hatad}^\hatad_{\sts\too}}{2}
\Biggr|
\le C''\epsilon^{3},
\lb{ECL03C2}
\en
where $\alpha=(\beta,\epsilon,\nu,\kappa)$.
Note that the system with $\kappa'=0$ is nothing but the equilibrium system with parameters $\beta$ and $\nu$, and therefore $S(\beta,\nu)$ is the equilibrium entropy (see \rlb{Rrigorous}).
If we suppose that we know everything about equilibrium states, the extended Clausius relation \rlb{ECL03C2} implies that one can determine the nonequilibrium entropy $S(\alpha)$ with the precision of $O(\epsilon^2)$ only by (experimentally) measuring the entropy productions $\sbkt{\Theta^\hata}^\hata_{\sts\too}$ and $\sbkt{\Theta^\hatad}^\hatad_{\sts\too}$.

\paragraph*{Extended Clausius relation in terms of excess entropy production:}
Finally we comment on the extended Clausius relation which is written in terms of the excess entropy production.
As we have discussed in section~\ref{s:exClEnt}, the house-keeping entropy production $\Sigma^\hata_\mathrm{hk}$ is defined by \rlb{Thk}, where the entropy production rate $\sigma_\mathrm{st}^\alpha$ is defined by \rlb{sigmasta}.
In the case of stepwise protocol \rlb{manysteps}, the house-keeping entropy production \rlb{Thk}  becomes
\eq
\Sigma^\hata_\mathrm{hk}=
\sum_{j=0}^N\frac{2\tau}{N+1}\,\sigma_\mathrm{st}^{\alpha_j}.
\lb{Thk2}
\en
Then we have the following where the excess entropy production $\sbkt{\Theta^\hata}^\hata_{\sts\too}-\Sigma^\hata_\mathrm{hk}$ plays a central role.

\begin{theorem}[Extended Clausius relation in terms of excess entropy production]
\label{t:SSTexcess}
Under the same assumptions as in Theorem~\ref{t:SSTstep}, we have \eq
\Bigl|
S(\alpha')-S(\alpha)
+\lim_{N\up\infty}\lti\sbkt{\Theta_\mathrm{ex}^\hata}^\hata_{\sts\too}
\Bigr|
\le C\epsilon^{2}\,\delta,
\lb{ECL03D}
\en
for the ``quasi-static'' protocol \rlb{manysteps}.
We have defined the excess entropy production by
\eq
\Theta_\mathrm{ex}^\hata:=
\Theta^\hata-
\Sigma^\hata_\mathrm{hk}.
\lb{Thetaex}
\en
\end{theorem}
\bigskip

This theorem is proved in section~\ref{s:excess}.

Note that the divergence (as $\tau\up\infty$) of the total entropy production $\sbkt{\Theta^\hata}^\hata_{\sts\too}$ is precisely canceled by that of the house-keeping entropy production $\Sigma^\hata_\mathrm{hk}$.

\paragraph*{The inequality corresponding to the extended Clausius relation:}
As we have noted in section~\ref{s:SSTexample}, the standard Clausius relation $S(\alpha'_\mathrm{eq})-S(\alpha_\mathrm{eq})\simeq-\sbkt{\Theta^{\hata_\mathrm{eq}}}^{\hata_\mathrm{eq}}_{\mathrm{eq}\too}$, which becomes an exact equality in the quasi-static limit, is associated with the inequality $S(\alpha'_\mathrm{eq})-S(\alpha_\mathrm{eq})\ge-\sbkt{\Theta^{\hata_\mathrm{eq}}}^{\hata_\mathrm{eq}}_{\mathrm{eq}\too}$, which is valid for {\em any}\/ equilibrium protocol $\hata_\mathrm{eq}$.
We shall make some remarks regarding the inequality corresponding to our extended Clausius relation.

We first note that there can be no inequalities corresponding to the extended Clausius relation expressed in the forms \rlb{ExCl}, \rlb{ExCled}, or \rlb{ECL03B}.
To see this suppose that the inequality $S(\alpha')-S(\alpha)\gtrsim-({\sbkt{\Theta^\hata}^\hata-\sbkt{\Theta^\hatad}^\hatad})/{2}$ is valid for general protocols $\hata$.
But if one replaces $\hata$ by its time-reversal $\hatad$, this inequality becomes  $S(\alpha')-S(\alpha)\lesssim-({\sbkt{\Theta^\hata}^\hata-\sbkt{\Theta^\hatad}^\hatad})/{2}$.
This means that a general inequality is impossible (unless the equality is valid for any $\hata$).

If we write the extended Clausius relation by using the excess entropy production as in \rlb{ECL03D}, on the other hand, we can prove the following.

\begin{theorem}[Extended Clausius inequality]
\label{t:ClausiusIneq}
Let $\hata$ be an arbitrary protocol as defined in section~\ref{s:Markov}.
We fix $\tau_0$.
Then we have
\eq
S(\alpha')-S(\alpha)
\ge
\lti\Bigl\{
-\sbkt{\Theta_\mathrm{ex}^\hata}^\hata_{\sts\too}
-\tilde{C}\epsilon^{2}\int_{-\tau}^\tau dt\,\sum_{x\in\calS}|\dot{p}_x(t)|
\Bigr\}-2A\epsilon^3,
\lb{ECLIneq}
\en
where $\bsp(t)=(p_x(t))_{x\in\calS}$ is the probability distribution at time $t$, i.e., the solution of the master equation \rlb{master}, \rlb{master2} with the initial condition $\bsp(-\tau)=\bsrho^\alpha$, and $\tilde{C}$ is a constant which depends only on the class of models.
$A$ is the constant introduced in Theorem~\ref{t:SandS}.
\end{theorem}

We see that \rlb{ECLIneq} has the precise form that one expects as the inequality corresponding to the extended Clausius relation \rlb{ECL03D}.
It should be noted, however, that the error term $\tilde{C}\epsilon^{2}\int_{-\tau}^\tau dt\,\sum_{x\in\calS}|\dot{p}_x(t)|$ depends explicitly on the solution of the master equation \rlb{master}, \rlb{master2}.

The theorem is proved in section~\ref{s:ClIneq}, where we make use of the  standard inequality for Markov processes.

\section{Method based on the time reversal symmetry}
\label{s:proof1}
We prove our theorems in sections~\ref{s:proof1}, \ref{s:matrix}, \ref{s:KNS}, and \ref{s:ClIneq}.

In this section we prove Theorems~\ref{t:1} and \ref{t:Nakagawa} by using the time reversal symmetry of the path probability.
In the course we use some lemmas which are proved in section~\ref{s:matrix}.

\subsection{Some definitions and basic symmetry}
In addition to the expectation \rlb{fss}, we  define a new expectation
\eq
\sbkt{f}^{\hata}_{x\too}:=
\int\Dx\,f[\hatx]\,\delta_{x(-\tau),x}\,\calT^{\hata}[\hatx],
\lb{fx}
\en
in which the system starts from a specified initial state $x\in\calS$.

It is also convenient to define the ``unnormalized expectations'' in which the final state is specified as
\eq
[f]^{\hata}_{\sts\to y}:=
\int\Dx\,f[\hatx]\,\rho^\alpha_{x(-\tau)}\,\delta_{x(\tau),y}\,\calT^{\hata}[\hatx],
\lb{fsy}
\en
and
\eq
[f]^{\hata}_{\xty}:=
\int\Dx\,f[\hatx]\,\delta_{x(-\tau),x}\,\delta_{x(\tau),y}\,\calT^{\hata}[\hatx].
\lb{fxy}
\en
The corresponding normalized expectation will appear later in \rlb{<>sx}.

Let us derive the well-known symmetry relation, which will be a basis of the present work.
From the definitions \rlb{xdagger} of the time-reversed path and \rlb{Wdef} of the weight, one finds
\eqa
\calT^{\hatad}[\hatxd]&=
\prod_{j=1}^nR^{\alpha(t_j)}_{x_{j}\to x_{j-1}}\,
\prod_{j=0}^n\exp\Bigl[-\int_{t_j}^{t_{j+1}}dt\,\lambda_{x_j}^{\alpha(t)}\Bigr]
\notag
\intertext{which differs from  \rlb{Wdef} only in the subscripts of the transition rates.
By using the definition \rlb{thetadef} of $\thetaa$, we get
}
&=\Bigl(\prod_{j=1}^nR^{\alpha(t_j)}_{x_{j-1}\to x_{j}}\,\exp[{-\theta}^{\alpha(t_j)}_{x_{j-1}\to x_{j}}]\Bigr)
\prod_{j=0}^n\exp\Bigl[-\int_{t_j}^{t_{j+1}}dt\,\lambda_{x_j}^{\alpha(t)}\Bigr]
\nl
&=\exp\bigl[{-\Theta^{\hata}[\hatx]}\bigr]\,\calT^{\hata}[\hatx],
\lb{DFT}
\ena
where $\Theta^{\hata}[\hatx]$ is defined in \rlb{Thetadef}.
The relation \rlb{DFT}, which is quite standard in nonequilibrium physics, is the path version of the local detailed balance \rlb{thetadef}, and is sometimes called the ``detailed fluctuation theorem''.

\subsection{Proof of Theorem~\protect{\ref{t:1}}}
\label{s:t1proof}
By using \rlb{TWPH}, one can rewrite \rlb{DFT} as
\eq
e^{-\beta H^{\nu(-\tau)}_{x(-\tau)}-(\beta W^\hatn[\hatx]+\Psi^{\hata}[\hatx])/2}\,
\calT^{\hata}[\hatx]
=
e^{-\beta H^{\nu(\tau)}_{x(\tau)}-(\beta W^\hatnd[\hatxd]+\Psi^{\hatad}[\hatxd])/2}\,
\calT^{\hatad}[\hatxd],
\lb{DFTsym}
\en
where we made use of \rlb{Thetasym}.
By integrating over all path $\hatx$ such that $x(-\tau)=x$ and $x(\tau)=y$, and recalling the definition \rlb{fxy}, we get
\eq
e^{-\beta H^\nu_x}\,\bigl[e^{-(\beta W^\hatn+\Psi^\hata)/2}\bigr]_{\xty}^\hata
=
e^{-\beta H^{\nu'}_y}\,\bigl[e^{-(\beta W^\hatnd+\Psi^\hatad)/2}\bigr]_{y\to x}^\hatad.
\lb{DFT3new}
\en 

At this stage the following ``splitting lemma", which will be proved in section~\ref{s:splitproof}, is essential.
The lemma enables us to treat the expectation $\sbkt{\,\ldots\,}^{[-\tau/2,\tau/2],\hata}_{\sts\too}$ (where the initial distribution is chosen as the stationary distribution) without using the stationary distribution $\bsrho^\alpha$ explicitly.
Technically speaking the use of the ``splitting lemma" is one of the new points in the present work.
\begin{lemma}
\label{l:split}
Let $f[\hatx]$ be an arbitrary nonnegative (and nonvanishing) function of path which depends only on $x(t)$ with $t\in[-\tau/4,\tau/4]$.
Then for  arbitrary protocol $\hata$ and $x,y\in\calS$, one has
\eq
\lti
\frac{Y(\alpha)\,Y(\alpha')\,[\,f\,e^{-\Psi^\hata/2}\,]_{\xty}^\hata}
{\sbkt{\,e^{-\Psi^{(\alpha)}/2}\,}^{[-\tau,-\tau/2],(\alpha)}_{x\too}\,
\sbkt{\,f\,e^{-\Psi^\hata/2}\,}^{[-\tau/2,\tau/2],\hata}_{\sts\too}\,
[\,e^{-\Psi^{(\alpha')}/2}\,]^{[\tau/2,\tau],(\alpha')}_{\sts\to y}}
=1,
\lb{decompose}
\en
where $Y(\alpha)$ is a certain positive function.
We have defined the expectations in the limited time intervals (such as $[-\tau,-\tau/2]]$) by replacing the time interval $[-\tau,\tau]$ in \rlb{fss}, \rlb{fx}, \rlb{fsy}, and \rlb{fxy} with the specified ones.
We also see that the limit
\eq
\lti\frac{\sbkt{\,e^{-\Psi^{(\alpha)}/2}\,}^{[-\tau,-\tau/2],(\alpha)}_{x\too}}
{[\,e^{-\Psi^{(\alpha)}/2}\,]^{[\tau/2,\tau],(\alpha)}_{\sts\to x}}
\lb{decomposition2}
\en
exists for any $x\in\calS$.
\end{lemma}

With a slight abuse of notation, we have denoted here by $\Psi$ the entropy production in each subinterval.
Thus $\Psi^{(\alpha)}$ in $\sbkt{\cdots}^{[-\tau,-\tau/2],(\alpha)}_{x\too}$, for example, actually means $\Psi^{[-\tau,-\tau/2],(\alpha)}$.

By applying the decomposition \rlb{decompose} to the equality \rlb{DFT3new}, we have
\eq
\lti
\frac{
e^{-\beta H^\nu_x}\,
\sbkt{\,e^{-\Psi^{(\alpha)}/2}\,}^{[-\tau,-\tau/2],(\alpha)}_{x\too}\,
\sbkt{\,e^{-(\beta W^\hatn+\Psi^\hata)/2}\,}^{[-\tau/2,\tau/2],\hata}_{\sts\too}\,
[\,e^{-\Psi^{(\alpha')}/2}\,]^{[\tau/2,\tau],(\alpha')}_{\sts\to y}
}
{
e^{-\beta H^{\nu'}_y}\,
\sbkt{\,e^{-\Psi^{(\alpha')}/2}\,}^{[-\tau,-\tau/2],(\alpha')}_{y\too}\,
\sbkt{\,e^{-(\beta W^\hatnd+\Psi^\hatad)/2}\,}^{[-\tau/2,\tau/2],\hatad}_{\sts\too}\,
[\,e^{-\Psi^{(\alpha)}/2}\,]^{[\tau/2,\tau],(\alpha)}_{\sts\to x}
}
=1,
\lb{DFT5}
\en
which is an essential relation for us.

Let us first set $\hata=(\alpha)$, a constant protocol.
Then \rlb{DFT5} becomes
\eq
\lti
\frac{
e^{-\beta\Hn_x}\,
\sbkt{\,e^{-\Psi^{(\alpha)}/2}\,}^{[-\tau,-\tau/2],(\alpha)}_{x\too}\,
[\,e^{-\Psi^{(\alpha)}/2}\,]^{[\tau/2,\tau],(\alpha)}_{\sts\to y}
}
{
e^{-\beta\Hn_y}\,
\sbkt{\,e^{-\Psi^{(\alpha)}/2}\,}^{[-\tau,-\tau/2],(\alpha)}_{y\too}\,[\,e^{-\Psi^{(\alpha)}/2}\,]^{[\tau/2,\tau],(\alpha)}_{\sts\to x}
}
=1.
\lb{DFT6}
\en
Since the limit \rlb{decomposition2} exists, this  implies an interesting relation
\eq
e^{-\beta\Hn_x}\,
\lti\frac{\sbkt{\,e^{-\Psi^{(\alpha)}/2}\,}^{[-\tau,-\tau/2],(\alpha)}_{x\too}}
{[\,e^{-\Psi^{(\alpha)}/2}\,]^{[\tau/2,\tau],(\alpha)}_{\sts\to x}}
=
e^{-\beta\Hn_y}\,
\lti\frac{\sbkt{\,e^{-\Psi^{(\alpha)}/2}\,}^{[-\tau,-\tau/2],(\alpha)}_{y\too}}
{[\,e^{-\Psi^{(\alpha')}/2}\,]^{[\tau/2,\tau],(\alpha)}_{\sts\to y}}
\lb{DFT7}
\en
for any $x,y\in\calS$.
In other words, the quantity equated in \rlb{DFT7} is independent of $x$ (or of $y$).
Let us examine this quantity for $\alpha$ which corresponds to equilibrium, where we have $\alpha=(\beta,\nu)$, and $\Psi^{(\alpha)}=0$.
Since $\lti[1]^{[\tau/2,\tau],(\alpha)}_{\sts\to x}=\rho^{\beta,\nu}_x=e^{\beta\{F(\beta,\nu)-\Hn_x\}}$, where $F(\beta,\nu)$ is the standard equilibrium free energy, we see that \rlb{DFT7} is equal to $e^{-\beta F(\beta,\nu)}$.
This motivates us to define for general $\alpha$ the corresponding free energy $F(\alpha)$ by
\eq
e^{-\beta F(\alpha)}:=e^{-\beta\Hn_x}\,
\lti\frac{\sbkt{\,e^{-\Psi^{(\alpha)}/2}\,}^{[-\tau,-\tau/2],(\alpha)}_{x\too}}
{[\,e^{-\Psi^{(\alpha)}/2}\,]^{[\tau/2,\tau],(\alpha)}_{\sts\to x}}.
\lb{Fdef}
\en

By considering a general protocol $\hata$, and combining  \rlb{DFT5} with \rlb{Fdef}, we get
\eq
\lti\frac{
e^{-\beta F(\alpha)}\,
\sbkt{\,e^{-(\beta W^\hatn+\Psi^\hata)/2}\,}^{[-\tau/2,\tau/2],\hata}_{\sts\too}
}{
e^{-\beta F(\alpha')}\,
\sbkt{\,e^{-(\beta W^\hatnd+\Psi^\hatad)/2}\,}^{[-\tau/2,\tau/2],\hatad}_{\sts\too}
}
=1,
\lb{FF1id}
\en
which (after replacing $\tau/2$ with $\tau$) is the desired equality \rlb{SSTJ1} in Theorem~\ref{t:1}.

\paragraph*{Remark:}
The equality  \rlb{SSTJ4} can be proved  essentially in the same manner as  \rlb{SSTJ1}.
One simply starts from the symmetry relation \rlb{DFT} and repeats the same procedure.

In the derivation, the quantity $\tiS(\alpha)$ is naturally defined as
\eq
\tiS(\alpha):=
\lti\log\frac{\sbkt{\,e^{-\Theta^{(\alpha)}/2}\,}^{[-\tau,-\tau/2],(\alpha)}_{x\too}}
{[\,e^{-\Theta^{(\alpha)}/2}\,]^{[\tau/2,\tau],(\alpha)}_{\sts\to x}}.
\lb{tSdef}
\en
In an equilibrium state, this definition gives
\eq
\tiS(\alpha)=
\log Z(\alpha)+
\log\frac{\bbkt{e^{\beta H/2}}_\mathrm{eq}}
{\bbkt{e^{-\beta H/2}}_\mathrm{eq}},
\lb{tiSeq}
\en
which is different from the standard equilibrium entropy
$S(\alpha)=\log Z(\alpha)+\beta\,\sbkt{H}_\mathrm{eq}$.

\subsection{Proof of Theorem~\protect{\ref{t:Nakagawa}}}
\label{s:tNakproof}
By using \rlb{TWPH}, we now rewrite \rlb{DFT} as
\eq
e^{-\beta H^{\nu(-\tau)}_{x(-\tau)}-\beta W^\hatn[\hatx]-\Psi^{\mathrm{b},\hata}[\hatx]/2}\,
\calT^{\hata}[\hatx]
=
e^{-\beta H^{\nu(\tau)}_{x(\tau)}-\Psi^{\mathrm{i},\hatad}[\hatxd]
-\Psi^{\mathrm{b},\hatad}[\hatxd]/2}\,
\calT^{\hatad}[\hatxd],
\en
where we have decomposed the entropy production as $\Psi^\hata[\hatx]=\Psi^{\mathrm{b},\hata}[\hatx]+\Psi^{\mathrm{i},\hata}[\hatx]$, where $\Psi^{\mathrm{b},\hata}[\hatx]:=\Psi^{[-\tau,-\tau/4],\hata}[\hatx]+\Psi^{[\tau/4,\tau],\hata}[\hatx]$ and $\Psi^{\mathrm{i},\hata}[\hatx]:=\Psi^{[-\tau/4,\tau/4],\hata}[\hatx]$.
Again we integrate over $\hatx$ with $x(-\tau)=x$ and $x(\tau)=y$ to get
\eq
e^{-\beta\Hn_x}\,
[e^{-\beta W^\hatn-\Psi^{\mathrm{b},\hata}/2}]^\hata_{\xty}
=
e^{-\beta H^{\nu'}_y}\,
[e^{-\Psi^{\mathrm{i},\hatad}
-\Psi^{\mathrm{b},\hatad}/2}]^\hatad_{y\to x},
\en
By using the decomposition \rlb{decompose} as before, this implies
\eq
\lti
\frac{
e^{-\beta\Hn_x}\,
\sbkt{\,e^{-\Psi^{(\alpha)}/2}\,}^{[-\tau,-\tau/2],(\alpha)}_{x\too}\,
\sbkt{\,e^{-\beta W^\hatn-\Psi^{\mathrm{b}',\hata}/2}\,}^{[-\tau/2,\tau/2],\hata}_{\sts\too}\,
[\,e^{-\Psi^{(\alpha')}/2}\,]^{[\tau/2,\tau],(\alpha')}_{\sts\to y}
}
{
e^{-\beta H^{\nu'}_y}\,
\sbkt{\,e^{-\Psi^{(\alpha')}/2}\,}^{[-\tau,-\tau/2],(\alpha')}_{y\too}\,
\sbkt{\,e^{-\Psi^{\mathrm{i},\hatad}-\Psi^{\mathrm{b}',\hatad}/2}\,}^{[-\tau/2,\tau/2],\hatad}_{\sts\too}\,
[\,e^{-\Psi^{(\alpha)}/2}\,]^{[\tau/2,\tau],(\alpha)}_{\sts\to x}
}
=1,
\en
where $\Psi^{\mathrm{b}',\hata}[\hatx]:=\Psi^{[-\tau/2,-\tau/4],\hata}[\hatx]+\Psi^{[\tau/4,\tau/2],\hata}[\hatx]$.
We can use the definition \rlb{Fdef} of the free energy to rewrite this as
\eq
\lti
\frac{
e^{-\beta F(\alpha)}\,
\sbkt{\,e^{-\beta W^\hatn-\Psi^{\mathrm{b}',\hata}/2}\,}^{[-\tau/2,\tau/2],\hata}_{\sts\too}
}{
e^{-\beta F(\alpha')}\,
\sbkt{\,e^{-\Psi^{\mathrm{i},\hatad}-\Psi^{\mathrm{b}',\hatad}/2}\,}^{[-\tau/2,\tau/2],\hatad}_{\sts\too}
}
=1.
\en
Now, by rewriting $\tau/2$ as $\tau$ (which is allowed because we let $\tau\up\infty$),  this precisely becomes
\eq
\lti
\frac{
e^{-\beta F(\alpha)}\,\sbkt{e^{-\beta\,W^\hatn}}^{\hata}_{\text{\rm$\Psi$-mod}}
}{
e^{-\beta F(\alpha')}\,\bbkt{\exp[-\Psi^{[-\tau/2,\tau/2],\hatad}]}^{\hatad}_{\text{\rm$\Psi$-mod}}
}
=1,
\lb{Nakagawapre}
\en
where we used $\Psi$-modified expectation defined in \rlb{Psimod}.
Finally by using the relation \rlb{NkagawaFF} below, we get the desired \rlb{Nakagawa} in Theorem~\ref{t:Nakagawa} from  \rlb{Nakagawapre}. 

\begin{lemma}
\label{l:Nakagawa}
Let $f[\hatx]$ be a function which depends only on $x(t)$ with $t\in[-\tao,\tao]$.
Then
\eq
\lti\sbkt{f}^{\hata}_{\text{\rm$\Psi$-mod}}=\lti\sbkt{f}^{\hata}_{\sts\too}
\lb{NkagawaFF}
\en
\end{lemma}

\section{Method based on modified rate matrix}
\label{s:matrix}
Here we prove the technically important splitting lemma (Lemma~\ref{l:split}), and our main results  about SST summarized in Theorems~\ref{t:SSTstep} and \ref{t:SSTexcess}.
The expressions of  various quantities in terms of matrices obtained by modifying the rate matrix $\sfR^\alpha$ play essential roles\footnote{
Such a technique is common, for example, in the large deviation theory \cite{Dembo}.
Similar technique was used for steady state thermodynamics in \cite{SagawaHayakwa}.
}.
The desired results then follow from the Perron-Frobenius theorem.

\subsection{Modified rate matrix and its eigenvectors}
\label{s:modifiedratematrix}
In what follows we treat vectors whose components are indexed by $x\in\calS$.
We use boldface symbols like $\bsv=(v_x)_{x\in\calS}$ for column vectors, and arrowed symbols like $\vec{u}=(u_x)_{x\in\calS}$ for row vectors.
We use the standard notation where $\vec{u}\bsv=\sum_{x\in\calS}u_xv_x$ denotes the scalar product, and $\bsv\vec{u}$ denotes the matrix whose $xy$-entry is $v_xu_y$ (i.e., the Kronecker product).
For a matrix $\sfA$, we write $\vec{u}\sfA\bsv=\sum_{x,y\in\calS}u_x(\sfA)_{xy}v_y$.
We let  $\vone:=(1,1,\ldots,1)$ the row vector whose components are all 1.
Finally $\bsdelta^{(x)}$ and $\vdelta^{(x)}$ denote the column and row vectors, respectively, whose $x$-component is 1 and all the other components are 0.

Since $\sfR^\alpha$ is a transition rate matrix, there is a unique positive vector $\bsrho^\alpha$ such that $\vone\bsrho^\alpha=1$ and $\sfR^\alpha\bsrho^\alpha=0$.
Here 0 is the Perron-Frobenius eigenvalue of $\sfR^\alpha$, and the corresponding left eigenvector is $\vone$.

We define a modified matrix $\tR^\alpha$ by specifying its entries as
\eq
(\tR^\alpha)_{yx}:=R^\alpha_{yx}\,e^{-\psi^\alpha_\xty/2}.
\en
Although  $\tR^\alpha$ is no longer a transition rate matrix, it still satisfies the conditions for the Perron-Frobenius theorem.
We denote by $\mu^\alpha\in\bbR$ the Perron-Frobenius eigenvalue of  $\tR^\alpha$, and by $\bsphi^\alpha=(\varphi^\alpha_x)_{x\in\calS}$ and $\vxi^\alpha=(\xi^\alpha_x)_{x\in\calS}$ the corresponding right and left eigenvectors, respectively, i.e.,
\eq
\tR^\alpha\bsphi^\alpha=\mu^\alpha\bsphi^\alpha,\quad
\vxi^\alpha\,\tR^\alpha=\mu^\alpha\vxi^\alpha.
\en
We can assume $\varphi^\alpha_x>0$ and $\xi^\alpha_x>0$ for any $x\in\calS$.
We normalize these vectors so that $\vone\bsphi^\alpha=1$ and $\vxi^\alpha\bsphi^\alpha=1$ (more precisely, $\sum_{x\in\calS}\varphi^\alpha_x=1$ and $\sum_{x\in\calS}\xi^\alpha_x\varphi^\alpha_x=1$).

From \rlb{thetadef} and \rlb{psidef}, we see that $(\tR^\alpha)_{xy}=(\tR^\alpha)_{yx}\,e^{\beta(\Hn_y-\Hn_x)}$.
From the eigenvalue equation for $\vxi^\alpha$, we see
\eq
\mu^\alpha\,\xi^\alpha_y=\sum_{x\in\calS}\xi^\alpha_x(\tR^\alpha)_{xy}
=e^{\beta\Hn_y}
\sum_{x\in\calS}(\tR^\alpha)_{yx}e^{-\beta\Hn_x}\xi^\alpha_x,
\en
which implies that $(e^{-\beta\Hn_x}\xi^\alpha_x)_{x\in\calS}$ is the right eigenvector of $\tR^\alpha$.
We can therefore write
\eq
\varphi^\alpha_x=\frac{e^{-\beta\Hn_x}\,\xi^\alpha_x}{Z(\alpha)},
\lb{phiZ}
\en
with a certain function $Z(\alpha)$ of the parameters $\alpha$.
In equilibrium system, it coincides with the canonical partition function  since $\vxi^\alpha=\vone$.
We note in passing that the relations $\vone\bsphi^\alpha=1$ and $\vxi^\alpha\bsphi^\alpha=1$ imply $Z(\alpha)=\sum_{x\in\calS}e^{-\beta\Hn_x}\xi^\alpha_x$ and $Z(\alpha)=\sum_{x\in\calS}e^{-\beta\Hn_x}(\xi^\alpha_x)^2$, respectively.

A straightforward consequence of the Perron-Frobenius theorem is that there exist constants $C,C'>0$ and $\gamma>0$, and one has
\eq
\normi{e^{t\,\tR^\alpha}}\le C\,e^{\mu^\alpha\,t},
\lb{b1}
\en
and
\eq
\normi{e^{t\,\tR^\alpha}-e^{\mu^\alpha\,t}\bsphi^\alpha\vxi^\alpha}\le 
C'\,e^{\mu^\alpha\,t}\,e^{-\gamma\,t},
\lb{b2}
\en
for any $t\ge0$.
We can assume that the constants $C,C',\gamma$ are independent of $\alpha$.
Here we used the matrix norm $\Vert\sfA\Vert_\infty:=\max_{x,y\in\calS}|(\sfA)_{xy}|$.
Note that $\bsphi^\alpha\vxi^\alpha$ is the (non-orthogonal) projection onto the Perron-Frobenius eigenvector $\bsphi^\alpha$.

\subsection{Splitting Lemma}
\label{s:splitproof}
Let us prove Lemma~\ref{l:split}, which played an essential role in the proof of the equalities.
The key for the proof is the following bound.
\begin{lemma}
Let $Y(\alpha)=\vxi^\alpha\bsrho^\alpha$.
Then for any $t_1,t_2\ge0$, one has
\eq
\normi{Y(\alpha)\,e^{(t_1+t_2)\tR^\alpha}-e^{t_1\tR^\alpha}\bsrho^\alpha\vone\,e^{t_2\tR^\alpha}}\le C''\,e^{\mu^\alpha\,t}(e^{-\gamma\,t_1}+e^{-\gamma\,t_2}),
\lb{b3}
\en
for any $\alpha$, where $C''>0$ is a constant.
\end{lemma}

\noindent{\em Proof:}\/
From \rlb{b1}, one roughly finds that
\eq
e^{(t_1+t_2)\tR^\alpha}\simeq e^{\mu^\alpha\,(t_1+t_2)}\bsphi^\alpha\vxi^\alpha,
\en
and
\eq
e^{t_1\tR^\alpha}\bsrho^\alpha\vone\,e^{t_2\tR^\alpha}\simeq
e^{\mu^\alpha\,t_1}\bsphi^\alpha\vxi^\alpha
\bsrho^\alpha\vone\,
e^{\mu^\alpha\,t_2}\bsphi^\alpha\vxi^\alpha
=Y(\alpha)\,e^{\mu^\alpha\,(t_1+t_2)}\bsphi^\alpha\vxi^\alpha.
\en
It is a standard exercise to make this into a rigorous estimate by using \rlb{b1} and \rlb{b2}.~\qedm

\bigskip

We denote by $\tF$ the matrix whose entry $(\tF)_{yx}$ is equal to $[\,f\,e^{-\Psi^\hata/2}\,]_{\xty}^{[-\tau/4,\tau/4],\hata}$.
Then one can write
\eqg
[\,f\,e^{-\Psi^\hata/2}\,]_{\xty}^\hata=
\vdelta^{(y)}e^{(3\tau/4)\tR^{\alpha'}}\,\tF\,e^{(3\tau/4)\tR^{\alpha}}\,\bsdelta^{(x)},
\lb{[F]xy]}
\\
\sbkt{\,e^{-\Psi^{(\alpha)}/2}\,}^{[-\tau,-\tau/2],(\alpha)}_{x\too}=
\vone\,e^{(\tau/2)\tR^{\alpha}}\,\bsdelta^{(x)},
\lb{denom1}
\\
\sbkt{\,f\,e^{-\Psi^\hata/2}\,}^{[-\tau/2,\tau/2],\hata}_{\sts\too}=
\vone\,e^{(\tau/4)\tR^{\alpha'}}\,\tF\,e^{(\tau/4)\tR^{\alpha}}\bsrho^\alpha,
\lb{denom2}
\\
[\,e^{-\Psi^{(\alpha')}/2}\,]^{[\tau/2,\tau],(\alpha')}_{\sts\to y}=
\vdelta^{(y)}\,e^{(\tau/2)\tR^{\alpha'}}\bsrho^\alpha.
\lb{denom3}
\eng
We apply \rlb{b3} to \rlb{[F]xy]} and split $3\tau/4$ into $\tau/2$ and $\tau/4$ as
\eq
Y(\alpha)\,Y(\alpha')\,[\,f\,e^{-\Psi^\hata/2}\,]_{\xty}^\hata
\simeq
\vdelta^{(y)}e^{(\tau/2)\tR^{\alpha'}}\bsrho^{\alpha'}\vone\,e^{(\tau/4)\tR^{\alpha'}}\,\tF\,e^{(\tau/4)\tR^{\alpha}}\bsrho^{\alpha}\vone\,e^{(\tau/2)\tR^{\alpha}}\,\bsdelta^{(x)}
\en
where the right-hand side is equal to the product of \rlb{denom1}, \rlb{denom2}, and \rlb{denom3}.
This roughly shows that 
\eq
\Bigl|\,(\text{numerator of \rlb{decompose}})-(\text{denominator of \rlb{decompose}})\,\Bigr|
\le \tilde{C}\,e^{(3/4)(\mu^{\alpha'}+\mu^\alpha)\tau}\,e^{-\gamma\,\tau/4}\,\normi{\tF},
\en
where $\tilde{C}$ is a constant.
On the other hand, we can also show that
\eq
\Bigl|\,(\text{denominator of \rlb{decompose}})\,\Bigr|
\ge\tilde{C}'\,e^{(3/4)(\mu^{\alpha'}+\mu^\alpha)\tau}\,\normi{\tF}
\en
by noting that the Perron-Frobenius theorem ensures that, in the expression in the right-hand side of \rlb{denom2}, the entry $(\tF)_{yx}$ for any $x,y$ has a uniformly nonvanishing contribution.
This proves \rlb{decompose}, which is the main claim in Lemma~\ref{l:split}.

The existence of the limit \rlb{decomposition2} and the proof of Lemma~\ref{l:Nakagawa} are easy.

\subsection{Description of key quantities in terms of the eigenvectors}
We shall examine how the two important quantities in our theory can be written in terms the language of the present section.
Although these observations are not directly used in the proof (and thus can be omitted) they may shed light on mathematical structures behind our thermodynamic relations.

We first examine the nonequilibrium free energy defined in \rlb{Fdef}.
Clearly we can rewrite \rlb{Fdef} as
\eqa
e^{-\beta F(\alpha)}&=e^{-\beta\Hn_x}\,
\lti\frac{\vone\,e^{(\tau/2)\tR^\alpha}\bsdelta_x}
{\vdelta_x\,e^{(\tau/2)\tR^\alpha}\bsrho^\alpha}
=e^{-\beta \Hn_x}\,
\frac{\vone\bsphi^\alpha\vxi^\alpha\bsdelta_x}
{\vdelta_x\bsphi^\alpha\vxi^\alpha\bsrho^\alpha}
\notag
\intertext{where we used \rlb{b2}.
By recalling $Y(\alpha)=\vxi^\alpha\bsrho^\alpha$, this becomes}
&=e^{-\beta\Hn_x}\,\frac{\xi^\alpha_x}{\varphi^\alpha_x\,Y(\alpha)}
=\frac{Z(\alpha)}{Y(\alpha)},
\lb{FRep}
\ena
where we used \rlb{phiZ}.
Thus the free energy is neatly expressed in terms of the quantities obtained from $\bsphi^\alpha$ and $\vxi^\alpha$ as 
\eq
F(\alpha)=\frac{1}{\beta}\bigl\{\log Y(\alpha)-\log Z(\alpha)\bigr\}.
\lb{FYZ}
\en

Next we examine the right-hand side of our main equality \rlb{SSTJ1} in the special case of the step protocol \rlb{step}, which was treated in section~\ref{s:SSTrigorous}.
We of course know (rigorously) that it is equal to $F(\alpha')-F(\alpha)$, it might be interesting to analyze the right-hand side as it is.

Let us define the work matrix $\sfW^\hatn$ by
\eq
(\sfW^\hatn)_{yx}:=
\begin{cases}
H^{\nu'}_x-\Hn_x&\text{if $x=y$};\\
0&\text{if $x\ne y$}.
\end{cases}
\lb{workmatrix}
\en
Then the right-hand side of \rlb{SSTJ1} (for the step protocol) is rewritten as
\eqa
-\frac{1}{\beta}\lti\log&\frac{\bbkt{\,\exp[-(\beta\,W^\hatn+\Psi^{\hata})/2]\,}^{\hata}_{\sts\too}}
{\bbkt{\,\exp[-(\beta\,W^{\hatnd}+\Psi^{\hatad})/2]\,}^{\hatad}_{\sts\too}}
=
-\frac{1}{\beta}\lti\log\frac{\vone\,e^{\tau\tR^{\alpha'}}e^{-\beta\sfW^\hatn/2}\,
e^{\tau\tR^{\alpha}}\bsrho^\alpha}
{\vone\,e^{\tau\tR^{\alpha}}e^{-\beta\sfW^\hatnd/2}\,
e^{\tau\tR^{\alpha'}}\bsrho^{\alpha'}}
\notag
\intertext{By using \rlb{b2}, this becomes}
&=
-\frac{1}{\beta}\lti\log\frac{
e^{(\mu^\alpha+\mu^{\alpha'})\tau}\vone\,\bsphi^{\alpha'}\vxi^{\alpha'}
e^{-\beta\sfW^\hatn/2}\bsphi^{\alpha}\vxi^{\alpha}\bsrho^{\alpha}
+(\text{remainder})
}{
e^{(\mu^\alpha+\mu^{\alpha'})\tau}\vone\,\bsphi^{\alpha}\vxi^{\alpha}
e^{-\beta\sfW^\hatnd/2}\bsphi^{\alpha'}\vxi^{\alpha'}\bsrho^{\alpha'}
+(\text{remainder})
}
\nl&
=
-\frac{1}{\beta}\log\frac{
Y(\alpha)\,\vxi^{\alpha'}
e^{-\beta\sfW^\hatn/2}\bsphi^{\alpha}
}{
Y(\alpha')\,\vxi^{\alpha}
e^{-\beta\sfW^\hatnd/2}\bsphi^{\alpha'}
}
\nl&
=
\frac{1}{\beta}\{\log Y(\alpha')-\log Y(\alpha)\}
-\frac{1}{\beta}\log\frac{
\vxi^{\alpha'}
e^{-\beta\sfW^\hatn/2}\bsphi^{\alpha}
}{
\vxi^{\alpha}
e^{-\beta\sfW^\hatnd/2}\bsphi^{\alpha'}
}.
\lb{SSTJ1B1}
\ena
Recalling \rlb{SSTJ1} and \rlb{FYZ}, this implies
\eq
-\frac{1}{\beta}\{\log Z(\alpha')-\log Z(\alpha)\}
=
-\frac{1}{\beta}\log\frac{
\vxi^{\alpha'}
e^{-\beta\sfW^\hatn/2}\bsphi^{\alpha}
}{
\vxi^{\alpha}
e^{-\beta\sfW^\hatnd/2}\bsphi^{\alpha'}
}.
\en

\subsection{Thermodynamic relations}
\label{s:SSTproof}
We can now prove Theorem~\ref{t:SSTstep}, which mathematically states our thermodynamic relations for NESS including the extended Clausius relation.
Our strategy here is to justify the heuristic error estimate in the expansion \rlb{SSTJC2n} for the case of the step protocol \rlb{step},  i.e.,  $\hata$ with $\alpha(t)=\alpha$ for $t\in[-\tau,0]$, and $\alpha(t)=\alpha'$ for $t\in(0,\tau]$.

A key observation for the proof has already been made in \rlb{SSTJ1B1}.
This expression shows that, for the step protocol, the right-hand side of the equality \rlb{SSTJ1} can be compactly represented in terms of the modified rate matrix and its eigenvectors, and converges exponentially as $\tau\up\infty$.
Since the modified matrix and its eigenvectors can be expanded in convergent power series of $\epsilon$ and $\delta$, we can compare them with the heuristic power estimate in section~\ref{s:SSThigher}.
We shall make this idea more precise.

\paragraph*{Basic setup:}
To generate a power series corresponding to \rlb{SSTJC2n}, we introduce a new expansion parameter $\zeta\in[-\eta,1+\eta]$ where $\eta>0$ is a small fixed constant.
Then we define
\eq
\Xi(\hata,\zeta,\tau):=-\frac{1}{\beta}\log\frac{\bbkt{\,\exp[-\zeta(\beta\,W^\hatn+\Psi^{\hata})/2]\,}^{\hata}_{\sts\too}}
{\bbkt{\,\exp[-\zeta(\beta\,W^{\hatnd}+\Psi^{\hatad})/2]\,}^{\hatad}_{\sts\too}}.
\lb{Xitheta}
\en
Note that $\lti\Xi(\hata,1,\tau)$ is nothing but the right-hand side of our main equality \rlb{SSTJ1}, and thus equal to $F(\alpha')-F(\alpha)$.
We also find from the definition of cumulant \rlb{cumulant} that
\eq
\sqbk{\frac{\partial^k\Xi(\hata,\zeta,\tau)}{\partial\zeta^k}}_{\zeta=0}
=-\frac{(-1)^k}{\beta\,2^k}
\Bigl\{
{}^\mathrm{c}\!
\bbkt{(\beta\,W^\hatn+\Psi^{\hata})^k}^{\hata}_{\sts\too}
-{}^\mathrm{c}\!
\bbkt{(\beta\,W^{\hatnd}+\Psi^{\hatad})^k}^{\hatad}_{\sts\too}
\Bigr\}.
\lb{dkXi}
\en

For each $n=1,2,\ldots$, consider the (rigorous) expansion
\eq
\Xi(\hata,1,\tau)=
\sum_{k=0}^{2n}\frac{1}{k!}
\sqbk{\frac{\partial^k\Xi(\hata,\zeta,\tau)}{\partial\zeta^k}}_{\zeta=0}
+\frac{1}{(2n+1)!}
\sqbk{\frac{\partial^{2n+1}\Xi(\hata,\zeta,\tau)}{\partial\zeta^{2n+1}}}_{\zeta=\tilde{\zeta}_n(\hata,\tau)}
,
\en
where $\tilde{\zeta}_n(\hata,\tau)\in(0,1)$, and rearrange it as
\eq
\Xi(\hata,1,\tau)=
\sum_{k=0}^{2n-1}\frac{1}{k!}
\sqbk{\frac{\partial^k\Xi(\hata,\zeta,\tau)}{\partial\zeta^k}}_{\zeta=0}
+R_1+R_2,
\lb{Xi1exp}
\en
where we have set
\eqa
R_1:&=\frac{1}{(2n)!}\sqbk{\frac{\partial^{2n}\Xi(\hata,\zeta,\tau)}{\partial\zeta^{2n}}}_{\zeta=0}
\nl
&=-\frac{1}{\beta\,2^{2n}(2n)!}
\Bigl\{
{}^\mathrm{c}\!
\bbkt{(\beta\,W^\hatn+\Psi^{\hata})^{2n}}^{\hata}_{\sts\too}
-{}^\mathrm{c}\!
\bbkt{(\beta\,W^{\hatnd}+\Psi^{\hatad})^{2n}}^{\hatad}_{\sts\too}
\Bigr\},
\lb{R1}\\
R_2&:=\frac{1}{(2n+1)!}
\sqbk{\frac{\partial^{2n+1}\Xi(\hata,\zeta,\tau)}{\partial\zeta^{2n+1}}}_{\zeta=\tilde{\zeta}_n(\hata,\tau)}.
\lb{R2}
\ena

Recalling $\Xi(\hata,1,\tau)=F(\alpha')-F(\alpha)$ and \rlb{dkXi}, we see that the expression \rlb{Xi1exp} is nothing but the desired expansion \rlb{SSTJC2n} if we can show that $R_1+R_2=O(\epsilon^{2n}\delta)+O(\delta^2)$.

\paragraph*{Perturbation:}
We now turn to a rigorous perturbative estimate of the remainder $R_1+R_2$ in the expansion \rlb{Xi1exp}.

Let us first define a new matrix $\tR^{\alpha,\zeta}$ by
\eq
(\tR^{\alpha,\zeta})_{yx}:=
\begin{cases}
R^\alpha_{xx}&\text{if $x=y$};\\
R^\alpha_{yx}\,e^{-\zeta\psi^\alpha_\xty/2}&\text{if $x\ne y$},
\end{cases}
\en
and denote by $\mu^{\alpha,\zeta}$, $\bsphi^{\alpha,\zeta}$, and $\vxi^{\alpha,\zeta}$ its Perron-Frobenius eigenvalue and the corresponding right and left eigenvectors, respectively.
We use the same normalization as in section~\ref{s:modifiedratematrix} for the eigenvectors.

Let us fix the equilibrium transition rate $\Rbn_\xty$ that appears in \rlb{Rrigorous}.
Then, according to the expression \rlb{Rrigorous}, the rate $\Ra_\xty$ can be regarded as a perturbation to $\Rbn_\xty$ where $\epsilon$ is the parameter of perturbation.
Likewise the rate $R^{\alpha'}_\xty$ after the step can be regarded as a perturbation to $\Rbn_\xty$ where $\epsilon$ and $\delta$ are the parameters of the perturbation.
The same is true for the matrices  $\tR^{\alpha,\zeta}$ and  $\tR^{\alpha',\zeta}$ for a fixed $\zeta$.

We imagine that every quantity is obtained by a perturbation around the corresponding quantity with $\epsilon=\delta=0$.
Since the unperturbed matrix $\sfR^{(\beta,\nu),\zeta}$ has a nonvanishing spectral gap below its Perron-Frobenius eigenvalue, one can express the perturbed eigenvalues and eigenvectors of  $\tR^{\alpha,\zeta}$ and  $\tR^{\alpha',\zeta}$ in convergent power series of $\epsilon$ and $\delta$ provided that $\epsilon$ and $\delta$ are small enough.
Our task here is to rearrange these series and compare the result with the heuristic series \rlb{SSTJC}.

By using the matrices (the work matrix $\sfW^{\hatn}$ is defined in \rlb{workmatrix}), the key quantity \rlb{Xitheta} can be rewritten as
\eqa
\Xi(\hata,\zeta,\tau)&
=
-\frac{1}{\beta}\log\frac{\vone\,e^{\tau\tR^{\alpha',\zeta}}e^{-\zeta\beta\sfW^{\hatn}/2}\,
e^{\tau\tR^{\alpha,\zeta}}\bsrho^{\alpha,\zeta}}
{\vone\,e^{\tau\tR^{\alpha,\zeta}}e^{-\zeta\beta\sfW^\hatnd/2}\,
e^{\tau\tR^{\alpha',\zeta}}\bsrho^{\alpha',\zeta}}
\nl&
=
-\frac{1}{\beta}\log\frac{
e^{(\mu^{\alpha,\zeta}+\mu^{\alpha',\zeta})\tau}\,
\vone\,\bsphi^{\alpha',\zeta}\vxi^{\alpha',\zeta}
e^{-\zeta\beta\sfW^\hatn/2}
\bsphi^{\alpha,\zeta}\vxi^{\alpha,\zeta}\bsrho^{\alpha}
+(\text{remainder})
}{
e^{(\mu^{\alpha,\zeta}+\mu^{\alpha',\zeta})\tau}\,
\vone\,\bsphi^{\alpha,\zeta}\vxi^{\alpha,\zeta}
e^{-\zeta\beta\sfW^\hatnd/2}
\bsphi^{\alpha',\zeta}\vxi^{\alpha',\zeta}\bsrho^{\alpha'}
+(\text{remainder})
}
\nl&
=
-\frac{1}{\beta}\log\frac{
Y(\alpha,\zeta)\,\vxi^{\alpha',\zeta}
e^{-\zeta\beta\sfW^\hatn/2}\bsphi^{\alpha,\zeta}+D(\hata,\zeta,\tau)
}{
Y(\alpha',\zeta)\,\vxi^{\alpha,\zeta}
e^{-\zeta\beta\sfW^\hatnd/2}\bsphi^{\alpha',\zeta}+D(\hatad,\zeta,\tau)
},
\lb{XiYYDD}
\ena
where we have used \rlb{b2} (which of course holds for the matrices with $\zeta$).
The time dependent parts $D(\hata,\zeta,\tau)$, $D(\hatad,\zeta,\tau)$ decay exponentially in $\tau$ uniformly in $\zeta$, and also has convergent series expansions in $\epsilon$ and $\delta$ for any $\tau$.

We are now ready to evaluate $R_1$ and $R_2$ defined by \rlb{R1} and \rlb{R2}, respectively.
We first observe from \rlb{XiYYDD} that $R_1$ can be expanded into a power series $\sum_{\ell,m} C^{(\tau)}_{\ell,m}\epsilon^\ell\,\delta^m$, which converges uniformly in $\tau>0$.
We wish to argue that $C^{(\tau)}_{\ell,1}=0$ for $\ell<2n$.
For this purpose we recall the heuristic estimate \rlb{formalerrorestimate} which is valid only for finite $\tau$ and sufficiently small $\epsilon$ and $\delta$.
Nevertheless the heuristic estimates shows rigorously that $C^{(\tau)}_{\ell,1}=0$ for $\ell<2n$ for sufficiently small $\tau$.
But, when we recall that \rlb{XiYYDD} admits a nice expansion, that $C^{(\tau)}_{\ell,1}=0$ for a finite interval of $\tau$ implies that $C^{(\tau)}_{\ell,1}=0$ for any $\tau$.

The quantity $R_2$ is much more complicated but can be treated in a similar manner.
We first note that 
$[{\partial^{2n+1}\Xi(\hata,\zeta,\tau)}/{\partial\zeta^{2n+1}}]_{\zeta=\tilde{\zeta}_n(\hata,\tau)}$ reduces to a kind of cumulant 
${}^\mathrm{c}\!
\langle\langle(\beta\,W^\hatn+\Psi^{\hata})^{2n+1}\rangle\rangle^{\hata}_{\sts\too}-{}^\mathrm{c}\!
\langle\langle(\beta\,W^{\hatnd}+\Psi^{\hatad})^{2n+1}\rangle\rangle^{\hatad}_{\sts\too}
$, where ${}^\mathrm{c}\!
\langle\langle\cdots\rangle\rangle^{\hata}_{\sts\too}$ is defined by replacing the expectation $\sbkt{\cdots}^{\hata}_{\sts\too}$ (in the standard cumulant) by $\sbkt{(\cdots)\,e^{-\tilde{\zeta}_n(\hata,\tau)(\beta\,W^\hatn+\Psi^\hata)/2}}^{\hata}_{\sts\too}$.
Now one expands $e^{-\tilde{\zeta}_n(\hata,\tau)(\beta\,W^\hatn+\Psi^\hata)/2}$ in a power series, and make a heuristic order estimate as in section~\ref{s:heuristicerror}.
A good news here is that one does not have to invoke the complicated estimate using the time-reversal symmetry;
only naive expansion and power counting is enough.
For example the dominant term can be evaluated as
\eqa
&\bbkt{(\beta\,W^\hatn+\Psi^{\hata})^{2n+1}}^{\hata}
-\bbkt{(\beta\,W^{\hatnd}+\Psi^{\hatad})^{2n+1}}^{\hatad}
\nl
&=
\bbkt{(\Psi^{(\alpha)}+\Di\Phi^\hata)^{2n+1}(1+\Di\Gamma^\hata)}^{(\alpha)}
-\bbkt{(\Psi^{(\alpha)}+\Di\Phi^\hatad)^{2n+1}(1+\Di\Gamma^\hatad)}^{(\alpha)}
\nl
&=O(\epsilon^{2n}\delta)+O(\epsilon^{2n-1}\delta^2).
\ena
This heuristic estimate can be converted into a rigorous estimate as before by relying on the representation \rlb{XiYYDD}.

This proves the desired Theorem~\ref{t:SSTstep}.

\subsection{Excess entropy production}
\label{s:excess}

Let us prove Theorem~\ref{t:SSTexcess}, which states that the extended Clausius relation can be written in terms of the excess entropy production.

We first define the entropy production rate as a function of $x\in\calS$ by
\eq
\sigma^\alpha_x:=\sum_{y\in\calS}\thetaa_\xty\Ra_\xty.
\lb{sigmaax}
\en
Let $\hata=(\alpha(t))_\ttt$ be an arbitrary protocol, and $\bsp^\hata(t)$ be the corresponding solution of the master equation \rlb{master} with the  initial condition $\bsp^\hata(-\tau)=\bsrho^{\alpha(-\tau)}$.
Then from the expression \rlb{Wdef} of the path weight, the definition \rlb{fss} of the path average, and the definition \rlb{Thetadef} of the total entropy production, one finds
\eq
\sbkt{\Theta^\hata}^\hata_{\sts\too}=\int_{-\tau}^\tau dt\sum_{x\in\calS}
\sigma^{\alpha(t)}_xp^\hata_x(t).
\en
This in particular implies $\sbkt{\Theta^\ca}^\ca_{\sts\too}=2\tau\sum_{x}\sigma^\alpha_x\rho^\alpha_x$, which shows that the right-hand side of \rlb{sigmasta} is independent of $\tau$, and $\sigma^\alpha_\sts=\sum_{x}\sigma^\alpha_x\rho^\alpha_x$.

For simplicity we first consider the single step protocol \rlb{step}, i.e.,  $\hata$ with $\alpha(t)=\alpha$ for $t\in[-\tau,0]$, and $\alpha(t)=\alpha'$ for $t\in(0,\tau]$.
Then we have
\eq
\sbkt{\Theta^\hata}^\hata_{\sts\too}=\tau\,\sigma^\alpha_\sts
+\int_0^\tau dt\sum_{x\in\calS}\sigma^{\alpha'}_x(e^{t\sfR^{\alpha'}}\bsrho^\alpha)_x,
\en
where we noted that the probability distribution at $t=0$ is $\bsrho^\alpha$.
On the other hand, noting that $\bsrho^{\alpha'}=e^{t\sfR^{\alpha'}}\bsrho^{\alpha'}$, we have
\eq
\tau\,\sigma^{\alpha'}_\sts
=\int_0^\tau dt\sum_{x\in\calS}\sigma^{\alpha'}_x(e^{t\sfR^{\alpha'}}\bsrho^{\alpha'})_x.
\en
Since the house-keeping entropy production is $\Sigma^\hata_\mathrm{hk}=\tau(\sigma^{\alpha}_\sts+\sigma^{\alpha'}_\sts)$ in this case, we find
\eq
\sbkt{\Theta^\hata}^\hata_{\sts\too}-\Sigma^\hata_\mathrm{hk}
=
\int_0^\tau dt\sum_{x\in\calS}\sigma^{\alpha'}_x
\bigl(e^{t\sfR^{\alpha'}}(\bsrho^\alpha-\bsrho^{\alpha'})\bigr)_x.
\lb{exEPsym1}
\en
Noting that $e^{t\sfR^{\alpha'}}(\bsrho^\alpha-\bsrho^{\alpha'})$ converges exponentially to zero as $t\up\infty$, we conclude that the excess entropy production $\sbkt{\Theta^\hata}^\hata_{\sts\too}-\Sigma^\hata_\mathrm{hk}$ has a finite $\tau\up\infty$ limit.

For the reversed protocol $\hatad$, we similarly have
\eqa
\sbkt{\Theta^\hatad}^\hatad_{\sts\too}-\Sigma^\hatad_\mathrm{hk}
&=
\int_0^\tau dt\sum_{x\in\calS}\sigma^{\alpha}_x
\bigl(e^{t\sfR^{\alpha}}(\bsrho^{\alpha'}-\bsrho^{\alpha})\bigr)_x
\nl
&=\int_0^\tau dt\sum_{x\in\calS}\sigma^{\alpha'}_x
\bigl(e^{t\sfR^{\alpha'}}(\bsrho^{\alpha'}-\bsrho^{\alpha})\bigr)_x+O(\delta^2),
\lb{exEPsym2}
\ena
where the final estimate follows from $\sigma^{\alpha}_x-\sigma^{\alpha'}_x=O(\delta)$, $\sfR^{\alpha}-\sfR^{\alpha'}=O(\delta)$, and $\bsrho^\alpha-\bsrho^{\alpha'}=O(\delta)$.
This is easily made into a rigorous estimate by using the exponential convergence.
By comparing \rlb{exEPsym1} and \rlb{exEPsym2}, we find
\eq
\sbkt{\Theta^\hata}^\hata_{\sts\too}-\Sigma^\hata_\mathrm{hk}
=-\bigl\{
\sbkt{\Theta^\hatad}^\hatad_{\sts\too}-\Sigma^\hatad_\mathrm{hk}
\bigr\}+O(\delta^2).
\en
Thus the excess entropy production is (nearly) antisymmetric with respect to time reversal.
Since the house-keeping entropy production is clearly symmetric, i.e., $\Sigma^\hata_\mathrm{hk}=\Sigma^\hatad_\mathrm{hk}$, one gets
\eq
\sbkt{\Theta^\hata}^\hata_{\sts\too}-\Sigma^\hata_\mathrm{hk}
=\frac{1}{2}\Bigl\{
\sbkt{\Theta^\hata}^\hata_{\sts\too}-\sbkt{\Theta^\hatad}^\hatad_{\sts\too}
\Bigr\}+O(\delta^2),
\en
which is a rigorous estimate for the single step protocol \rlb{step}.

It is clear that the same estimate applies to the $N$-step protocol $\hata$ defined in \rlb{manysteps}, \rlb{manysteps2}, and we get
\eq
\lti\Bigl\{
\sbkt{\Theta^\hata}^\hata_{\sts\too}-\Sigma^\hata_\mathrm{hk}
\Bigr\}
=\lti\frac{1}{2}\Bigl\{
\sbkt{\Theta^\hata}^\hata_{\sts\too}-\sbkt{\Theta^\hatad}^\hatad_{\sts\too}
\Bigr\}+N\times O\biggl(\Bigl(\frac{\delta}{N}\Bigr)^2\biggr).
\en
By letting $N\up\infty$, we see that the two limits coincide.
Given \rlb{ECL03C}, this proves the desired \rlb{ECL03D}.
%

\section{Representation of the stationary distribution and entropy}
\label{s:KNS}
Here we prove Theorem~\ref{t:SandS} for the nonequilibrium entropy.
For this purpose we prove the representation \rlb{KNrep} for the stationary distribution of NESS \cite{KN,KNST3}, which is interesting in its own light.

\bigskip

In the present section, we only consider a constant protocol $(\alpha)$, i.e., $\alpha(t)=\alpha$ for the whole time interval $\ttt$.
We denote the corresponding expectation $\sbkt{\cdots}^{(\alpha)}_{x\to}$ and the unnormalized expectation $[\cdots]^{(\alpha)}_{\sts\to x}$  (see \rlb{fx} and \rlb{fxy}) as $\sbkt{\cdots}^{\tau,(\alpha)}_{x\to}$ and  $[\cdots]^{\tau,(\alpha)}_{\sts\to x}$, respectively, to emphasize the dependence on $\tau$.
We also define a new conditional expectation by
\eq
\sbkt{\cdots}^{\tau,(\alpha)}_{\sts\to x}=
\frac{[\cdots]^{\tau,(\alpha)}_{\sts\to x}}{\rho_x^\alpha},
\lb{<>sx}
\en
which is normalized.

Let us define $\tau$-dependent free energy $F_x^\tau(\alpha)$ by
\eq
F_x^\tau(\alpha):=
e^{-\beta\,H_x^\alpha}\,
\frac{\sbkt{e^{-\Psi^{(\alpha)}/2}}^{\tau,(\alpha)}_{x\to}}
{[e^{-\Psi^{(\alpha)}/2}]^{\tau,(\alpha)}_{\sts\to x}}.
\lb{Ftau}
\en
By comparing this with \rlb{Fdef}, we find that $F_x^\tau(\alpha)\to F(\alpha)$ as $\tau\up\infty$.
Note that the $x$ dependence vanishes in the limit.

By using \rlb{<>sx} and \rlb{Ftau}, we can also write
\eq
\rho^\alpha_x=e^{\beta\,F_x^\tau(\alpha)-\beta\,H_x^\alpha}
\frac{\sbkt{e^{-\Psi^{(\alpha)}/2}}^{\tau,(\alpha)}_{x\to}}
{\sbkt{e^{-\Psi^{(\alpha)}/2}}^{\tau,(\alpha)}_{\sts\to x}}.
\lb{KNformal}
\en
Applying formal cumulant expansion to this expression, one has
\eq
\log\rho^\alpha_x=\beta\,F_x^\tau(\alpha)-\beta\,H_x^\alpha
+\sum_{k=1}^\infty\frac{(-1)^k}{k!\,2^k}\Bigl\{
{}^\mathrm{c}\!\bbkt{(\Psi^{(\alpha)})^k}^{\tau,(\alpha)}_{x\to}
-{}^\mathrm{c}\!\bbkt{(\Psi^{(\alpha)})^k}^{\tau,(\alpha)}_{\sts\to x}
\Bigr\}.
\lb{KNfullexp}
\en

As in section~\ref{s:heuristicerror}, we fix a finite $\tau$, and expand around the equilibrium constant protocol $(\alpha_\mathrm{eq})$ by using $\sbkt{\cdots}^{\tau,(\alpha)}=\sbkt{\cdots(1+\Di\Omega^{(\alpha)})}^{\tau,(\alpha_\mathrm{eq})}$ with $\Di\Omega^{(\alpha)}=O(\epsilon)$.
Then, by recalling $\Psi^{(\alpha)}=O(\epsilon)$ the term with $k=2$ in \rlb{KNfullexp} is evaluated as
\eq
{}^\mathrm{c}\!\bbkt{(\Psi^{(\alpha)})^2}^{\tau,(\alpha)}_{x\to}
-{}^\mathrm{c}\!\bbkt{(\Psi^{(\alpha)})^2}^{\tau,(\alpha)}_{\sts\to x}
=
{}^\mathrm{c}\!\bbkt{(\Psi^{(\alpha)})^2}^{\tau,(\alpha_\mathrm{eq})}_{x\to}
-{}^\mathrm{c}\!\bbkt{(\Psi^{(\alpha)})^2}^{\tau,(\alpha_\mathrm{eq})}_{\sts\to x}
+O(\epsilon^3)=O(\epsilon^3),
\lb{KNOec}
\en
where we noted that ${}^\mathrm{c}\!\bbkt{(\Psi^{(\alpha)})^2}^{\tau,(\alpha_\mathrm{eq})}_{x\to}
={}^\mathrm{c}\!\bbkt{(\Psi^{(\alpha)})^2}^{\tau,(\alpha_\mathrm{eq})}_{\sts\to x}$ by the time-reversal symmetry \rlb{eqsym}. 
We noted that $(\Psi^{(\alpha)})^\dagger=-\Psi^{(\alpha)}$, and that we here have $\alpha_\mathrm{eq}=\alpha'_\mathrm{eq}$ and $W=0$.

By substituting the error estimate \rlb{KNOec} into the expansion \rlb{KNfullexp}, we get
\eq
\log\rho^\alpha_x=\beta\,F_x^\tau(\alpha)-\beta\,H_x^\alpha
-\frac{1}{2}\Bigl\{
\bbkt{\Psi^{(\alpha)}}^{\tau,(\alpha)}_{x\to}
-\bbkt{\Psi^{(\alpha)}}^{\tau,(\alpha)}_{\sts\to x}
\Bigr\}+O(\epsilon^3).
\lb{KNwithtau}
\en
By formally letting $\tau\up\infty$, we get a formal but very suggestive representation \rlb{theKN}.
It was first written down by two of us (T.S.K. and N.N.) in \cite{KN}.

We now note that this representation can be turned into a rigorous estimate.
The key observation is that the quantity in the right-hand side of 
\rlb{KNformal} is written as
\eq
\frac{\sbkt{e^{-\Psi^{(\alpha)}/2}}^{\tau,(\alpha)}_{x\to}}
{\sbkt{e^{-\Psi^{(\alpha)}/2}}^{\tau,(\alpha)}_{\sts\to x}}
=\rho^\alpha_x\,\frac{\vone\,e^{\tau\tR}\bsdelta_x}
{\vdelta^{(x)}\,e^{\tau\tR}\bsrho^\alpha},
\en
where we used the same quantities as in section~\ref{s:matrix}.
Then by using the machinery in section~\ref{s:matrix}, it is not hard to show that the above quantity admits a series expansion in $\epsilon$ which converges uniformly in $\tau$, and each coefficient in the expansion converges as $\tau\up\infty$.
We then get the following theorem.

\begin{theorem}[Representation of the probability distribution for NESS]
\label{t:KNrep}
Consider a class of models introduced in the beginning of section~\ref{s:SSTrigorous}.
Then for any set of parameters $\alpha$, the corresponding stationary distribution $\rho^\alpha=(\rho^\alpha_x)_{x\in\calS}$ satisfies
\eq
\Biggl|\log\rho^\alpha_x-\biggl\{\beta\,F(\alpha)-\beta\,H^\alpha_x-\frac{1}{2}\lim_{\tau\up\infty}\Bigl\{
\bbkt{\Psi^{(\alpha)}}^{\tau,(\alpha)}_{x\to}
-\bbkt{\Psi^{(\alpha)}}^{\tau,(\alpha)}_{\sts\to x}
\Bigr\}\biggr\}\Biggr|\le C_0\,\epsilon^3,
\lb{KNrep}
\en
where $C_0$ is a (model dependent) positive constant.
\end{theorem}

To prove Theorem~\ref{t:SandS}, let us observe that, for any $\tau>0$,
\eq
\sum_{x\in\calS}\rho^\alpha_x\Bigl\{
\bbkt{\Psi^{(\alpha)}}^{\tau,(\alpha)}_{x\to}
-\bbkt{\Psi^{(\alpha)}}^{\tau,(\alpha)}_{\sts\to x}
\Bigr\}
=
\bbkt{\Psi^{(\alpha)}}^{\tau,(\alpha)}_{\sts\to}
-\bbkt{\Psi^{(\alpha)}}^{\tau,(\alpha)}_{\sts\to}
=0,
\en
which is a direct consequence of the definition.
Then from the representation \rlb{KNwithtau} (with $\tau\up\infty$), we see that
\eqa
S_\mathrm{Sh}[\bsrho^\alpha]&=-\sum_{x\in\calS}\rho^\alpha_x\,\log\rho^\alpha_x
\nl
&=-\beta\,F(\alpha)+\beta\sum_{x\in\calS}\rho^\alpha_x\,H^\alpha_x+O(\epsilon^3).
\ena
By recalling the relation \rlb{S=U-F} between the nonequilibrium free energy and entropy, we find
\eq
S_\mathrm{Sh}[\bsrho^\alpha]=S(\alpha)+O(\epsilon^3).
\en
By using Theorem~\ref{t:KNrep}, this becomes a rigorous estimate, and we get the desired Theorem~\ref{t:SandS}.

\section{Proof of the extended Clausius inequality}
\label{s:ClIneq}
Here we shall prove the extended Clausius inequality stated in theorem~\ref{t:ClausiusIneq}.

For $\ttt$, let $\bsp(t)=(p_x(t))_{x\in\calS}$ be the solution of the master equation \rlb{master}, \rlb{master2} with the initial condition $\bsp(-\tau)=\bsrho^\alpha$.
Then it is well-known that the ``second law of thermodynamics'' or the ``H-theorem''
\eq
\fdt{}D[\bsp(t)|\bsrho^\alpha]\Bigr|_{\alpha=\alpha(t)}\le0
\lb{dDdt}
\en
holds,
where 
\eq
D[\bsp|\bsq]:=\sum_{x\in\calS}p_x\log\frac{p_x}{q_x}
\lb{Ddef}
\en
is the relative entropy (or the Kullback-Leibler divergence) of the two probability distributions $\bsp=(p_x)_{x\in\calS}$ and $\bsq=(q_x)_{x\in\calS}$.
See, for example, \cite{CoverThomas}, and also Appendix C of \cite{SasaTasaki}.

By substituting \rlb{Ddef} into \rlb{dDdt} and recalling the definition \rlb{SSh} of the Shannon entropy, we get
\eq
-\fdt{}S_\mathrm{Sh}[\bsp(t)]
-\sum_{x\in\calS}\dot{p}_x(t)\log\rho_x^{\alpha(t)}\le0.
\en
By noting that $\lti p_x(\tau)=\rho_x^{\alpha'}$, this implies
\eq
S_\mathrm{Sh}[\bsrho^{\alpha'}]-S_\mathrm{Sh}[\bsrho^\alpha]\ge-\lti\int_{-\tau}^\tau dt\sum_{x\in\calS}\dot{p}_x(t)\log\rho_x^{\alpha(t)}.
\lb{SSIneq}
\en
In fact this is precisely identical to the Clausius inequality written by Hatano and Sasa \cite{HatanoSasa}.
See also \cite{Sasa2013}.

We need to rewrite \rlb{SSIneq} in terms of the excess entropy production to get \rlb{ECLIneq}.
We can make us of the following representation of the stationary distribution $\bsrho^\alpha$.

\begin{lemma}[Linear response formula for stationary distribution]
\label{l:LRrep}
Consider a class of models introduced in the beginning of section~\ref{s:SSTrigorous}.
Then for any set of parameters $\alpha$, the corresponding stationary distribution $\rho^\alpha=(\rho^\alpha_x)_{x\in\calS}$ satisfies
\eq
\biggl|\log\rho^\alpha_x-
\Bigl\{\beta\,F(\alpha)-\beta\,H^\alpha_x
-\lim_{\tau\up\infty}\bbkt{\Psi^{(\alpha)}}^{\tau,(\alpha_\mathrm{eq})}_{x\to}\Bigr\}
\biggr|\le \tilde{C}'\,\epsilon^2,
\lb{LRrep}
\en
and
\eq
\biggl|\log\rho^\alpha_x-
\Bigl\{-S(\alpha)
-\lim_{\tau\up\infty}\bbkt{\Theta_\mathrm{ex}^{(\alpha)}}^{\tau,(\alpha)}_{x\to}\Bigr\}
\biggr|\le \tilde{C}\,\epsilon^2,
\lb{LRrep2}
\en
where $\tilde{C}$, $\tilde{C}'$ are (model dependent) positive constants.
\end{lemma}

Note that these representations have simpler forms but larger errors than the previous representation \rlb{theKN}, \rlb{KNrep}.
In fact \rlb{LRrep} is a rigorous version of the linear response representation (see, e.g., \cite{MaesNetocny10,KNST3}), and \rlb{LRrep2} is its variant.

\bigskip\noindent
{\em Proof of Lemma~\ref{l:LRrep}:}\/
We shall give a heuristic argument, which can be turned into rigorous estimates.
We start from the more accurate representation \rlb{theKN}, \rlb{KNrep}, and first note that 
\eqa
\bbkt{\Psi^{(\alpha)}}^{\tau,(\alpha)}_{x\to}
-\bbkt{\Psi^{(\alpha)}}^{\tau,(\alpha)}_{\sts\to x}
&=
\bbkt{\Psi^{(\alpha)}}^{\tau,(\alpha_\mathrm{eq})}_{x\to}
-\bbkt{\Psi^{(\alpha)}}^{\tau,(\alpha_\mathrm{eq})}_{\sts\to x}+O(\epsilon^2)
\nl
&=
2\bbkt{\Psi^{(\alpha)}}^{\tau,(\alpha_\mathrm{eq})}_{x\to}
+O(\epsilon^2),
\ena
where the first equality follows by expanding around the equilibrium protocol (as in section~\ref{s:heuristicerror}) and the second equality follows from the time-reversal symmetry \rlb{eqsym}. 
See the remark after \rlb{KNOec}.
By substituting this into \rlb{theKN}, we get the linear response representation
\eq
\log\rho^\alpha_x=\beta\,F(\alpha)-\beta\,H^\alpha_x-\lim_{\tau\up\infty}
\bbkt{\Psi^{(\alpha)}}^{\tau,(\alpha)}_{x\to}+O(\epsilon^2).
\lb{LRtemp}
\en

We then note that
\eq
\bbkt{\Psi^{(\alpha)}}^{\tau,(\alpha_\mathrm{eq})}_{x\to}
=
\bbkt{\Psi_\mathrm{ex}^{(\alpha)}}^{\tau,(\alpha)}_{x\to}+O(\epsilon^2),
\en
where $\Psi_\mathrm{ex}^{(\alpha)}:=\Psi^{(\alpha)}-\Sigma^{(\alpha)}_\mathrm{hk}$.
It is crucial here that both the quantities have finite $\tau\up\infty$ limits\footnote{%
The difference between $\bbkt{\Psi^{(\alpha)}}^{\tau,(\alpha_\mathrm{eq})}_{x\to}$ and $\bbkt{\Psi^{(\alpha)}}^{\tau,(\alpha)}_{x\to}$ is proportional to $\epsilon^2$, but is also roughly proportional to $\tau$.
Thus the difference diverges as $\tau\up\infty$.
}.
We then use \rlb{TWPH} (with $W=0$) to get
\eq
\bbkt{\Psi_\mathrm{ex}^{(\alpha)}}^{\tau,(\alpha)}_{x\to}
=\bbkt{\Theta_\mathrm{ex}^{(\alpha)}}^{\tau,(\alpha)}_{x\to}
-\beta\,\Hn_x+\beta\sbkt{\Hn}^\alpha_\mathrm{st}.
\en
By recalling the definition \rlb{S=U-F} of the nonequilibrium entropy, \rlb{LRtemp} further reduces to
\eq
\log\rho^\alpha_x=-S(\alpha)-\lim_{\tau\up\infty}
\bbkt{\Theta_\mathrm{ex}^{(\alpha)}}^{\tau,(\alpha)}_{x\to}+O(\epsilon^2).
\lb{LRtemp2}
\en

These heuristic observation can be made into rigorous estimates by using the machinery of section~\ref{s:matrix}.
In particular all the error terms can be bounded uniformly in $\tau$.~\qedm

\bigskip

Let us examine what happens when we substitute the representation \rlb{LRrep2} into the right-hand side of the main inequality \rlb{SSIneq}.
Since 
\newline$\int_{-\tau}^\tau dt\sum_{x\in\calS}\dot{p}_x(t)\,S(\alpha)=\int_{-\tau}^\tau dt\,\frac{d}{dt}\{\sum_{x\in\calS}p_x(t)\}\,S(\alpha)=0$, we only need to investigate the contribution of $\bbkt{\Theta_\mathrm{ex}^{(\alpha)}}^{\tau,(\alpha)}_{x\to}$.
We rewrite the integral as
\eq
\int_{-\tau}^\tau dt\sum_{x\in\calS}\dot{p}_x(t)\,\bbkt{\Theta_\mathrm{ex}^{(\alpha)}}^{\tau,(\alpha(t))}_{x\to}
=
-\lim_{\Di t\dn0}\sum_t\sum_{x\in\calS}\{p_x(t)-p_x(t+\Di t)\}\,\bbkt{\Theta_\mathrm{ex}^{(\alpha(t))}}^{\tau,(\alpha(t))}_{x\to},
\en
where $t$ is summed over the integral multiples of $\Di t$ within the interval $[-\tau,\tau]$, and $(\alpha(t))$ denotes the protocol in which the parameters are fixed at $\alpha(t)$ (for the given $t$).
Now note that
\eq
\bbkt{\Theta_\mathrm{ex}^{(\alpha)}}^{[t,t+\Di t],(\alpha(t))}_{\bsp(t)\to}
:=
\sum_{x\in\calS}\{p_x(t)-p_x(t+\Di t)\}\,\bbkt{\Theta_\mathrm{ex}^{(\alpha(t))}}^{\tau,(\alpha(t))}_{x\to}
\en
is precisely the excess entropy production in the time interval $[t,t+\Di t]$, where the parameters are fixed at $\alpha(t)$ and the distribution at $t$ is given by $\bsp(t)$.
Then it follows that
\eq
\int_{-\tau}^\tau dt\sum_{x\in\calS}\dot{p}_x(t)\,\bbkt{\Theta_\mathrm{ex}^{(\alpha)}}^{\tau,(\alpha(t))}_{x\to}
=
-\lim_{\Di t\dn0}\sum_t\bbkt{\Theta_\mathrm{ex}^{(\alpha)}}^{[t,t+\Di t],(\alpha(t))}_{\bsp(t)\to}
=-\sbkt{\Theta_\mathrm{ex}^\hata}^\hata_{\sts\too}.
\en
The existence of the limits can be proved by using the materials in of section~\ref{s:matrix}.

We finally use \rlb{S=SSh2} to rewrite the Shannon entropy $S_\mathrm{Sh}[\bsrho^\alpha]$ in terms of our nonequilibrium entropy $S(\alpha)$.

\section{Models with momenta and symmetrized Shannon entropy}
\label{s:Ssym}

We have been so far studying models in which state variables are symmetric with respect to time reversal.
In the case of a system of $N$ particles, our state $x$ roughly corresponds to the collection $(\vecr_1,\ldots,\vecr_N)$, where $\vecr_j\in\bbR^3$ is the position of the $j$-th particle.
In a ``less coarse grained'' description of a particle system one also specifies the momenta of the  particles.
In this case the state  $x$ roughly corresponds to the collection $(\vecr_1,\ldots,\vecr_N,\vecp_1,\ldots,\vecp_N)$, where $\vecp_j\in\bbR^3$ denotes the momentum.
By the time reversal, this state is mapped to a different state $(\vecr_1,\ldots,\vecr_N,-\vecp_1,\ldots,-\vecp_N)$.
We denote the corresponding state as $x^*$.

Here we deal with a Markov jump process which, in some sense, mimics the structure of such a system with momenta.
We see that all but one of the results in the previous sections  remain valid if we properly modify the definition of entropy as in \rlb{Ssym}.
The only exception is the extended Clausius inequality \rlb{ECLIneq} in theorem~\ref{t:ClausiusIneq}, which can never be valid in the present setting.

\subsection{Setting and the main observation}
Let us give a precise and abstract definition.
We assume that to any state $x\in\calS$ there corresponds a state $x^*\in\calS$, and one has $(x^*)^*=x$.
We assume that the Hamiltonian is time reversal symmetric in the sense that $\Hn_x=\Hn_{x^*}$ for any $x\in\calS$.
With the above physical interpretation in mind, we should modify the detailed balance condition \rlb{dbeq} for equilibrium dynamics as
\eq
e^{-\beta \Hn_x}\Rbn_\xty=e^{-\beta \Hn_{y^*}}R^\bn_{y^*\to x^*},
\lb{dbeq*}
\en
and also assume that the escape rate (see \rlb{ladef}) $\lambda^\bn_x:=\sum_{y\in\calS\,(y\ne x)}\Rbn_\xty$ satisfies $\lambda^\bn_x=\lambda^\bn_{x^*}$.

As before let $\Ra_\xty$ denote the transition rate for a general model including a nonequilibrium one.
We assume that $\Ra_\xty\ne0$ for $x\ne y$ implies $\Ra_{y^*\to x^*}\ne0$.
The connectivity of $\calS$ by nonvanishing $\Ra_\xty$ is again assumed.
We still assume that the escape rate \rlb{ladef} has the symmetry\footnote{
It may be also reasonable to consider a model in which $\la_x\ne\la_{x^*}$.
In such a model, one should include the contribution from $\la_x/\la_{x^*}$ into the definition of $\Theta^{\hata}[\hatx]$ so as to keep the symmetry \rlb{DFT*} valid (see \cite{ItamiSasa}).
Then all the results in the present section remain valid.
} $\la_x=\la_{x^*}$.

Corresponding to \rlb{dbeq*}, the definition \rlb{thetadef0} of the entropy production should be modified as
\eq
\thetaa_\xty:=\log\frac{\Ra_\xty}{\Ra_{y^*\to x^*}}.
\lb{thetadef0*}
\en
All the other definitions are exactly the same as before, and we can develop the theory in an almost parallel manner.
One essential difference is that the fundamental time-reversal symmetry \rlb{DFT}
\eq
\calT^{\hatad}[\hatxd]=\exp\bigl[{-\Theta^{\hata}[\hatx]}\bigr]\,\calT^{\hata}[\hatx]
\lb{DFT*}
\en
is valid as it is, but for a given path $\hatx=(x(t))_\ttt$ we define its time-reversal as $\hatxd:=((x(-t))^*)_\ttt$.
This means that some relations that follows from \rlb{DFT} should be properly modified by putting $*$ on some variables.
The basic identity \rlb{DFT3new}, for example, now reads
\eq
e^{-\beta\Hn_x}\,\bigl[e^{-(\beta W^\hatn+\Psi^\hata)/2}\bigr]_{\xty}^\hata
=
e^{-\beta H^{\nu'}_{y^*}}\,\bigl[e^{-(\beta W^\hatnd+\Psi^\hatad)/2}\bigr]_{y^*\to x^*}^\hatad.
\lb{DFT3*}
\en
The most important change for us is that the definition \rlb{Fdef} should be modified as
\eq
e^{-\beta F(\alpha)}:=e^{-\beta\Hn_x}\,
\lti\frac{\sbkt{\,e^{-\Psi^{(\alpha)}/2}\,}^{[-\tau,-\tau/2],(\alpha)}_{x\too}}
{[\,e^{-\Psi^{(\alpha)}/2}\,]^{[\tau/2,\tau],(\alpha)}_{\sts\to x^*}}.
\lb{Fdef*}
\en
With this modification, the identity \rlb{FF1id} is valid as it is, and so are the (thermodynamic) relations we have discussed in sections~\ref{s:JarNESS}, \ref{s:SST}, and \ref{s:SSTrigorous}, except for the extended Clausius inequality \rlb{ECLIneq} of Theorem~\ref{t:ClausiusIneq}.
Here the nonequilibrium entropy $S(\alpha)$ is defined by \rlb{S=U-F} with the newly defined $F(\alpha)$.

As for the microscopic representation of the entropy $S(\alpha)$, we encounter a nontrivial and suggestive modification.
In  models with time reversal symmetry, we have shown that the nonequilibrium entropy $S(\alpha)$ coincides with the Shannon entropy of the stationary distribution (to the order $O(\epsilon^2)$) as in \rlb{S=SSh} or \rlb{S=SSh2} in Theorem~\ref{t:SandS}.
This is no longer valid in the present setting, and we have the following extension.

\begin{theorem}[Nonequilibrium entropy and the symmetrized Shannon entropy]
\label{t:SandSsym}
Take the same setting as in Theorem~\ref{t:SandS} but in the present class of models.
There is a constant $A'>0$, and one has
\eq
\Bigl|S(\alpha)-S_\mathrm{sym}[\bsrho^\alpha]\Bigr|\le A'\,\epsilon^3
\lb{S=SSsym}
\en
for any $\epsilon$, $\nu$, and $\kappa$.
Here $\bsrho^\alpha$ is the stationary probability distribution corresponding to the parameter $\alpha$.
For any probability distribution $\bsp$, we have defined the symmetrized Shannon entropy by
\eq
S_\mathrm{sym}[\bsp]=-\sum_{x\in\calS}p_x\log\sqrt{p_x\,p_{x^*}}.
\lb{Ssym}
\en
\end{theorem}

This theorem is proved in the next section.

Let us make a few remarks about the symmetrized Shannon entropy \rlb{Ssym}.
It is apparent that if a probability distribution $\bsp$ has a time-reversal symmetry in the sense that  $p_x=p_{x^*}$, then we have $S_\mathrm{sym}[\bsp]=S_\mathrm{Sh}[\bsp]$.
Since the equilibrium state $\rho^\bn_x=e^{-\beta \Hn_x}/Z(\beta)$ is symmetric, the Shannon and the symmetrized Shannon entropies coincide for the equilibrium state.

Note that one may rewrite \rlb{Ssym} in a suggestive form
\eq
S_\mathrm{sym}[\bsp]=-\sum_{x\in\calS}\frac{p_x+p_{x^*}}{2}
\log\sqrt{p_x\,p_{x^*}},
\lb{Ssym2}
\en
in which both the arithmetic mean  and  the geometric mean appear.
We also note that for a general probability distribution $\bsp$ that 
\eq
S_\mathrm{sym}[\bsp]=S_\mathrm{Sh}[\bsp]+\frac{1}{2}D[\bsp|\bsp^*]
\ge S_\mathrm{Sh}[\bsp],
\en
where $(\bsp^*)_x=p_{x^*}$, and $D[\bsp|\bsp']:=\sum_{x\in\calS}p_x\log(p_x/p'_x)$ is the relative entropy, which is in general nonnegative.

Recall that when we proved the extended Clausius inequality (Theorem~\ref{t:ClausiusIneq}) in section~\ref{s:ClIneq}, the Shannon entropy played an essential role.
This means that the proof does not extended to the present situation.
Indeed we know that the the extended Clausius inequality can never be valid in models with ``momenta'' since there are models in which it is explicitly violated.
We shall see such an example in section~\ref{s:toy}.

\subsection{Proof of Theorem~\protect\ref{t:SandSsym}}

We shall only present a heuristic argument, which can be made into a proof in the similar manner as in section~\ref{s:KNS}.

By following the derivation in section~\ref{s:KNS}, we can show the representation for the stationary distribution for NESS in the present setting, which is 
\eq
\log\rho^\alpha_x=\beta F(\alpha)-\beta\Hn_x
-\frac{1}{2}\lti\Bigl\{
\bbkt{\Psi^{(\alpha)}}^{\tau,(\alpha)}_{x^*\to}
-\bbkt{\Psi^{(\alpha)}}^{\tau,(\alpha)}_{\sts\to x}
\Bigr\}+O(\epsilon^3).
\lb{KNwithtau*}
\en

We thus find
\eq
\log\sqrt{\rho^\alpha_x\rho^\alpha_{x^*}}=
\beta F(\alpha)-\beta\Hn_x
+\frac{1}{4}\bigl\{
\sbkt{\Psi}_{\sts\to x}-\sbkt{\Psi}_{x^*\to}+
\sbkt{\Psi}_{\sts\to x^*}-\sbkt{\Psi}_{x\to}
\bigl\}+O(\epsilon^3),
\en
which means
\eq
\beta\{\sbkt{\Hn}^\alpha_\sts-F(\alpha)\}-S_\mathrm{sym}[\bsrho^\alpha]
=\frac{1}{4}\sum_{x\in\calS}\rho^\alpha_x\bigl\{
\sbkt{\Psi}_{\sts\to x}-\sbkt{\Psi}_{x^*\to}+
\sbkt{\Psi}_{\sts\to x^*}-\sbkt{\Psi}_{x\to}
\bigl\}+O(\epsilon^3).
\en
Since the left-hand side is $S(\alpha)-S_\mathrm{sym}[\bsrho^\alpha]$, we shall show that the right-hand side is $O(\epsilon^3)$.
To do this, rewrite the right-hand side as
\eqa
\frac{1}{4}\sum_{x\in\calS}&
\,\rho^\alpha_x\bigl\{
\sbkt{\Psi}_{\sts\to x}-\sbkt{\Psi}_{x^*\to}+
\sbkt{\Psi}_{\sts\to x^*}-\sbkt{\Psi}_{x\to}
\bigl\}
\nl
&=\frac{1}{4}\sum_{x\in\calS}\rho^\alpha_x\bigl\{
-\sbkt{\Psi}_{\sts\to x}-\sbkt{\Psi}_{x^*\to}+
\sbkt{\Psi}_{\sts\to x^*}+\sbkt{\Psi}_{x\to}
\bigl\}
\nl
&=
\frac{1}{4}\sum_{x\in\calS}(\rho^\alpha_x-\rho^\alpha_{x^*})
\bigl\{\sbkt{\Psi}_{x\to}+\sbkt{\Psi}_{\sts\to x^*}\bigl\}.
\ena
To get the second line, we noted that $(1/2)\sum_x\rho^\alpha_x\bigl\{\sbkt{\Psi}_{\sts\to x}-\sbkt{\Psi}_{x\to}\bigl\}=0$ for any $\tau$, and subtracted this from the first line.
We clearly have $\rho^\alpha_x-\rho^\alpha_{x^*}=O(\epsilon)$.
To bound the other term, we note that the corresponding equilibrium process satisfies
\eq
\sbkt{\Psi^{(\alpha)}}^{(\alpha_\mathrm{eq})}_{x\to}+\sbkt{\Psi^{(\alpha)}}_{\sts\to x^*}^{(\alpha_\mathrm{eq})}=0,
\en
which is an easy consequence of (properly rewritten version of) \rlb{DFTsym}.
Since $\Psi=O(\epsilon)$, we see that $\sbkt{\Psi}_{x\to}+\sbkt{\Psi}_{\sts\to x^*}=O(\epsilon^2)$.
This leads us to the desired (heuristic) estimate $S(\alpha)-S_\mathrm{sym}[\bsrho^\alpha]=O(\epsilon^3)$.

\subsection{An illustrative example}
\label{s:toy}
In order to demonstrate that one can never expect a general extended Clausius inequality, we here discuss an oversimplified model of heat conduction.
The analysis also illustrates  the role of the symmetrized Shannon entropy in the extended Clausius relation.

We consider a system of a single particle on a chain of length $L$ whose  left and  right ends are attached to heat baths with the inverse temperatures $\beta_1$ and $\beta_2$, respectively.
The particle performs a back-and-forth motion between the two ends, carrying energy from one end to the other.

The state of the model is specified as $x=(j,v,k)\in\calS$, where $j\in\{1,\ldots,L\}$ denotes the position, $v\in\{+,-\}$ the velocity, and $k\in\{1,\ldots,n\}$ the internal degree of freedom of the particle.
We define $x^*=(j,-v,k)$.
The particle has the internal energy $u_k$ when it is in the state $k$.

\begin{figure}[btp]
\begin{center}
\includegraphics[width=10cm]{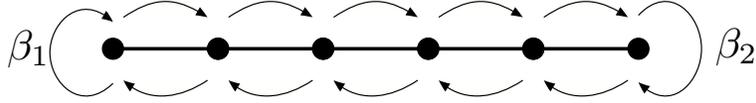}
\end{center}
\caption[dummy]{
The transition rates for the simple model of heat conduction.
The particle performs a back-and-forth stochastic motion between the two ends of the chain which are attached to distinct heat baths.}
\label{fig:toy}
\end{figure}

The model parameter is $\alpha=\beot$.
We define the transition rates as follows (Fig.~\ref{fig:toy}).
Let $\lambda>0$.
We first set
\eqg
R^\beot_{(j,+,k)\to(j+1,+,k)}=\lambda,
\quad(j=1,\ldots,L-1),
\\
R^\beot_{(j,-,k)\to(j-1,-,k)}=\lambda,
\quad(j=2,\ldots,L),
\eng
for any $k$, which describe the one-way-motion of the particle.
Note that $k$ does not change when the particle moves.
At the ends of the chain we set
\eqg
R^\beot_{(0,-,k)\to(0,+,\ell)}=
\lambda\,\frac{e^{-\beta_1u_\ell}}{Z(\beta_1)},
\\
R^\beot_{(L,+,k)\to(L,-,\ell)}=
\lambda\,\frac{e^{-\beta_2u_\ell}}{Z(\beta_2)},
\eng
for any $k$ and $\ell$. 
These represent the processes where the particle is bounced back and thermalized at the ends.
We defined $Z(\beta)=\sum_{k=1}^ne^{-\beta u_k}$.
We set $R^\beot_\xty=0$ for other combinations.
Then the escape rate \rlb{ladef} is $\lambda^\beot_x=\lambda$ for all $x\in\calS$.

From the definition \rlb{thetadef0*}, we see that the only nonvanishing components of the entropy production $\theta^\beot_\xty$ are
\eq
\theta^\beot_{(0,-,k)\to(0,+,\ell)}=
\log\frac{R^\beot_{(0,-,k)\to(0,+,\ell)}}
{R^\beot_{(0,-,\ell)\to(0,+,k)}}
=\beta_1\,(u_k-u_\ell),
\en
and
\eq
\theta^\beot_{(L,+,k)\to(L,-,\ell)}=\beta_2\,(u_k-u_\ell).
\en
Note that $u_k-u_\ell$ is the energy transferred from the particle to the bath.

It must be clear that the model is almost trivial.
The stationary distribution is readily obtained as
\eq
\rho^\beot_{(j,v,k)}=\frac{1}{2L}\biggl\{
\delta_{v,+}\,\frac{e^{-\beta_1u_k}}{Z(\beta_1)}
+\delta_{v,-}\,\frac{e^{-\beta_2u_k}}{Z(\beta_2)}
\biggr\}.
\lb{rhotoy}
\en
Here the position of the particle is distributed uniformly on the chain, and the internal degree of freedom has the inverse temperature $\beta_1$ or $\beta_2$ when the particle is moving to the right or to the left, respectively.

Let the average (internal) energy at the inverse temperature $\beta$ be
\eq
u(\beta):=
\frac{1}{Z(\beta)}\sum_{k=1}^n u_k\,e^{-\beta u_k}.
\en
In the NESS, the average (internal) energy of the particle is $u(\beta_1)$ or $u(\beta_2)$ when it is moving to the right or to the left, respectively.
The average heat (or energy) current from the bath to the particle at $L=1$ is thus $\{u(\beta_1)-u(\beta_2)\}\lambda/(2L)$, where $\lambda/(2L)$ is the rate by which the particle is bounced back at $L=1$.
By also considering the current at $j=L$, the entropy production rate (in the baths) in NESS \rlb{sigmasta} is found to be
$\sigma^\beot_\mathrm{st}=-(\beta_1-\beta_2)\{u(\beta_1)-u(\beta_2)\}\lambda/(2L)\ge0$.

From the stationary distribution \rlb{rhotoy}, we can explicitly compute its Shannon and symmetrized Shannon entropies as
\eq
S_\mathrm{Sh}[\bsrho^\beot]=\frac{\beta_1\,u(\beta_1)+\beta_2\,u(\beta_2)}{2}+\frac{\log Z(\beta_1)+\log Z(\beta_2)}{2}+\log(2L),
\lb{Shtoy}
\en
and
\eq
S_\mathrm{sym}[\bsrho^\beot]=\frac{\beta_1+\beta_2}{2}\,
\frac{u(\beta_1)+u(\beta_2)}{2}+\frac{\log Z(\beta_1)+\log Z(\beta_2)}{2}+\log(2L).
\lb{Ssymtoy}
\en

Let us consider the step protocol \rlb{step} with $\alpha=\beot$ and $\alpha'=(\beta_1',\beta_2')$.
The excess entropy production in the step protocol is easily evaluated by comparing the entropy productions in this process and another process that starts at $t=0$ with the stationary distribution corresponding to $(\beta_1',\beta_2')$.
The two processes differ only in the distribution of the internal energy of the particle at $t=0$, and they become identical after the particle is bounced back by one of the ends for the first time.
From this observation one immediately finds that the excess entropy production is given by\footnote{
One can also compute the expectation value of the entropy production rate explicitly as
$\sum_x\sigma_x^{(\beta_1',\beta_2')}\,p_x(t)
=\sigma_\mathrm{st}^{(\beta_1',\beta_2')}
+[\beta_1'\{u(\beta_1)-u(\beta_1')\}
+\beta_2'\{u(\beta_2)-u(\beta_2')\}]
(2L)^{-1}\sum_{j=1}^L\lambda^jt^{j-1}e^{-\lambda t}$ for $t\ge0$.
It decays exponentially to the steady value.
}
\eq
\sbkt{\Theta_\mathrm{ex}^\hata}^\hata_{\sts\too}
=
\frac{1}{2}
\Bigl[\beta_1'\bigl\{u(\beta_2)-u(\beta_2')\bigr\}
+\beta_2'\bigl\{u(\beta_1)-u(\beta_1')\bigr\}\Bigr].
\lb{Thetaextoy}
\en

Let us start by confirming the validity of the extended Clausius relation \rlb{ExClexcess} or \rlb{ECL03D}.
Writing $\beta'_k=\beta_k+\Db_k$ for $k=1,2$, and $\delta=\max\{|\Db_1|,|\Db_2|\}$, we find from \rlb{Thetaextoy} that
\eqa
\sbkt{\Theta_\mathrm{ex}^\hata}^\hata_{\sts\too}
=&-\frac{1}{2}\bigl\{\beta_2\,u'(\beta_1)\,\Db_1+\beta_1\,u'(\beta_2)\,\Db_2\bigr\}+O(\delta^2)
\nl
=&-\frac{1}{2}\,\beta\,u'(\beta)\,(\Db_1+\Db_2)
+\frac{1}{4}\bigl\{u'(\beta)-\beta u''(\beta)\bigr\}\,
(\Db_1-\Db_2)\,\epsilon
\nl&+\oesd+O(\delta^2),
\lb{Textoyexp}
\ena
where, we wrote $u'(\beta)=du(\beta)/d\beta$.
In the second line, we set $\beta=(\beta_1+\beta_2)/2$ and $\epsilon=\beta_1-\beta_2$.
The expression \rlb{Textoyexp} should be compared with the difference between the entropies (defined as $S\beot:=S_\mathrm{sym}[\bsrho^\beot]$), which is
\eqa
S(\beta_1',\beta_2')-S\beot
=&\frac{1}{4}\Bigl[
-\bigl\{u(\beta_1)-u(\beta_2)\bigr\}(\Db_1-\Db_2)
\nl&\hspace{11pt}+
(\beta_1+\beta_2)\bigl\{u'(\beta_1)\,\Db_1+u'(\beta_2)\,\Db_2\bigr\}
\Bigr]+O(\delta^2)
\nl
=&\frac{1}{2}\,\beta\,u'(\beta)\,(\Db_1+\Db_2)
-\frac{1}{4}\bigl\{u'(\beta)-\beta u''(\beta)\bigr\}\,
(\Db_1-\Db_2)\,\epsilon
\nl&+\oesd+O(\delta^2).
\ena
We thus have\footnote{
The term $\oesd$ is nonvanishing.
} $S(\beta_1',\beta_2')-S\beot=-\sbkt{\Theta_\mathrm{ex}^\hata}^\hata_{\sts\too}+\oesd+O(\delta^2)$, which is the extended Clausius relation \rlb{ExClexcess}, \rlb{ECL03D}.
We note that the difference between the Shannon entropies
\eq
S_\mathrm{Sh}[\bsrho^{(\beta_1',\beta_2')}]-S_\mathrm{Sh}[\bsrho^\beot]
=
\frac{1}{2}\bigl\{
\beta_1\,u'(\beta_1)\,\Db_1+\beta_2\,u'(\beta_2)\,\Db_2
\bigr\}+O(\delta^2),
\en
differs from $-\sbkt{\Theta_\mathrm{ex}^\hata}^\hata_{\sts\too}$ by $O(\epsilon\,\delta)$.

To see that the corresponding {\em in}\/equality is impossible we go onto compute the $O(\delta^2)$ term explicitly to get
\eqa
S(\beta_1',\beta_2')-S\beot+\sbkt{\Theta_\mathrm{ex}^\hata}^\hata_{\sts\too}
=&-\frac{1}{2}\,u'(\beta)\,\Db_1\,\Db_2+
\frac{u''(\beta)}{8}\bigl\{(\Db_1)^2-(\Db_2)^2\bigr\}\epsilon
\nl&
+\oesd+O(\delta^3).
\ena
Let us set, for simplicity, $|\Db_1|=|\Db_2|$.
Then the right-hand side becomes
\newline
$-(1/2)\,u'(\beta)\,\Db_1\,\Db_2+\oesd+O(\delta^3)$.

We first note that the term $-(1/2)\,u'(\beta)\,\Db_1\,\Db_2$ vanishes if we use the quasi-static protocol (which connects $\beot$ and $(\beta_1',\beta_2')$) as in \rlb{manysteps}, \rlb{manysteps2}.
This term therefore represents the effect of the sudden operation, i.e., the step protocol.
Now recall that the  extended Clausius inequality \rlb{ECLIneq} would imply 
$S(\beta_1',\beta_2')-S\beot+\sbkt{\Theta_\mathrm{ex}^\hata}^\hata_{\sts\too}\ge\oesd+O(\delta^3)$.
But, since $u'(\beta)<0$, we have $-(1/2)\,u'(\beta)\,\Db_1\,\Db_2<0$ provided that $\Db_1\,\Db_2<0$.
This provides a concrete counterexample to the extended Clausius inequality.

\bigskip\bigskip\bigskip
 
It is a pleasure to thank
Hisao Hayakawa,
Masato Itami,
Nobuyasu Ito,
Chris Jarzynski,
Gianni Jona-Lasinio,
Joel Lebowitz,
Christian Maes,
Karel Netocny,
Yoshi Oono,
Glenn Paquette,
Takahiro Sagawa,
Keiji Saito,
Herebert Spohn,
and
Akira Shimizu
for valuable discussions.
The present study was supported by KAKENHI Nos. 22340109, 23540435, and 25103002,
by the JSPS Core-to-Core program ``Non-equilibrium dynamics of soft-matter and information'', and partially by JSPS and Leading Research
Organizations, namely NSERC, ANR, DFG, RFBR, RCUK and NSF as Partner
Organizations under the G8 Research Councils Initiative for Multilateral
Research Funding.



\begin{thebibliography}{10}

\bibitem{OP}
Y. Oono and M. Paniconi,
{\em Steady state thermodynamics}\/,
Prog. Theor. Phys. Suppl.
\textbf{130}, 29-44 (1998).


\bibitem{KNST1}
T. S. Komatsu, N. Nakagawa, S.-I. Sasa and H. Tasaki,
{\em Steady State Thermodynamics for Heat Conduction --- Microscopic Derivation}\/,
Phys. Rev. Lett. {\bf 100}, 230602 (2008).\\ {\tt arXiv:0711.0246}

\bibitem{KNST2}
T. S. Komatsu, N. Nakagawa, S.-I. Sasa and H. Tasaki,
{\em Entropy and Nonlinear Nonequilibrium Thermodynamic Relation for Heat Conducting Steady States}\/,
J. Stat. Phys. {\bf 142}, 127--153 (2011).\\ {\tt arXiv:1009.0970}

\bibitem{Landauer}
R. Landauer, {\em $dQ= TdS$ far from equilibrium}\/, Phys. Rev. A18, 255-266 (1978).

\bibitem{Ruelle}
D. Ruelle,
{\em Extending the definition of entropy to nonequilibrium steady states}\/,
Proc. Natl. Acad. Sci. U.S.A.  {\bf 100}, 3054--3058 (2003).\\
{\tt arXiv:cond-mat/0303156}


\bibitem{SaitoTasaki}
K. Saito and H. Tasaki,
{\em Extended Clausius Relation and Entropy for Nonequilibrium Steady States in Heat Conducting Quantum Systems}\/,
J. Stat. Phys. {\bf 145}, 1275--1290 (2011).\\
{\tt arXiv:1105.2168}

\bibitem{SagawaHayakwa}
T. Sagawa and H. Hayakawa,
{\em Geometrical expression of excess entropy production}\/,
Phys. Rev. E {\bf 84}, 051110 (2011).\\
{\tt arXiv:1109.0796}

\bibitem{YugeSagawaSugitaHayakawa}
T. Yuge, T. Sagawa, A. Sugita, and H. Hayakawa,
{\em Geometrical Excess Entropy Production in Nonequilibrium Quantum Systems}, J. Stat. Phys. {\bf 153}, 412--441 (2013).\\
{\tt arXiv:1305.5026}

\bibitem{HatanoSasa}
T. Hatano and S.-I. Sasa,
{\em Steady-State Thermodynamics of Langevin Systems}\/,
Phys. Rev. Lett. {\bf 86}, 3463 (2001).\\
{\tt arXiv:cond-mat/0010405}

\bibitem{BGJLL2012}
L. Bertini, D. Gabrielli, G. Jona-Lasinio, and C. Landim,
{\em Thermodynamic transformations of nonequilibrium states}\/,
J. Stat. Phys. {\bf 149}, 773--802 (2012).\\
{\tt arXiv:1206.2412}

\bibitem{BGJLL2013}
L. Bertini, D. Gabrielli, G. Jona-Lasinio, and C. Landim,
{\em Clausius Inequality and Optimality of Quasistatic Transformations for Nonequilibrium Stationary States}\/,
Phys. Rev. Lett. {\bf 110}, 020601 (2013).\\
{\tt arXiv:1208.1872}



\bibitem{JonaLasinio2014}
G. Jona-Lasinio,
{\em Thermodynamics of stationary states}\/,
J. Stat. Mech.  P02004 (2014).




\bibitem{MaesNetocny2012}
C. Maes and K. Netocny,
{\em A nonequilibrium extension of the Clausius heat theorem}\/,
J. Stat. Phys. {\bf 154}, 188--203 (2014).\\
{\tt arXiv:1206.3423}

\bibitem{Sasa2013}
S.-I. Sasa,
{\em Possible extended forms of thermodynamic entropy}\/,
J. Stat. Mech.  P01004 (2014).\\
{\tt arXiv:1309.7131}

\bibitem{SasaTasaki}
S.-I. Sasa and Hal Tasaki,
{\em Steady State Thermodynamics}\/,
J. Stat. Phys. {\bf 125}, 125--224 (2006).\\
{\tt arXiv:cond-mat/0411052}



\bibitem{PradhanAmannSeifert2010}
P. Pradhan, C.P. Amann, and U. Seifert,
{\em Nonequilibrium Steady States in Contact: Approximate Thermodynamic Structure and Zeroth Law for Driven Lattice Gases}\/,
Phys. Rev. Lett. {\bf 105}, 150601 (2010).\\
{\tt arXiv:1002.4349}


\bibitem{PradhanAmannSeifert2011}
P. Pradhan, C.P. Amann, and U. Seifert,
{\em Approximate thermodynamic structure for driven lattice gases in contact}\/,
Phys. Rev. E {\bf 84}, 041104 (2011).\\
{\tt arXiv:1107.5434}


\bibitem{DickmanMotai}
R. Dickman and R. Motai,
{\em Inconsistencies in steady state thermodynamics}\/, 
preprint (2014).\\
{\tt arXiv:1401.1678}





\bibitem{Boksenbojm}
E. Boksenbojm, C. Maes, K. Netocny, and J. Pesek,
{\em Heat capacity in nonequilibrium steady states}\/,
Euro Phys. Lett. {\bf 96}, 40001 (2011).\\
{\tt arXiv:1109.3054}

\bibitem{Mandal}
D. Mandal,
{\em Nonequilibrium heat capacity}\/,
Phys. Rev. E {\bf 88}, 062135 (2013).\\
{\tt arXiv:1311.7176}


\bibitem{Nakagawa2014}
N. Nakagawa,
{\em Universal exact expression for adiabatic pumping in terms of non-equilibrium steady states}\/,
preprint (2014).\\
{\tt arXiv:1401.4242}


\bibitem{BSGJLL01}
{L. Bertini, A. De Sole, D. Gabrielli, G. Jona-Lasinio, and C. Landim},
{\em Fluctuations in Stationary Nonequilibrium States of Irreversible Processes}\/,
Phys. Rev. Lett. {\bf 87}, 040601 (2001).\\
{\tt arXiv:cond-mat/0104153}

\bibitem{BSGJLL06}
L. Bertini, A. De Sole, D. Gabrielli, G. Jona-Lasinio, and C. Landim,
{\em Large deviation approach to non equilibrium processes in stochastic lattice gases}\/,
Bull. Braz. Math. Soc. {\bf 37}, 611--643 (2006).\\
{\tt arXiv:math/0602557}

\bibitem{BSGJLL09}
L. Bertini, A. De Sole, D. Gabrielli, G. Jona-Lasinio, and C. Landim,
{\em Towards a Nonequilibrium Thermodynamics: A Self-Contained Macroscopic Description of Driven Diffusive Systems}\/,
J. Stat. Phys.  {\bf 135}, 857--872  (2009).\\
{\tt arXiv:0807.4457}

\bibitem{JonaLasinio10}
G. Jona-Lasinio,
{\em From Fluctuations in Hydrodynamics to Nonequilibrium Thermodynamics}\/,
Prog. Theor. Phys. Suppl. {\bf 184}, 262 (2010).\\
{\tt arXiv:1003.4164}


\bibitem{DLS01}
B. Derrida, J. L. Lebowitz, and E. R. Speer,
{\em Free Energy Functional for Nonequilibrium Systems: An Exactly Solvable Case}\/,
Phys. Rev. Lett. {\bf 87}, 150601 (2001).\\
{\tt arXiv:cond-mat/0105110}

\bibitem{DLS03}
B. Derrida, J. L. Lebowitz, and E. R. Speer,
{\em Exact Large Deviation Functional of a Stationary Open Driven Diffusive System: The Asymmetric Exclusion Process}\/,
J. Stat. Phys. {\bf 110}, 775--810 (2003).\\
{\tt arXiv:cond-mat/0205353}

\bibitem{BD}
T. Bodineau and B. Derrida,
{\em Current Fluctuations in Nonequilibrium Diffusive Systems: An Additivity Principle}\/,
Phys. Rev. Lett. {\bf 92}, 180601 (2004).\\
{\tt arXiv:cond-mat/0402305}

\bibitem{LiebYngvason}
E. H. Lieb and J. Yngvason,
{\em The entropy concept for non-equilibrium states}\/, 
Proc. R. Soc. A 2013 {\bf 469}, 20130408 (2013).\\
{\tt http://rspa.royalsocietypublishing.org/content/469/2158/20130408}

\bibitem{Seifert12}
U. Seifert, 
{\em Stochastic thermodynamics, fluctuation theorems, and molecular machines}\/,
Rep. Prog. Phys. {\bf 75}, 126001 (2012).\\
{\tt arXiv:1205.4176}



\bibitem{SpinnyFord21012}
R. E. Spinny and I. J. Ford,
{\em Non-equilibrium thermodynamic systems with odd and even variables}\/,
Phys. Rev. Lett. {\bf 108}, 170603 (2012).\\
{\tt arXiv:1201.0904}



\bibitem{Jarzynski}
{C. Jarzynski},
{\em Nonequilibrium Equality for Free Energy Differences}\/,
Phys. Rev. Lett. {\bf 78}, 2690 (1997).\\
{\tt arXiv:cond-mat/9610209}


\bibitem{Crooks99}
G.E. Crooks,
{\em Entropy production fluctuation theorem and the nonequilibrium work relation for free energy differences}\/,
Phys. Rev. E {\bf 60}, 2721 (1999).\\
{\tt arXiv:cond-mat/9901352}

\bibitem{Crooks00}
G.E. Crooks,
{\em Path-ensemble averages in systems driven far from equilibrium}\/,
Phys. Rev. E {\bf 61}, 2361, (2000).\\
{\tt arXiv:cond-mat/9908420}


\bibitem{Seifert05}
U. Seifert,
{\em Entropy Production along a Stochastic Trajectory and an Integral Fluctuation Theorem}\/,
Phys. Rev. Lett. {\bf 95}, 040602 (2005).\\
{\tt arXiv:cond-mat/0503686}


\bibitem{MaesNetocny03}
C. Maes and K. Netocny,
{\em Time-Reversal and Entropy}\/,
J. Stat. Phys. {\bf 110}, 269--310 (2003).\\
{\tt arXiv:cond-mat/0202501}

\bibitem{BaiesiMaesWynants}
M. Baiesi, C. Maes, and B. Wynants,
{\em Fluctuations and response of nonequilibrium states}\/,
Phys. Rev. Lett. {\bf 103}, 010602 (2009).\\
{\tt arXiv:0902.3955}

\bibitem{BaertsBasuMaesSafaverdi}
P. Baerts, U. Basu, C. Maes, and S. Safaverdi,
{\em The frenetic origin of negative differential response}\/,
Phys. Rev. E {\bf 88}, 052109 (2013).\\
{\tt arXiv:1308.5613}


\bibitem{Nakagawa}
N. Nakagawa,
{\em Work Relation and the Second Law of Thermodynamics in Nonequilibrium Steady States}\/,
Phys. Rev. E {\bf 85}, 051115 (2012).\\
{\tt arXiv:1109.1374}

\bibitem{KN}
{T. S. Komatsu and N. Nakagawa},
{\em An expression for stationary distribution in nonequilibrium steady state}\/,
Phys. Rev. Lett. {\bf 100}, 030601 (2008).\\
{\tt arXiv:0708.3158}



\bibitem{MaesNetocny10}
C. Maes and K. Netocny,
{\em Rigorous meaning of McLennan ensembles}\/,
J. Math. Phys. {\bf 51}, 015219 (2010).\\
{\tt arXiv:0911.1032}



\bibitem{Dembo}
A. Dembo and O. Zeitouni,
{\em Large Deviations Techniques and Applications}\/,
(Springer, New York, 1998).



\bibitem{KNST3}
T. S. Komatsu, N. Nakagawa, S.-I. Sasa and H. Tasaki,
{\em Representation of Nonequilibrium Steady States in Large Mechanical Systems}\/,
J. Stat. Phys. {\bf 134}, 401--423 (2009).\\
{\tt arXiv:0805.3023}

\bibitem{CoverThomas}
T. M. Cover and J. A. Thomas,
{\em Elements of Information Theory}\/,
(Wiley-Interscience, 2006)


\bibitem{ItamiSasa}
M. Itami and S.-I. Sasa, 
{\em Nonequilibrium Statistical Mechanics for Adiabatic
Piston Problem}\/, in preparation.

\end{thebibliography}
\end{document}